\documentclass[%
reprint,
preprintnumbers,
amsmath,amssymb,
aps,
prb,
]{revtex4-2}

\usepackage{orcidlink}

\usepackage{graphicx}
\usepackage{dcolumn}
\usepackage{bm}
\usepackage{hyperref}
\usepackage{amsmath,amssymb}
\hypersetup{colorlinks=true, linkcolor=blue, citecolor=blue, filecolor=blue,	urlcolor=blue}
\usepackage[normalem]{ulem}
\usepackage{makecell} 
\usepackage{placeins}
\begin{document}

\title{Spectroscopic evidence of Kondo resonance in 3$d$ van der Waals ferromagnets}

\author{Deepali Sharma\orcidlink{0009-0000-2309-4386}}
\affiliation{Department of Physics, Indian Institute of Science Education and Research Bhopal, Bhopal Bypass Road, Bhauri, Bhopal 462066, India}%

\author{Neeraj Bhatt\orcidlink{0009-0002-8012-139X}}
\affiliation{Department of Physics, Indian Institute of Science Education and Research Bhopal, Bhopal Bypass Road, Bhauri, Bhopal 462066, India}%

\author{Asif Ali\orcidlink{0000-0001-7471-8654}}
\affiliation{Department of Physics, Indian Institute of Science Education and Research Bhopal, Bhopal Bypass Road, Bhauri, Bhopal 462066, India}%

\author{Rajeswari Roy Chowdhury\orcidlink{0000-0001-9693-9255}}
\affiliation{Department of Physics, Indian Institute of Science Education and Research Bhopal, Bhopal Bypass Road, Bhauri, Bhopal 462066, India}%

\author{Chandan Patra\orcidlink{0009-0001-1338-4977}}
\affiliation{Department of Physics, Indian Institute of Science Education and Research Bhopal, Bhopal Bypass Road, Bhauri, Bhopal 462066, India}%

\author{Ravi Prakash Singh\orcidlink{0000-0003-2548-231X}}
\affiliation{Department of Physics, Indian Institute of Science Education and Research Bhopal, Bhopal Bypass Road, Bhauri, Bhopal 462066, India}%

\author{Ravi Shankar Singh\orcidlink{0000-0003-3654-2023}}
\email{rssingh@iiserb.ac.in}
\affiliation{Department of Physics, Indian Institute of Science Education and Research Bhopal, Bhopal Bypass Road, Bhauri, Bhopal 462066, India}%


\begin{abstract}

Two-dimensional van der Waals (vdW) ferromagnets drive the advancement in spintronic applications and enable the exploration of exotic magnetism in low-dimensional systems. The entanglement of dual $-$ localized and itinerant $-$ nature of electrons lies at the heart of the correlated electron systems giving rise to exotic ground state properties such as complex magnetism, heavy fermionic behavior, Kondo lattice formation, \textit{etc}. Through temperature-dependent electronic structure of vdW ferromagnets, (Co$_{x}$Fe$_{1-x}$)$_{3}$GeTe$_{2}$, probed using high-resolution photoemission spectroscopy and density functional theory combined with dynamical mean field theory (DFT+DMFT), we provide direct evidence of the emergence of Kondo resonance peak driven by complex interplay between localized and itinerant electrons. 
In overall agreement with experimental electronic structure and magnetic properties, DFT+DMFT also reveals finite spin band splitting well beyond $T_{C}$. Core levels, valence band photoemission spectra together with DFT+DMFT spectral functions reveal insignificant change across $T_{C}$ indicating non-Stoner magnetism in (Co$_{x}$Fe$_{1-x}$)$_{3}$GeTe$_{2}$.
Our results provide a way forward to the understanding of complex interplay between electronic structure, exotic magnetism and heavy fermionic behavior leading to Kondo scenerio in 3$d$ vdW ferromagnets.
 
\end{abstract}

\maketitle

\section{introduction}

Two-dimensional (2D) magnetic van der Waals (vdW) systems have emerged as a material platform for the exploration of fundamental physics due to intrinsic magnetic order persisting down to monolayer limit, tunable spin textures, and exotic quantum phases \cite{mak_probing_2019,*gibertini_magnetic_2019,*burch_magnetism_2018}. Tailoring of electronic and magnetic properties in these materials through external influences such as pressure, field, proximity effects, \textit{etc}. unlocks new possibilities for control and functionality in spintronic applications \cite{mak_probing_2019,*gibertini_magnetic_2019,*burch_magnetism_2018}. A diverse array of magnetic orders have been observed in vdW transition metal chalcogens, ranging from antiferromagnetic insulator$-$ (Mn/Fe/Ni)PS$_{3}$ \cite{MnPs3,FePS3,NiPS3}; ferrimagnetic semiconductor$-$ Mn$_{3}$Si$_{2}$Te$_{6}$ \cite{Mn3⁢Si2⁢Te6}; ferromagnetic semiconductor$-$ VSe$_{2}$ \cite{bonilla_strong_2018}, Cr$_{2}$Ge$_{2}$Te$_{6}$ \cite{gong_discovery_2017}; ferromagnetic metal$-$ Fe$_{x}$GeTe$_{2}$, (\textit{x} = 3, 4, 5) \cite{chen_magnetic_2013,Fe4GeTe2,Fe5GeTe2}, Fe$_{3}$GaTe$_{2}$ \cite{Fe3GaTe2} \textit{etc}. In addition to exotic magnetism, Kondo behavior in heavy fermionic state, akin to that in \textit{f}-electron systems, arises due to interplay of localised moment and itinerant electrons in transition metal compounds, for example, YFe$_2$Ge$_2$ \cite{YFe2Ge2} and \textit{A}Fe$_2$As$_2$ (\textit{A} = K, Rb, Cs) \cite{AFe2} and also oxides CaCu$_3$Ru$_4$O${_{12}}$ \cite{CaCu3Ru4O12}, LiV$_2$O$_4$ \cite{LiV2O4,LiV2O4-2}, and chalcogenides VTe$_2$ \cite{VTe2}, FeTe \cite{FeTe}, and Fe$_3$GeTe$_2$ \cite{FGT,rana_spin-polarized_2022}, \textit{etc}.

Fe$_{3}$GeTe$_{2}$ (FGT) showcasing non-Stoner magnetism with intriguing interplay amongst itinerant and localised 3\textit{d}-electrons \cite{FGT,zhu_electronic_2016,xu_signature_2020,zhang_emergence_2018} along with Kondo lattice behavior in heavy fermionic state \cite{zhang_emergence_2018,zhao_kondo_2021,rana_spin-polarized_2022}, appears as fascinating system for the exploration of physics near the quantum critical point (QCP). QCP separates the Kondo lattice ground state from the long range magnetically ordered state by varying the non-thermal parameters such as field, pressure, substitution, \textit{etc}. \cite{Coleman2005,*QCPNT,*QCP1}. Signature of Kondo behavior has been observed from transport exhibiting a resistivity upturn, along with Fano-resonance feature in the tunneling spectroscopy, at low-temperatures (\textit{T}) \cite{zhao_kondo_2021,rana_spin-polarized_2022}. 
Kondo singlet ground state, arising due to the screening of the localized moment by the conduction electrons, manifests as Kondo resonance peak in the local density of states (DOS) at Fermi energy ($E_{F}$), much below the characteristic Kondo temperature, $T_{K}$ \cite{kondoCaCu,*kondoCeCu}. The strength of hybridization between local moments and conduction electrons drives the system from long range magnetically ordered state to the Kondo singlet ground state through a QCP \cite{swapnil1,*Patil_2010,*Patil_2012}. While this can also be achieved by varying the local moment or conduction electron density. Co substitution at Fe site in FGT offers an exciting platform to explore the quantum criticality, since, the local moment decreases and conduction electron density increases with increase in Co concentration as revealed by magnetic and transport measurements \cite{chowdhury_unconventional_2021,FGTarxiv,Cobalt_doping, *FeCo3GeTe2}. Thus, (Co$_{x}$Fe$_{1-x}$)$_{3}$GeTe$_{2}$ (Co$_{x}$FGT) provides an exciting platform to investigate the coexistence and competition between complex magnetism and Kondo singlet ground sates.

\begin{figure*}
\centerline{\includegraphics[width=0.82\textwidth]{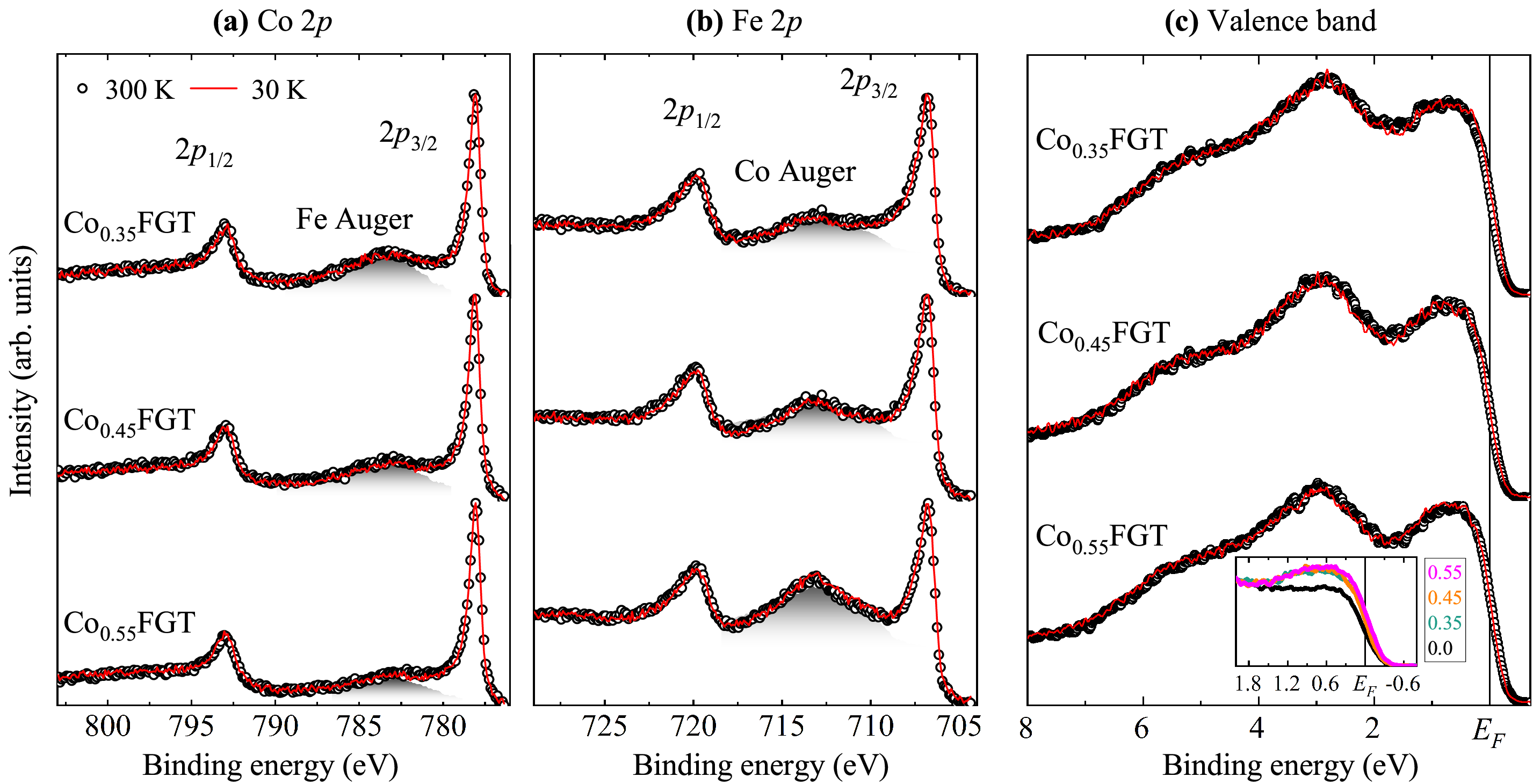}}
	\caption{Photoemission spectra of (a) Co 2$p$ and (b) Fe 2$p$ core levels, and (c) valence band, collected at 300 K (black symbols) and 30 K (red lines) using Al $K_{\alpha}$ for Co$_{x}$FGT (\textit{x} = 0.35, 0.45 and 0.55), in vertically stacked order. Shaded regions in (a) and (b) schematically depicts intensity of Fe and Co Auger, respectively. Inset of (c) shows the comparison of valence band spectra in the vicinity of $E_{F}$ at 300 K for Co$_{x}$FGT (\textit{x} = 0, 0.35, 0.45 and 0.55).}\label{fig:fig1}
\end{figure*}

In this Letter, through high-resolution photoemission spectroscopy and density functional theory combined with dynamical mean field theory (DFT+DMFT), we unveil the emergence of Kondo resonance peak around $E_{F}$ with lowering-\textit{T} in Co$_{x}$FGT. We provide the spectroscopic evidence of the evolution of Kondo peak demonstrating the system approaching QCP with increasing \textit{x}. Further, insignificant change in the core levels and overall valence band photoemission spectra across magnetic phase transition in conjunction with electronic and magnetic properties along with evolution of spin bands obtained within $T$-dependent DFT+DMFT reveal non-Stoner magnetism in Co$_{x}$FGT.

  \section{methodology} 
  
High-quality single crystals of Co$_{x}$FGT where, \textit{x} = 0, 0.35, 0.45 and 0.55, were prepared using chemical vapor transport method with I$_{2}$ as a transport agent \cite{chowdhury_unconventional_2021}. Direction dependent magnetic measurements reveal the average Curie temperature, $T_{C}$ to be 206 $\pm$ 4 K, 80 $\pm$ 1 K, 60 $\pm$ 1 K and 35 $\pm$ 1 K, for FGT, Co$_{0.35}$FGT, Co$_{0.45}$FGT and Co$_{0.55}$FGT, respectively. Photoemission spectroscopic measurements were performed using Al $K_{\alpha}$ (1486.6 eV) and He {\scriptsize I} (21.2 eV) radiations (energy), with total instrumental resolutions of about 300 meV and 5 meV, respectively. The Fermi level ($E_{F}$) position and energy resolution for the radiations were obtained by measuring the Fermi-edge of a clean polycrystalline silver sample at 30 K. Multiple single crystals of Co$_x$FGT were cleaved \textit{in-situ} at base pressure better than 4$\times$10$^{-11}$ mbar, to ascertain the cleanliness of the sample surface and reproducibility of the data.
Electronic structure calculations were performed using experimental lattice parameters of FGT having two formula units (f.u.) per unit cell \cite{FGT,chowdhury_unconventional_2021}. Full-potential linearized augmented plane wave method as implemented in \textsc{Wien2k} \cite{Wien2k2020} was used for the DFT calculations. Generalized gradient approximation of Purdew-Burke-Ernzerhof \cite{perdew_generalized_1996} was employed for the exchange correlation functional. We modified our calculations to account for the effect of Co substitution at Fe in FGT. In particular we change atomic mass and electron count to 26 + $x$ for Fe site in FGT to mimic Co$_x$FGT \cite{Liu,Lee,C9NR10171C}. 18 $\times$ 18 $\times$ 3 $k$-mesh within the first Brillouin zone was used for the self-consistent calculations. The energy and charge convergence criteria were set to $10^{-4}$ eV and $10^{-4}$ electronic charge per f.u., respectively. eDMFT code \cite{haule_dynamical_2010} was used for the charge self-consistent DFT+DMFT calculations, with two impurity problems for Fe/Co I and Fe/Co II, and all five Fe/Co 3$d$ orbitals (forming three non-degenerate groups, $d_{{z}^{2}}$, $d_{{{x}^{2}}-{{y}^{2}}}$/$d_{xy}$ and $d_{xz}$/$d_{yz}$) were considered into correlated sub-space. The continuous-time quantum Monte Carlo (CTQMC) impurity solver \cite{haule_quantum_2007} was used with ‘\textit{exact}’ as the double counting correction method \cite{haule_exact_2015}. The Hubbard-$U$ = 5.0 eV and Hund's coupling-$J$ = 0.9 eV was opted for both the Fe/Co I and Fe/Co II, in accordance with the earlier reports \cite{FGT,kim_large_2018,*ghosh_unraveling_2023,zhu_electronic_2016}. Analytical continuation was performed using maximum entropy method \cite{haule_dynamical_2010} to calculate self-energy on the real axis. 

\section{results and discussions}

The crystal structure of FGT consists of vdW bonded layers where, each layer comprises of Fe$_{3}$Ge slab (with inequivalent two Fe I and one Fe II) sandwiched between Te layers \cite{FGT}. Theoretical calculations, based on DFT, suggest preferential occupation at the Fe II site for substitution of Co in FGT \cite{ZHANG2021127219} however, experimental observations reveal homogeneously substituted Co \cite{roy_chowdhury_modification_2022,FeCo3GeTe2,CHANG2025177837}.
Core level photoemission spectra collected using Al $K_{\alpha}$ radiation at 300 K (open symbols) and 30 K (lines) are shown in Fig. \ref{fig:fig1} (a) and (b) corresponding to Co 2\textit{p} and Fe 2\textit{p} spectral regions, respectively.
Co 2\textit{p} core level spectra of Co$_{0.35}$FGT consists of two sharp peaks corresponding to 2\textit{p}$_{3/2}$ and 2\textit{p}$_{1/2}$ appearing at 778 eV and 793 eV binding energy (BE), respectively. Evident asymmetric lineshape towards higher BE is typically observed in metallic systems due to electron-hole excitations at the $E_{F}$ \cite{SDoniach_1970}. A broad feature at $\sim$ 783 eV BE corresponding to Fe Auger (L$_{3}$M$_{45}$M$_{45}$) is visible and can be seen in FGT as well (see Fig. S1 of Supplemental Material (SM) \cite{SM}). The peak position and width of Co 2\textit{p}$_{3/2}$ (and 2\textit{p}$_{1/2}$) remain very similar with the increasing \textit{x} (see Fig. S2 (a) of SM \cite{SM}). As expected the intensity of the Fe Auger feature decreases while going from Co$_{0.35}$FGT to Co$_{0.55}$FGT. The Fe 2\textit{p} core level spectral region for Co$_{0.35}$FGT shown in top panel of Fig. \ref{fig:fig1} (b) consists of two asymmetric peaks corresponding to spin orbit split 2\textit{p}$_{3/2}$ (706.7 eV BE) and 2\textit{p}$_{1/2}$ (719.8 eV BE) components alongwith a broad feature at around 713 eV BE corresponding to Co Auger (L$_{2}$M$_{23}$M$_{45}$) \cite{Handbook}, whose intensity gradually increases with increasing \textit{x} in Co$_{x}$FGT. Similar to Co, Fe 2\textit{p} core level spectra also do not show any change in the peak position and width with the increasing \textit{x} in Co$_{x}$FGT (see Fig. S2 (b)  of SM \cite{SM}). Notably, the widths of Fe 2\textit{p}$_{3/2}$ and Co 2\textit{p}$_{3/2}$ core level spectra are found to be increased while going from Co$_{x}$FGT to iron and cobalt, respectively (see Fig. S2 of SM \cite{SM}). FGT has lower saturation moment (1.56 $\mu_{B}$/Fe) as compared to ferromagnetic iron (2.2 $\mu_{B}$/Fe). Although ferromagnetic cobalt has saturation moment of about 1.6 $\mu_{B}$/Co, the saturation moment (as well as $T_{C}$) decreases drastically with increasing \textit{x} in Co$_{x}$FGT (0.6 $\mu_{B}$/Fe(Co) for \textit{x} = 0.55) suggests that Co acts as a non-magnetic substitution in FGT \cite{roy_chowdhury_modification_2022,Coexchangesplitt,FGT}. These observations are also commensurate with substantially smaller width of Co 2\textit{p} feature in Co$_{x}$FGT than that of Co metal, since, larger width of the core levels arises due to finite exchange splitting resulting in observed dichroism in the core level spectra in the case of magnetic systems \cite{FGT,PhysRevB.55.11488,PhysRevResearch.3.013151,IndranilSarkar,Coexchangesplitt}. Interestingly, no appreciable change was observed in both the core levels for all \textit{x} values while going across the magnetic transition, from 300 K to 30 K, suggesting that the exchange splitting does not change appreciably across $T_{C}$.

The valence band spectra collected using Al $K_{\alpha}$ radiation at 300 K and 30 K for Co$_{x}$FGT have been shown in Fig. \ref{fig:fig1} (c). Three discernible features at around 5.5 eV, 3 eV and below 2 eV BE with large intensity at $E_{F}$ is clearly evident for Co$_{x}$FGT in 300 K spectra (open symbols), similar to FGT \cite{FGT}. While various feature positions in the valence band spectra remain very similar, the spectral weight within 2 eV BE enhances when going from FGT to Co$_{x}$FGT for all \textit{x} (see inset). Since the Fe 3\textit{d} as well as Co 3\textit{d} states primarily appear below 2 eV BE, enhancement in the spectral weight with increasing \textit{x} can be due to larger photoionization cross-section ratio ($\sim$ 1.68) of Co 3\textit{d} with Fe 3\textit{d} states \cite{yeh_atomic_1985}. As shown in the inset of Fig. \ref{fig:fig1} (c), the integrated intensity below 2 eV BE increases by $\sim$ 16.5 \%, $\sim$ 18 \% and $\sim$ 20 \% while going from \textit{x} = 0 to 0.35, 0.45 and 0.55, respectively. This non-uniform increase in the intensity of the 3\textit{d} states deviate from the expected photoionization cross-section ratio, indicating intrinsic electronic origin. Interestingly, similar to the core level spectra the overall valence band spectra also do not change appreciably across $T_{C}$ while going from 300 K to 30 K (lines) for all the \textit{x} values suggesting that the non-Stoner magnetism may also be applicable in Co$_{x}$FGT, similar to its parent compound FGT \cite{FGT}. 

For further understanding of the evolution of electronic structure with increasing \textit{x} in Co$_{x}$FGT, we employed DFT where the electron counts were increased to account for Co substitution in FGT. Non-magnetic DFT results reveal a gradual shift of total DOS towards higher BE alongwith decrease in DOS($E_{F}$) (see Fig. S3 of SM \cite{SM}), which are in sharp contrast with experimental observations where feature positions remain similar and spectral intensity at $E_{F}$ increases upon Co substitution.
\begin{figure}
\centerline{\includegraphics[width=0.5\textwidth]{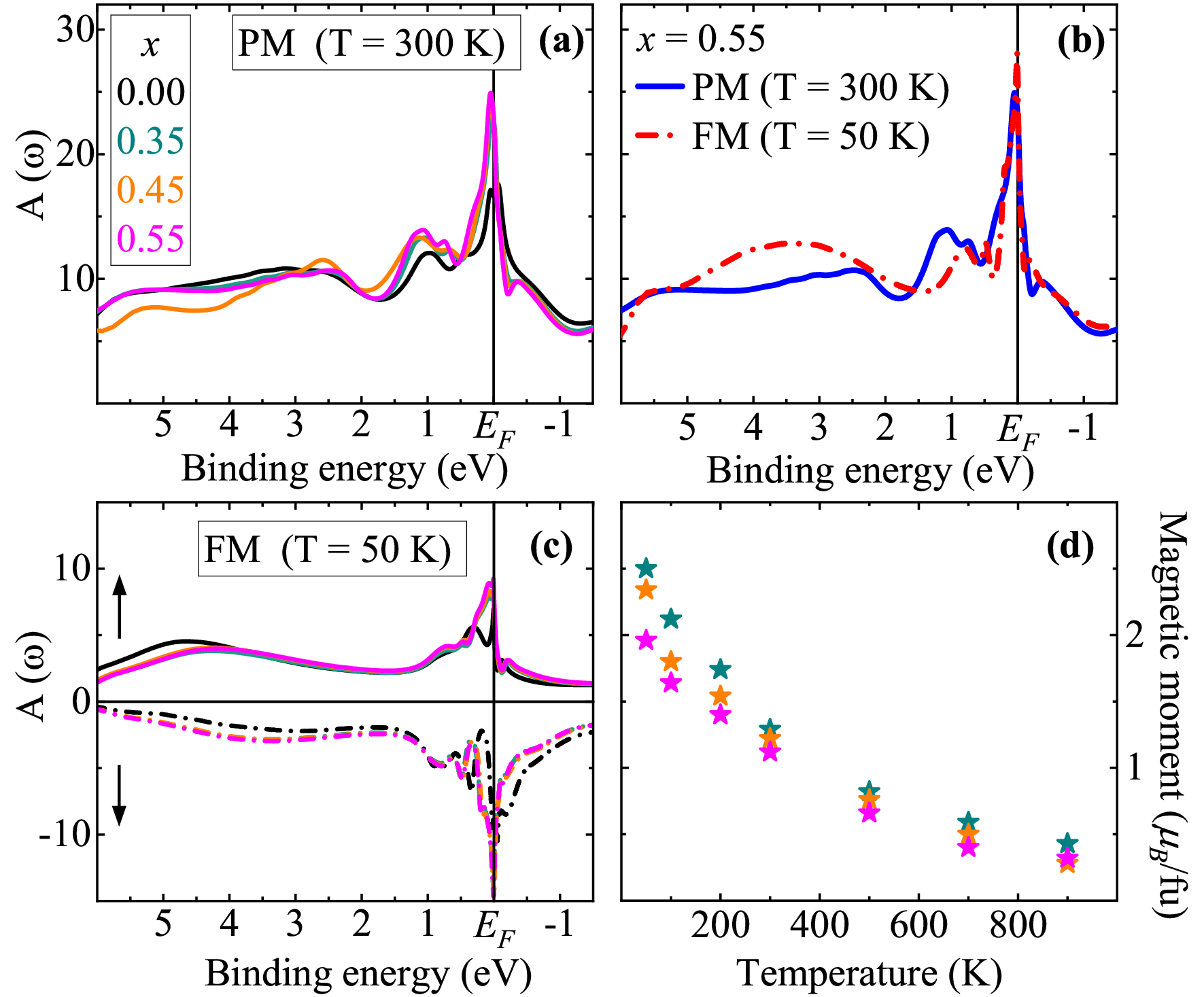}}
	\caption{ DFT+DMFT results. (a) Total spectral functions for Co$_{x}$FGT (\textit{x} = 0, 0.35, 0.45 and 0.55) in PM phase. (b) Comparison of total spectral function for Co$_{0.55}$FGT in the PM and FM phases. (c) Spin-up (solid  lines) and spin-down (dot-dash lines) spectral functions of Fe/Co in FM phase. (d) Magnetic moment at various $T$ within FM phase.}\label{fig:fig2}
\end{figure}

\begin{figure}
\centerline{\includegraphics[width=0.5\textwidth]{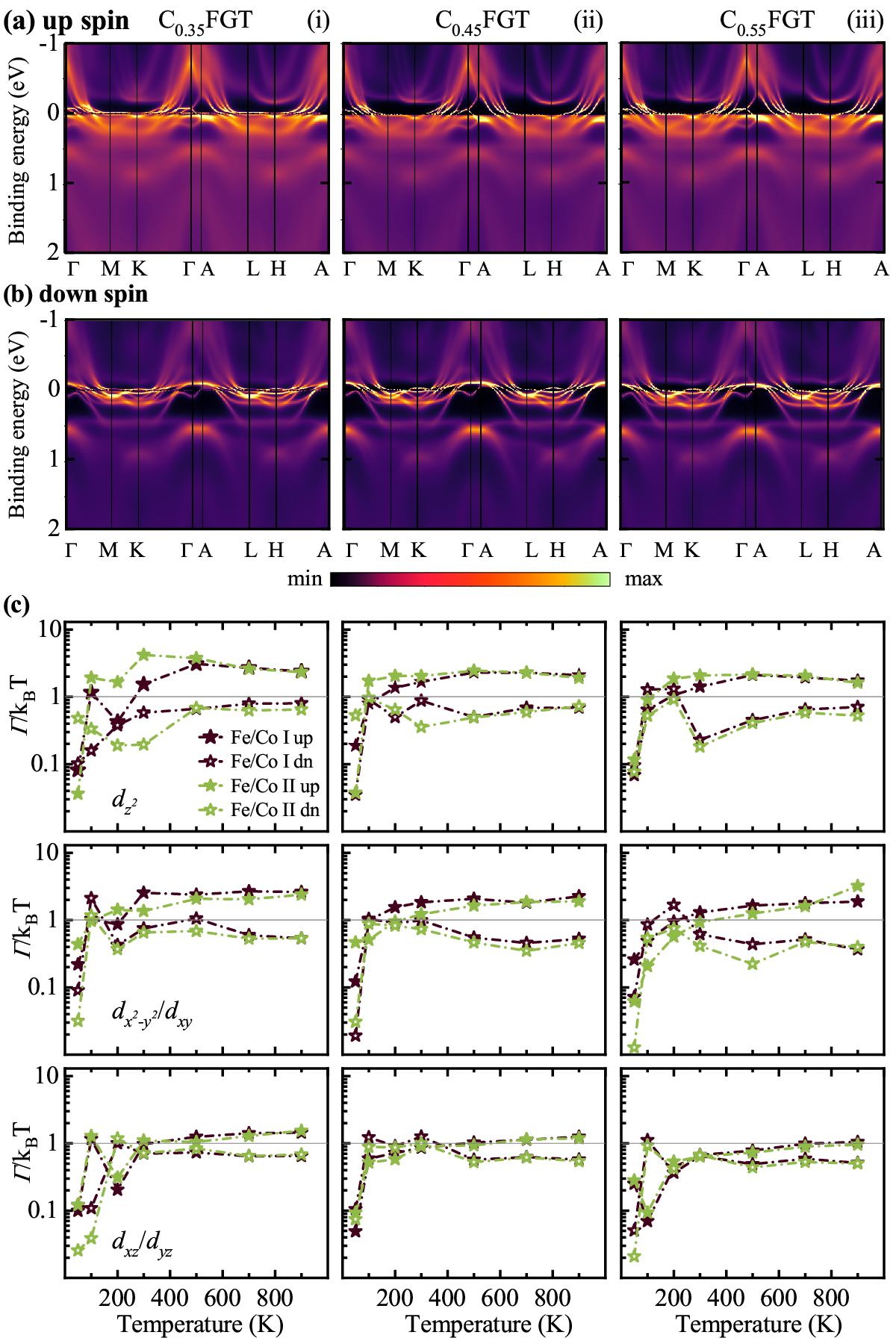}}
	\caption{$k$-resolved spectral functions for (a) spin-up and (b) spin-down, along high symmetry directions at \textit{T} = 50 K obtained using FM DFT+DMFT calculations for Co$_{x}$FGT (\textit{x} = 0.35, 0.45, and 0.55). (c) $\varGamma$/\textit{k}$_{B}T$ for \textit{d}$_{z^2}$, $d_{{{x}^{2}}-{{y}^{2}}}$/$d_{xy}$ and $d_{xz}$/$d_{yz}$ orbitals (top to bottom) for both the spin channels and for both the non-degenerate Fe/Co sites at various \textit{T} for Co$_{x}$FGT. Spin-up and spin-down are shown in closed markers and open markers, respectively for Fe/Co I (brown) and Fe/Co II (green).}\label{fig:fig3}
\end{figure} 

Since, DFT+DMFT has been quite successful in accurately describing the complex interplay of electronic structure and magnetism in strongly correlated FGT, thus we show the DFT+DMFT results within paramagnetic (PM) and ferromagnetic (FM) phases in Fig. \ref{fig:fig2} for Co$_{x}$FGT. Total spectral function (A($\omega$)) obtained from PM DFT+DMFT at \textit{T} = 300 K (${\beta}$ = 38.68 eV$^{-1}$) for Co$_{x}$FGT have been shown in Fig. \ref{fig:fig2} (a). For FGT, strongly renormalized 3\textit{d} states appear between $\pm$ 2 eV BE with a peak around $E_{F}$ and a hump at $\sim$ 1 eV BE while states corresponding to Te and Ge appear at higher BE \cite{FGT}. The feature positions remain very similar with a large enhancement of the spectral function at $E_{F}$ for Co$_{0.35}$FGT in comparison to FGT. Interestingly, the spectral weight for both the features do not significantly change while going from Co$_{0.35}$FGT to Co$_{0.45}$FGT and further to Co$_{0.55}$FGT, in a good agreement with the experimental observations (inset of Fig. \ref{fig:fig1} (c)). The local fluctuating moment within PM DFT+DMFT framework is calculated using ${\sum}_{i}~2P{_i} |S{_i}$$^{z}|$ where, $P{_i}$ and $|S{_i}$$^{z}|$ represent the probability and absolute spin moment, respectively. The local fluctuating moment per f.u. was found to be 5.16 $\mu_{B}$, 4.85 $\mu_{B}$, 4.65 $\mu_{B}$, and 4.54 $\mu_{B}$, for FGT, Co$_{0.35}$FGT, Co$_{0.45}$FGT, and Co$_{0.55}$FGT, respectively.
Total spectral function obtained from FM DFT+DMFT results for \textit{T} = 50 K (${\beta}$ = 232.08 eV$^{-1}$) show very similar spectral lineshape with PM DFT+DMFT (\textit{T} = 300 K) result for Co$_{0.55}$FGT, as shown in Fig. \ref{fig:fig2} (b) (see Fig S4. of SM for other values of \textit{x} in Co$_{x}$FGT \cite{SM}). These observations are similar to FGT \cite{FGT} and aligns well with no significant change observed in the overall valence band spectra across the magnetic transition, as seen in Fig. \ref{fig:fig1} (c). Spin-polarised  3\textit{d} spectral function obtained from FM DFT+DMFT has been shown in Fig. \ref{fig:fig2} (c). Similar to the results obtained in PM phase, the spin-polarised spectral functions exhibit a large enhancement in the spectral weight near $E_{F}$ while going from FGT to Co$_{0.35}$FGT, with no significant changes for further increase in Co concentration. Further, the spin-down spectral function shows dominant contribution in the vicinity of $E_{F}$. Interestingly, the magnetic moment as well as exchange splitting obtained from FM DFT+DMFT, reduces approximately to half while going from FGT to Co$_{0.35}$FGT. The drastic change in the experimental valence band spectra as well as magnetic moment with the introduction of 35$\%$ Co in FGT could be successfully captured within DFT+DMFT framework.
The obtained magnetic moment per f.u. (for 50 K FM DFT+DMFT) of about 4.52 $\mu_B$, 2.50 $\mu_B$, 2.34 $\mu_B$, and 1.96 $\mu_B$, for FGT, Co$_{0.35}$FGT, Co$_{0.45}$FGT, and Co$_{0.55}$FGT, respectively, are in close agreement with experimental saturation moment \cite{chowdhury_unconventional_2021}. The magnetic moment decreases monotonically and remains finite even upto 900 K, as shown in Fig. \ref{fig:fig3} (d) due to non-degenerate spin-polarised spectral functions for all the compounds (see Fig S5. of SM \cite{SM}). It is to note that the total energy difference per f.u. between PM and FM phases in DFT+DMFT framework at higher-\textit{T} are less than 3 meV, which is much smaller compared to the thermal energy (e.g., at 900 K, $k_BT$ $\sim$ 75 meV). Similar to the parent compound FGT, overestimation of $T_{C}$ within DFT+DMFT for all Co$_{x}$FGT, where experimental $T_{C}$ ranges from $\sim$ 80 K to $\sim$ 35 K further supports non-Stoner magnetism, where, the temporal and spatial thermal fluctuation leads to disordered moment (itinerant and local both) thereby destroying the long range magnetic order beyond $T_{C}$ \cite{maiti_finite_2002,*APL,FGT}.

\begin{figure*}
\centerline{\includegraphics[width=0.98\textwidth]{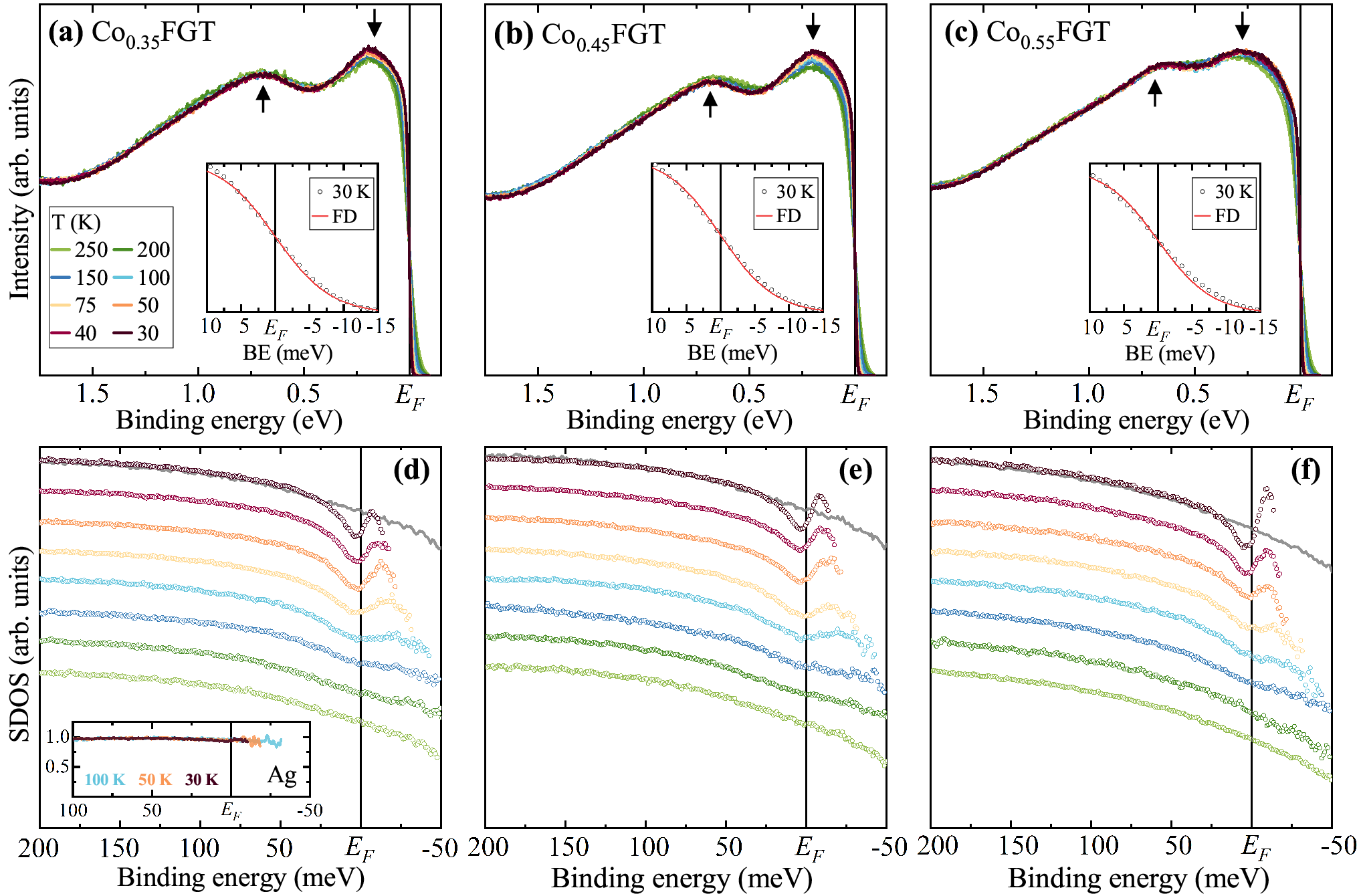}}
	\caption{(a-c) $T$-dependent high-resolution valence band spectra of Co$_{x}$FGT for (a) \textit{x} = 0.35, (b) \textit{x}= 0.45, and (c) \textit{x} = 0.55, while the insets show the comparison of 30 K spectra with resolution broadened FD function around $E_{F}$. (d-f) Corresponding SDOS has been shown (upto 5$k_BT$) in (d), (e), and (f), respectively. SDOS for \textit{T} = 250 K is overlaid (grey lines) with SDOS for \textit{T} = 30 K for all \textit{x}. The inset in  (d) shows the $T$-dependent spectra of Silver as a reference.}\label{fig:fig4}
\end{figure*}

Additionally, $k$-resolved spin-polarized spectral functions along high symmetry directions obtained from FM DFT+DMFT calculations at \textit{T} = 50 K for Co$_{x}$FGT, are shown in Fig. \ref{fig:fig3} (a) and (b). Large number of dispersive bands along with flat bands around $E_F$ are present in both the spin channels for Co$_{x}$FGT. Interestingly, bands in the close vicinity of $E_{F}$ show significant spin dependent evolution with increasing Co concentration in Co$_{x}$FGT, where the spin-up bands around $E_F$ shift toward lower BE whereas, the spin-down bands shift toward higher BE, resulting in an overall decrease in exchange splitting (see Fig S5. of SM \cite{SM}) and hence magnetic moment (Fig. \ref{fig:fig2} (d)), in agreement with experimental observations \cite{chowdhury_unconventional_2021}. Notably, the bands become sharper/coherent in the vicinity of $E_F$ for both the spin channels with increasing \textit{x} in Co$_{x}$FGT. Quasiparticle scattering rate, $\varGamma$ (inverse of the lifetime), was obtained for Co$_{x}$FGT for both the spins of inequivalent magnetic sites, using $\varGamma$$_{l,s}$ = $-(m^{*}$/$m_{_{b}}$)$_{l,s}$ $^{-1}$ Im${\Sigma}$$_{l,s}$ ($i{\omega}$${\rightarrow 0^{+}}$), where ( $m^{*}$/$m_{_{b}}$)$_{l,s}$ = 1 $-~{\partial}$ Im${\Sigma}$$_{l,s}$($i{\omega}$)/${\partial}$(${\omega}$)$|_{\omega}$$_{\rightarrow 0^{+}}$ for $l$ orbitals, and \textit{s} spins \cite{FGT} and is shown in Fig. \ref{fig:fig3} (c) for all orbitals. The imaginary part of self energy at the zero frequency limit and its derivative were obtained using fourth order polynomial fit for the first six data points. It is to be noted that, (\textit{i}) {$\varGamma$} approaches to zero faster with decreasing-\textit{T} for the spin-up channel in comparison to the spin-down channel, for both the Fe/Co sites in all the doped compounds and (\textit{ii}) {$\varGamma$} overall decreases with increase in Co concentration. Further, the $\varGamma$ for spin-down channels remain below $k_BT$ (thermal energy) for both the inequivalent Fe/Co sites irrespective of the \textit{T} in all the compounds, suggesting the coherent scenario throughout, while the spin-up channel tends to achieve an incoherent-coherent crossover with lowering-\textit{T} for all the orbitals, as shown in Fig. \ref{fig:fig3} (c), therefore, implying significant influence of spin-differentiated electron correlation in all the doped compounds. Calculated mass enhancement factor, $\frac{m^*}{m_{\text{DFT}}}$, and Sommerfeld coefficient$, \gamma_{_{\text{DMFT}}}$, unveil an increasing trend with lowering-\textit{T}, suggesting Co$_{x}$FGT to be a heavy fermionic system (see Fig. S7 of SM \cite{SM}), similar to FGT \cite{FGT}.

\begin{figure}
\centerline{\includegraphics[width=0.42\textwidth]{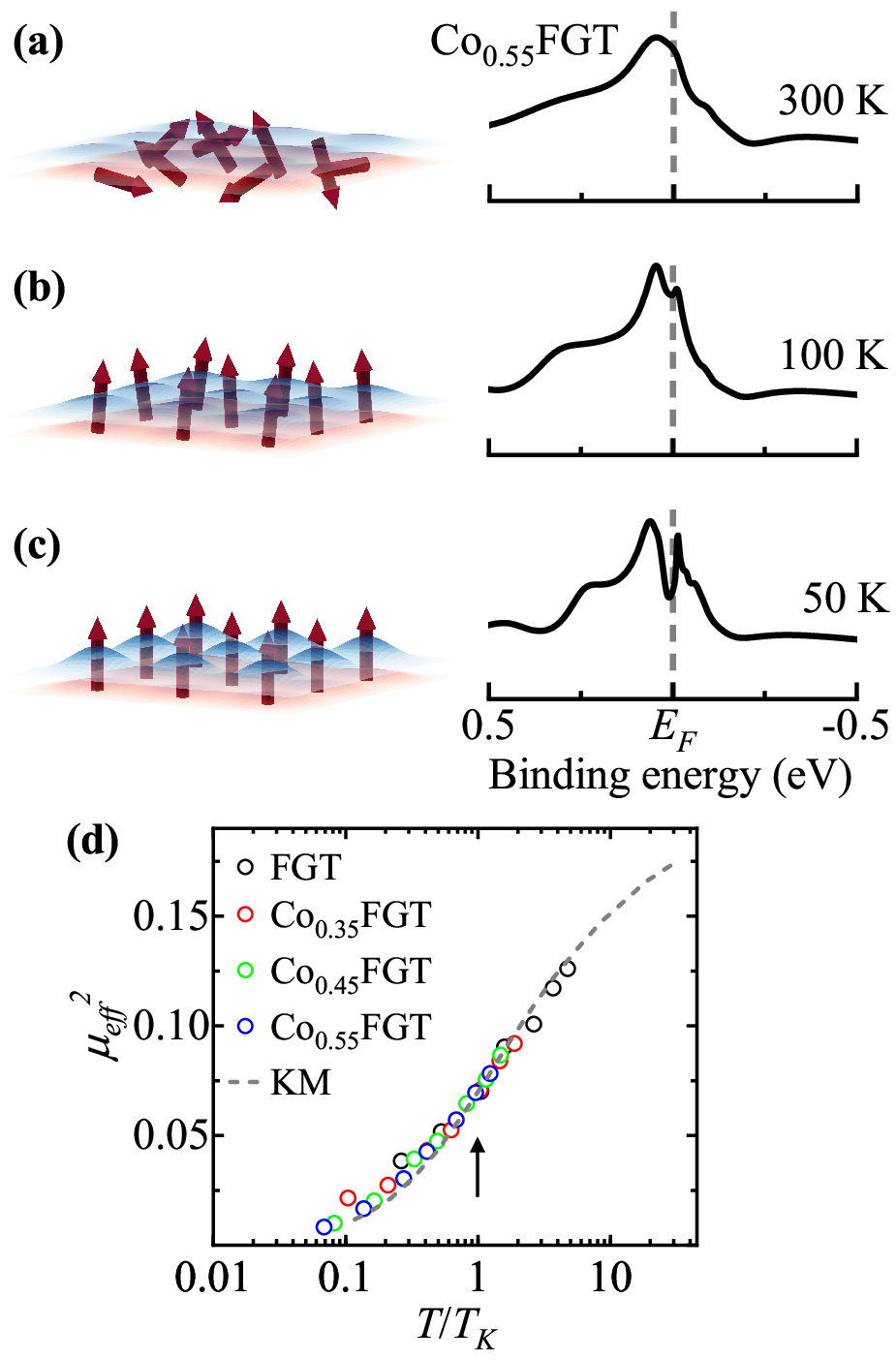}}
	\caption{(a)-(c) Schematic representation of \textit{T}-dependent evolution of the Kondo lattice along with DFT+DMFT spectral functions for Co$_{0.55}$FGT. Red and blue sheets represent up spin and down spin itinerant electrons and thick arrows represent localized moments. (d)  Effective local moment, $\mu_{eff}^2$ obtained from DFT+DMFT for Co$_{x}$FGT (where, $x$ = 0, 0.35, 0.45 and 0.55) at various $T$. The $T_K$ is obtained from the fit to the universal dependence for the Kondo model (KM, grey dash lines) \cite{PhysRevB.21.1003}.}\label{fig:fig5}
\end{figure}

To dwelve deeper insights into the \textit{T}-dependent behavior of electronic states, we show the high-resolution valence band spectra around $E_F$, for Co$_{x}$FGT at various \textit{T} collected using He {\scriptsize I} radiation in Fig. \ref{fig:fig4} (a-c). All the spectra have been collected in the normal emission and are normalized by the total integrated intensity below 2.0 eV BE. The high-\textit{T} spectra for Co$_{0.35}$FGT exhibit an intense feature near $E_F$ and a broad feature at $\sim$ 0.7 eV BE along with a shoulder structure at around 1.2 eV BE, as shown in Fig. \ref{fig:fig4} (a). Interestingly, while going from Co$_{0.35}$FGT to Co$_{0.45}$FGT and further to Co$_{0.55}$FGT, the intense feature near $E_F$ (indicated by down arrow) shifts toward higher BE whereas the broad feature (indicated by up arrow) remains at similar energy position, resulting in an overall decrease in the energy separation between the features. The \textit{T}-dependent spectra of all three samples remain very similar with respect to feature positions suggesting insignificant change in the overall electronic structure across $T_{C}$. The spectral intensity in the vicinity of $E_F$ exhibit a  Fermi Dirac (FD) function like behavior with varying \textit{T}. Interestingly, when normalized at $E_F$ the 30 K spectra unveils finitely lower (higher) intensity than the instrumental resolution broadened FD function in occupied (unoccupied) region, as shown in the insets of Fig. \ref{fig:fig4} (a-c). The deviation in the occupied region remains similar while it increases in unoccupied region with increasing \textit{x} in Co$_{x}$FGT suggesting a peak like behavior of the spectral function. To recover the intrinsic spectral DOS (SDOS) in the vicinity of $E_F$ we obtained the SDOS by dividing the photoemission intensity (normalized at 200 meV BE) with the resolution broadened FD function at respective-\textit{T} \cite{FGT,LSNO,*Reddy_2019}. It is to note here that the features such as Kondo resonance peak in SDOS appearing above $E_F$ in various 3\textit{d} and 4\textit{f} based heavy fermionic systems are accessible only after removing the contribution of FD function \cite{CaCu3Ru4O12,LiV2O4-2,Hufner2007}.
Thus obtained \textit{T}-dependent SDOS for Co$_{0.35}$FGT, shown in Fig. \ref{fig:fig4} (d) (stacked along ordinate axis and presented up to 5$k_BT$ above $E_F$), remains very similar in larger BE range ($>$ 25 meV) while a dip like structure evolves monotonously around $E_F$ with lowering-\textit{T}. Interestingly, for \textit{T} $\leq$ 100 K the minima of the dip moves below $E_F$ and an increasing trend of the SDOS at $E_F$ is observed, demonstrating an intriguing evolution of a peak-like structure positioned at 7 $\pm$ 1 meV above $E_F$ at 30 K. For reference, similarly obtained $T$-dependent SDOS for Ag remain constant for all the $T$, as shown in the inset. A very similar scenario of monotonously increasing dip below $E_F$ and peak evolving above $E_F$ for \textit{T} $\leq$ 100 K is observed for Co$_{0.45}$FGT as well as for Co$_{0.55}$FGT, as shown in Fig. \ref{fig:fig4} (e) and (f), respectively. It is to note that the observed peak becomes significantly pronounced at 30 K for the largest Co concentration \textit{i.e.} Co$_{0.55}$FGT. The suppression of the SDOS with lowering-\textit{T} in the vicinity of $E_F$ could primarily be attributed to the intrinsic disorder present in Co substituted FGT, as also seen in various disordered correlated metals \cite{LSNO,*Reddy_2019,Disordered,*Electronic,*Altshuler,*ALTSHULER1979115}. Interestingly, within the disorder induced dip a Kondo-like peak evolves at lower-\textit{T} and gets pronounced for higher Co substitution in FGT. 

While quite common in the rare-earth and actinide intermetallics, heavy fermionic or Kondo behavior is rarely encountered in \textit{d}-electron systems \cite{Heavy-fermion,*f-electron,*heavyfermionsreview}. 
FGT, inspite being a 3\textit{d} system, shows coexistence of localized and itinerant electrons, driving it to a heavy-Fermionic state at low-$T$ \cite{FGT,zhang_emergence_2018}. Further, a resistivity upturn and a Fano resonance feature in the scanning tunneling spectra, supports the Kondo scenario with $T_K$ $\sim$ 190 K \cite{zhang_emergence_2018,zhao_kondo_2021}. 
At high-$T$ the localized moments remain unscreened, interacting weakly with the conduction electrons, as schematically depicted in the Fig. \ref{fig:fig5} (a) with representative spectral function for Co$_{0.55}$FGT at high-\textit{T}. For \textit{T} $\lesssim$  $T_K$, a partially screened state is formed (Fig. \ref{fig:fig5} (b)) where the spectral function exhibits a peak like structure just above $E_F$, which further evolves to exhibit a sharp Kondo resonance peak for \textit{T} $\ll$ $T_K$ \cite{kondoCaCu,*kondoCeCu}, where the localized moment is largely screened by the conduction electrons, forming a many-body Kondo singlet state within the partially ordered state in case of Co$_{x}$FGT (Fig. \ref{fig:fig5} (c)). This scenario is well captured in experimental results where increasing \textit{x} in Co$_{x}$FGT, leads to decreased moment due to effective screening and hence, increased Kondo temperature, as manifested by pronounced Kondo resonance peak at low-\textit{T} in case of Co$_{0.55}$FGT. In addition, the effective local moment, $\mu_{eff}^2$ ($T$*$\chi$, where, $\chi$ is the electronic susceptibility) was obtained from PM DFT+DMFT calculations. Figure \ref{fig:fig5} (d) shows the $\mu_{eff}^2$ scaled with respect to the Kondo model \cite{PhysRevB.21.1003} (grey dash line)  as a function of $T$/$T_K$ for Co$_{x}$FGT ($x$ = 0, 0.35, 0.45, and 0.55). A good agreement of DFT+DMFT results are obtained with Kondo model for different values of $x$ in   Co$_{x}$FGT. The obtained $T_K$ of $\sim$ 190 K for FGT is in very good agreement with experimental observation \cite{zhang_emergence_2018}. The $T_K$ increases further with increasing Co substitution to $\sim$ 480 K, $\sim$ 610 K, and $\sim$ 730 K, for Co$_{0.35}$FGT, Co$_{0.45}$FGT and Co$_{0.55}$FGT, respectively.

Notably, atomic-scale contacts of Fe, Co, and Ni, despite being strong bulk FM, reveal Kondo behavior in electronic transport with $T_K ^{Fe}$ $\textless$ $T_K ^{Co}$ $\textless$ $T_K ^{Ni}$ as the screened magnetic moment decreases \cite{kondoFCN}. Interestingly, with increasing \textit{x} in Co$_{x}$FGT beyond 0.55, the magnetic moment and $T_C$ reduces further \cite{tian_domain_2019} and the system approaches PM state towards Co$_{3}$GeTe$_{2}$ (CGT). DFT+DMFT results reveal very similar ground state energy for PM and FM states with extremely small exchange splitting (also magnetic moment) for CGT (details in SM \cite{SM}). Our results suggest that QCP to be achieved with increasing \textit{x} in Co$_{x}$FGT while going from magnetically ordered to paramagnetic Kondo singlet state, and it would be interesting to explore CGT experimentally and possibly with further electron doping.\\


\section{conclusion}

We comprehensively investigate the electronic structure of Co$_{x}$FGT using photoemission spectroscopy and DFT+DMFT. Unchanged core levels and valence band spectra across $T_{C}$ and DFT+DMFT obtained spin band splitting (also magnetic moment) of the 3\textit{d} magnetic ion persisting well beyond $T_{C}$, collectively indicates non-Stoner magnetism in Co$_{x}$FGT. The overall valence band spectra along with increase in spectral intensity near $E_{F}$ with increasing \textit{x} in Co$_{x}$FGT are well captured within DFT+DMFT. Monotonic evolution of disorder induced dip in the SDOS around $E_{F}$ with lowering-\textit{T} is manifested in the high-resolution spectra for all the systems. Below 100 K, a peak like structure evolves at 7 $\pm$ 1 meV above $E_{F}$, providing the spectroscopic evidence of the emergence of Kondo resonance peak driven by complex interplay between localized and itinerant electrons and further corroborated within DFT+DMFT framework. Interestingly, pronounced Kondo resonance peak indicative of increased $T_{K}$ together with decreasing magnetic moment, suggest that Co$_{x}$FGT approaches QCP with increasing \textit{x}. Our results provide a pathway to understand the rich interplay between strongly correlated electronic structure, complex magnetism, and Kondo physics in the vicinity of QCP in 3\textit{d} vdW ferromagnets.\\

\section*{acknowledgments}

D. S. and N. B. acknowledge the Council of Scientific and Industrial Research (CSIR), Government of India, for financial support with Award No. 09/1020(0198)/2020-EMR-I and 09/1020(0177)/2019-EMR-I, respectively. R.R.C. acknowledges the Department of Science and Technology (DST), Government of India, for financial support. R.P.S. acknowledges the SERB, India, for Core Research Grant No. CRG/2023/000817. We gratefully acknowledge the use of HPC facility and CIF at IISER Bhopal.  


\begin{thebibliography}{71}%
\makeatletter
\providecommand \@ifxundefined [1]{%
 \@ifx{#1\undefined}
}%
\providecommand \@ifnum [1]{%
 \ifnum #1\expandafter \@firstoftwo
 \else \expandafter \@secondoftwo
 \fi
}%
\providecommand \@ifx [1]{%
 \ifx #1\expandafter \@firstoftwo
 \else \expandafter \@secondoftwo
 \fi
}%
\providecommand \natexlab [1]{#1}%
\providecommand \enquote  [1]{``#1''}%
\providecommand \bibnamefont  [1]{#1}%
\providecommand \bibfnamefont [1]{#1}%
\providecommand \citenamefont [1]{#1}%
\providecommand \href@noop [0]{\@secondoftwo}%
\providecommand \href [0]{\begingroup \@sanitize@url \@href}%
\providecommand \@href[1]{\@@startlink{#1}\@@href}%
\providecommand \@@href[1]{\endgroup#1\@@endlink}%
\providecommand \@sanitize@url [0]{\catcode `\\12\catcode `\$12\catcode `\&12\catcode `\#12\catcode `\^12\catcode `\_12\catcode `\%12\relax}%
\providecommand \@@startlink[1]{}%
\providecommand \@@endlink[0]{}%
\providecommand \url  [0]{\begingroup\@sanitize@url \@url }%
\providecommand \@url [1]{\endgroup\@href {#1}{\urlprefix }}%
\providecommand \urlprefix  [0]{URL }%
\providecommand \Eprint [0]{\href }%
\providecommand \doibase [0]{https://doi.org/}%
\providecommand \selectlanguage [0]{\@gobble}%
\providecommand \bibinfo  [0]{\@secondoftwo}%
\providecommand \bibfield  [0]{\@secondoftwo}%
\providecommand \translation [1]{[#1]}%
\providecommand \BibitemOpen [0]{}%
\providecommand \bibitemStop [0]{}%
\providecommand \bibitemNoStop [0]{.\EOS\space}%
\providecommand \EOS [0]{\spacefactor3000\relax}%
\providecommand \BibitemShut  [1]{\csname bibitem#1\endcsname}%
\let\auto@bib@innerbib\@empty
\bibitem [{\citenamefont {Mak}\ \emph {et~al.}(2019)\citenamefont {Mak}, \citenamefont {Shan},\ and\ \citenamefont {Ralph}}]{mak_probing_2019}%
  \BibitemOpen
  \bibfield  {author} {\bibinfo {author} {\bibfnamefont {K.~F.}\ \bibnamefont {Mak}}, \bibinfo {author} {\bibfnamefont {J.}~\bibnamefont {Shan}},\ and\ \bibinfo {author} {\bibfnamefont {D.~C.}\ \bibnamefont {Ralph}},\ }\bibfield  {title} {\bibinfo {title} {{Probing and controlling magnetic states in {2D} layered magnetic materials}},\ }\href {https://www.nature.com/articles/s42254-019-0110-y} {\bibfield  {journal} {\bibinfo  {journal} {Nat. Rev. Phys.}\ }\textbf {\bibinfo {volume} {1}},\ \bibinfo {pages} {646} (\bibinfo {year} {2019})}\BibitemShut {NoStop}%
\bibitem [{\citenamefont {Gibertini}\ \emph {et~al.}(2019)\citenamefont {Gibertini}, \citenamefont {Koperski}, \citenamefont {Morpurgo},\ and\ \citenamefont {Novoselov}}]{gibertini_magnetic_2019}%
  \BibitemOpen
  \bibfield  {author} {\bibinfo {author} {\bibfnamefont {M.}~\bibnamefont {Gibertini}}, \bibinfo {author} {\bibfnamefont {M.}~\bibnamefont {Koperski}}, \bibinfo {author} {\bibfnamefont {A.~F.}\ \bibnamefont {Morpurgo}},\ and\ \bibinfo {author} {\bibfnamefont {K.~S.}\ \bibnamefont {Novoselov}},\ }\bibfield  {title} {\bibinfo {title} {{Magnetic {2D} materials and heterostructures}},\ }\href {https://www.nature.com/articles/s41565-019-0438-6} {\bibfield  {journal} {\bibinfo  {journal} {Nat. Nanotechnol.}\ }\textbf {\bibinfo {volume} {14}},\ \bibinfo {pages} {408} (\bibinfo {year} {2019})}\BibitemShut {NoStop}%
\bibitem [{\citenamefont {Burch}\ \emph {et~al.}(2018)\citenamefont {Burch}, \citenamefont {Mandrus},\ and\ \citenamefont {Park}}]{burch_magnetism_2018}%
  \BibitemOpen
  \bibfield  {author} {\bibinfo {author} {\bibfnamefont {K.~S.}\ \bibnamefont {Burch}}, \bibinfo {author} {\bibfnamefont {D.}~\bibnamefont {Mandrus}},\ and\ \bibinfo {author} {\bibfnamefont {J.-G.}\ \bibnamefont {Park}},\ }\bibfield  {title} {\bibinfo {title} {{Magnetism in two-dimensional van der {Waals} materials}},\ }\href {https://www.nature.com/articles/s41586-018-0631-z} {\bibfield  {journal} {\bibinfo  {journal} {Nature}\ }\textbf {\bibinfo {volume} {563}},\ \bibinfo {pages} {47} (\bibinfo {year} {2018})}\BibitemShut {NoStop}%
\bibitem [{\citenamefont {Shiomi}\ \emph {et~al.}(2017)\citenamefont {Shiomi}, \citenamefont {Takashima},\ and\ \citenamefont {Saitoh}}]{MnPs3}%
  \BibitemOpen
  \bibfield  {author} {\bibinfo {author} {\bibfnamefont {Y.}~\bibnamefont {Shiomi}}, \bibinfo {author} {\bibfnamefont {R.}~\bibnamefont {Takashima}},\ and\ \bibinfo {author} {\bibfnamefont {E.}~\bibnamefont {Saitoh}},\ }\bibfield  {title} {\bibinfo {title} {Experimental evidence consistent with a magnon nernst effect in the antiferromagnetic insulator {MnPS$_{3}$}},\ }\href {https://doi.org/10.1103/PhysRevB.96.134425} {\bibfield  {journal} {\bibinfo  {journal} {Phys. Rev. B}\ }\textbf {\bibinfo {volume} {96}},\ \bibinfo {pages} {134425} (\bibinfo {year} {2017})}\BibitemShut {NoStop}%
\bibitem [{\citenamefont {McCreary}\ \emph {et~al.}(2020)\citenamefont {McCreary}, \citenamefont {Simpson}, \citenamefont {Mai}, \citenamefont {McMichael}, \citenamefont {Douglas}, \citenamefont {Butch}, \citenamefont {Dennis}, \citenamefont {Aguilar},\ and\ \citenamefont {Walker}}]{FePS3}%
  \BibitemOpen
  \bibfield  {author} {\bibinfo {author} {\bibfnamefont {A.}~\bibnamefont {McCreary}}, \bibinfo {author} {\bibfnamefont {J.~R.}\ \bibnamefont {Simpson}}, \bibinfo {author} {\bibfnamefont {T.~T.}\ \bibnamefont {Mai}}, \bibinfo {author} {\bibfnamefont {R.~D.}\ \bibnamefont {McMichael}}, \bibinfo {author} {\bibfnamefont {J.~E.}\ \bibnamefont {Douglas}}, \bibinfo {author} {\bibfnamefont {N.}~\bibnamefont {Butch}}, \bibinfo {author} {\bibfnamefont {C.}~\bibnamefont {Dennis}}, \bibinfo {author} {\bibfnamefont {R.~V.}\ \bibnamefont {Aguilar}},\ and\ \bibinfo {author} {\bibfnamefont {A.~R.~H.}\ \bibnamefont {Walker}},\ }\bibfield  {title} {\bibinfo {title} {Quasi-two-dimensional magnon identification in antiferromagnetic {FePS$_{3}$} via magneto-raman spectroscopy},\ }\href {https://doi.org/10.1103/PhysRevB.101.064416} {\bibfield  {journal} {\bibinfo  {journal} {Phys. Rev. B}\ }\textbf {\bibinfo {volume} {101}},\ \bibinfo {pages} {064416} (\bibinfo {year} {2020})}\BibitemShut {NoStop}%
\bibitem [{\citenamefont {Belvin}\ \emph {et~al.}(2021)\citenamefont {Belvin}, \citenamefont {Baldini}, \citenamefont {Ozel}, \citenamefont {Mao}, \citenamefont {Po}, \citenamefont {Allington}, \citenamefont {Son}, \citenamefont {Kim}, \citenamefont {Kim}, \citenamefont {Hwang}, \citenamefont {Kim}, \citenamefont {Park}, \citenamefont {Senthil},\ and\ \citenamefont {Gedik}}]{NiPS3}%
  \BibitemOpen
  \bibfield  {author} {\bibinfo {author} {\bibfnamefont {C.~A.}\ \bibnamefont {Belvin}}, \bibinfo {author} {\bibfnamefont {E.}~\bibnamefont {Baldini}}, \bibinfo {author} {\bibfnamefont {I.~O.}\ \bibnamefont {Ozel}}, \bibinfo {author} {\bibfnamefont {D.}~\bibnamefont {Mao}}, \bibinfo {author} {\bibfnamefont {H.~C.}\ \bibnamefont {Po}}, \bibinfo {author} {\bibfnamefont {C.~J.}\ \bibnamefont {Allington}}, \bibinfo {author} {\bibfnamefont {S.}~\bibnamefont {Son}}, \bibinfo {author} {\bibfnamefont {B.~H.}\ \bibnamefont {Kim}}, \bibinfo {author} {\bibfnamefont {J.}~\bibnamefont {Kim}}, \bibinfo {author} {\bibfnamefont {I.}~\bibnamefont {Hwang}}, \bibinfo {author} {\bibfnamefont {J.~H.}\ \bibnamefont {Kim}}, \bibinfo {author} {\bibfnamefont {J.-G.}\ \bibnamefont {Park}}, \bibinfo {author} {\bibfnamefont {T.}~\bibnamefont {Senthil}},\ and\ \bibinfo {author} {\bibfnamefont {N.}~\bibnamefont {Gedik}},\ }\bibfield  {title} {\bibinfo {title} {Exciton-driven antiferromagnetic metal in a correlated van der waals
  insulator},\ }\href {https://doi.org/10.1038/s41467-021-25164-8} {\bibfield  {journal} {\bibinfo  {journal} {Nat. Commun.}\ }\textbf {\bibinfo {volume} {12}},\ \bibinfo {pages} {4837} (\bibinfo {year} {2021})}\BibitemShut {NoStop}%
\bibitem [{\citenamefont {May}\ \emph {et~al.}(2017)\citenamefont {May}, \citenamefont {Liu}, \citenamefont {Calder}, \citenamefont {Parker}, \citenamefont {Pandey}, \citenamefont {Cakmak}, \citenamefont {Cao}, \citenamefont {Yan},\ and\ \citenamefont {McGuire}}]{Mn3⁢Si2⁢Te6}%
  \BibitemOpen
  \bibfield  {author} {\bibinfo {author} {\bibfnamefont {A.~F.}\ \bibnamefont {May}}, \bibinfo {author} {\bibfnamefont {Y.}~\bibnamefont {Liu}}, \bibinfo {author} {\bibfnamefont {S.}~\bibnamefont {Calder}}, \bibinfo {author} {\bibfnamefont {D.~S.}\ \bibnamefont {Parker}}, \bibinfo {author} {\bibfnamefont {T.}~\bibnamefont {Pandey}}, \bibinfo {author} {\bibfnamefont {E.}~\bibnamefont {Cakmak}}, \bibinfo {author} {\bibfnamefont {H.}~\bibnamefont {Cao}}, \bibinfo {author} {\bibfnamefont {J.}~\bibnamefont {Yan}},\ and\ \bibinfo {author} {\bibfnamefont {M.~A.}\ \bibnamefont {McGuire}},\ }\bibfield  {title} {\bibinfo {title} {Magnetic order and interactions in ferrimagnetic {Mn$_{3}$Si$_{2}$Te$_{6}$}},\ }\href {https://doi.org/10.1103/PhysRevB.95.174440} {\bibfield  {journal} {\bibinfo  {journal} {Phys. Rev. B}\ }\textbf {\bibinfo {volume} {95}},\ \bibinfo {pages} {174440} (\bibinfo {year} {2017})}\BibitemShut {NoStop}%
\bibitem [{\citenamefont {Bonilla}\ \emph {et~al.}(2018)\citenamefont {Bonilla}, \citenamefont {Kolekar}, \citenamefont {Ma}, \citenamefont {Diaz}, \citenamefont {Kalappattil}, \citenamefont {Das}, \citenamefont {Eggers}, \citenamefont {Gutierrez}, \citenamefont {Phan},\ and\ \citenamefont {Batzill}}]{bonilla_strong_2018}%
  \BibitemOpen
  \bibfield  {author} {\bibinfo {author} {\bibfnamefont {M.}~\bibnamefont {Bonilla}}, \bibinfo {author} {\bibfnamefont {S.}~\bibnamefont {Kolekar}}, \bibinfo {author} {\bibfnamefont {Y.}~\bibnamefont {Ma}}, \bibinfo {author} {\bibfnamefont {H.~C.}\ \bibnamefont {Diaz}}, \bibinfo {author} {\bibfnamefont {V.}~\bibnamefont {Kalappattil}}, \bibinfo {author} {\bibfnamefont {R.}~\bibnamefont {Das}}, \bibinfo {author} {\bibfnamefont {T.}~\bibnamefont {Eggers}}, \bibinfo {author} {\bibfnamefont {H.~R.}\ \bibnamefont {Gutierrez}}, \bibinfo {author} {\bibfnamefont {M.-H.}\ \bibnamefont {Phan}},\ and\ \bibinfo {author} {\bibfnamefont {M.}~\bibnamefont {Batzill}},\ }\bibfield  {title} {\bibinfo {title} {{Strong room-temperature ferromagnetism in {VSe$_{2}$} monolayers on van der {Waals} substrates}},\ }\href {https://www.nature.com/articles/s41565-018-0063-9} {\bibfield  {journal} {\bibinfo  {journal} {Nat. Nanotechnol.}\ }\textbf {\bibinfo {volume} {13}},\ \bibinfo {pages} {289} (\bibinfo {year} {2018})}\BibitemShut
  {NoStop}%
\bibitem [{\citenamefont {Gong}\ \emph {et~al.}(2017)\citenamefont {Gong}, \citenamefont {Li}, \citenamefont {Li}, \citenamefont {Ji}, \citenamefont {Stern}, \citenamefont {Xia}, \citenamefont {Cao}, \citenamefont {Bao}, \citenamefont {Wang}, \citenamefont {Wang}, \citenamefont {Qiu}, \citenamefont {Cava}, \citenamefont {Louie}, \citenamefont {Xia},\ and\ \citenamefont {Zhang}}]{gong_discovery_2017}%
  \BibitemOpen
  \bibfield  {author} {\bibinfo {author} {\bibfnamefont {C.}~\bibnamefont {Gong}}, \bibinfo {author} {\bibfnamefont {L.}~\bibnamefont {Li}}, \bibinfo {author} {\bibfnamefont {Z.}~\bibnamefont {Li}}, \bibinfo {author} {\bibfnamefont {H.}~\bibnamefont {Ji}}, \bibinfo {author} {\bibfnamefont {A.}~\bibnamefont {Stern}}, \bibinfo {author} {\bibfnamefont {Y.}~\bibnamefont {Xia}}, \bibinfo {author} {\bibfnamefont {T.}~\bibnamefont {Cao}}, \bibinfo {author} {\bibfnamefont {W.}~\bibnamefont {Bao}}, \bibinfo {author} {\bibfnamefont {C.}~\bibnamefont {Wang}}, \bibinfo {author} {\bibfnamefont {Y.}~\bibnamefont {Wang}}, \bibinfo {author} {\bibfnamefont {Z.~Q.}\ \bibnamefont {Qiu}}, \bibinfo {author} {\bibfnamefont {R.~J.}\ \bibnamefont {Cava}}, \bibinfo {author} {\bibfnamefont {S.~G.}\ \bibnamefont {Louie}}, \bibinfo {author} {\bibfnamefont {J.}~\bibnamefont {Xia}},\ and\ \bibinfo {author} {\bibfnamefont {X.}~\bibnamefont {Zhang}},\ }\bibfield  {title} {\bibinfo {title} {{Discovery of intrinsic ferromagnetism in
  two-dimensional van der {Waals} crystals}},\ }\href {https://www.nature.com/articles/nature22060} {\bibfield  {journal} {\bibinfo  {journal} {Nature}\ }\textbf {\bibinfo {volume} {546}},\ \bibinfo {pages} {265} (\bibinfo {year} {2017})}\BibitemShut {NoStop}%
\bibitem [{\citenamefont {Chen}\ \emph {et~al.}(2013)\citenamefont {Chen}, \citenamefont {Yang}, \citenamefont {Wang}, \citenamefont {Imai}, \citenamefont {Ohta}, \citenamefont {Michioka}, \citenamefont {Yoshimura},\ and\ \citenamefont {Fang}}]{chen_magnetic_2013}%
  \BibitemOpen
  \bibfield  {author} {\bibinfo {author} {\bibfnamefont {B.}~\bibnamefont {Chen}}, \bibinfo {author} {\bibfnamefont {J.}~\bibnamefont {Yang}}, \bibinfo {author} {\bibfnamefont {H.}~\bibnamefont {Wang}}, \bibinfo {author} {\bibfnamefont {M.}~\bibnamefont {Imai}}, \bibinfo {author} {\bibfnamefont {H.}~\bibnamefont {Ohta}}, \bibinfo {author} {\bibfnamefont {C.}~\bibnamefont {Michioka}}, \bibinfo {author} {\bibfnamefont {K.}~\bibnamefont {Yoshimura}},\ and\ \bibinfo {author} {\bibfnamefont {M.}~\bibnamefont {Fang}},\ }\bibfield  {title} {\bibinfo {title} {Magnetic properties of layered itinerant electron ferromagnet {Fe$_3$GeTe$_2$}},\ }\href {https://journals.jps.jp/doi/abs/10.7566/JPSJ.82.124711} {\bibfield  {journal} {\bibinfo  {journal} {J. Phys. Soc. Jpn.}\ }\textbf {\bibinfo {volume} {82}},\ \bibinfo {pages} {124711} (\bibinfo {year} {2013})}\BibitemShut {NoStop}%
\bibitem [{\citenamefont {Mondal}\ \emph {et~al.}(2021)\citenamefont {Mondal}, \citenamefont {Khan}, \citenamefont {Mishra}, \citenamefont {Satpati},\ and\ \citenamefont {Mandal}}]{Fe4GeTe2}%
  \BibitemOpen
  \bibfield  {author} {\bibinfo {author} {\bibfnamefont {S.}~\bibnamefont {Mondal}}, \bibinfo {author} {\bibfnamefont {N.}~\bibnamefont {Khan}}, \bibinfo {author} {\bibfnamefont {S.~M.}\ \bibnamefont {Mishra}}, \bibinfo {author} {\bibfnamefont {B.}~\bibnamefont {Satpati}},\ and\ \bibinfo {author} {\bibfnamefont {P.}~\bibnamefont {Mandal}},\ }\bibfield  {title} {\bibinfo {title} {Critical behavior in the van der waals itinerant ferromagnet {Fe$_4$GeTe$_2$}},\ }\href {https://doi.org/10.1103/PhysRevB.104.094405} {\bibfield  {journal} {\bibinfo  {journal} {Phys. Rev. B}\ }\textbf {\bibinfo {volume} {104}},\ \bibinfo {pages} {094405} (\bibinfo {year} {2021})}\BibitemShut {NoStop}%
\bibitem [{\citenamefont {May}\ \emph {et~al.}(2019)\citenamefont {May}, \citenamefont {Ovchinnikov}, \citenamefont {Zheng}, \citenamefont {Hermann}, \citenamefont {Calder}, \citenamefont {Huang}, \citenamefont {Fei}, \citenamefont {Liu}, \citenamefont {Xu},\ and\ \citenamefont {McGuire}}]{Fe5GeTe2}%
  \BibitemOpen
  \bibfield  {author} {\bibinfo {author} {\bibfnamefont {A.~F.}\ \bibnamefont {May}}, \bibinfo {author} {\bibfnamefont {D.}~\bibnamefont {Ovchinnikov}}, \bibinfo {author} {\bibfnamefont {Q.}~\bibnamefont {Zheng}}, \bibinfo {author} {\bibfnamefont {R.}~\bibnamefont {Hermann}}, \bibinfo {author} {\bibfnamefont {S.}~\bibnamefont {Calder}}, \bibinfo {author} {\bibfnamefont {B.}~\bibnamefont {Huang}}, \bibinfo {author} {\bibfnamefont {Z.}~\bibnamefont {Fei}}, \bibinfo {author} {\bibfnamefont {Y.}~\bibnamefont {Liu}}, \bibinfo {author} {\bibfnamefont {X.}~\bibnamefont {Xu}},\ and\ \bibinfo {author} {\bibfnamefont {M.~A.}\ \bibnamefont {McGuire}},\ }\bibfield  {title} {\bibinfo {title} {Ferromagnetism near room temperature in the cleavable van der waals crystal {Fe$_5$GeTe$_2$}},\ }\href {https://doi.org/10.1021/acsnano.8b09660} {\bibfield  {journal} {\bibinfo  {journal} {ACS Nano}\ }\textbf {\bibinfo {volume} {13}},\ \bibinfo {pages} {4436} (\bibinfo {year} {2019})}\BibitemShut {NoStop}%
\bibitem [{\citenamefont {Zhang}\ \emph {et~al.}(2022)\citenamefont {Zhang}, \citenamefont {Guo}, \citenamefont {Wu}, \citenamefont {Wen}, \citenamefont {Yang}, \citenamefont {Jin}, \citenamefont {Zhang},\ and\ \citenamefont {Chang}}]{Fe3GaTe2}%
  \BibitemOpen
  \bibfield  {author} {\bibinfo {author} {\bibfnamefont {G.}~\bibnamefont {Zhang}}, \bibinfo {author} {\bibfnamefont {F.}~\bibnamefont {Guo}}, \bibinfo {author} {\bibfnamefont {H.}~\bibnamefont {Wu}}, \bibinfo {author} {\bibfnamefont {X.}~\bibnamefont {Wen}}, \bibinfo {author} {\bibfnamefont {L.}~\bibnamefont {Yang}}, \bibinfo {author} {\bibfnamefont {W.}~\bibnamefont {Jin}}, \bibinfo {author} {\bibfnamefont {W.}~\bibnamefont {Zhang}},\ and\ \bibinfo {author} {\bibfnamefont {H.}~\bibnamefont {Chang}},\ }\bibfield  {title} {\bibinfo {title} {Above-room-temperature strong intrinsic ferromagnetism in 2{D} van der waals {Fe$_3$GaTe$_2$} with large perpendicular magnetic anisotropy},\ }\href {https://doi.org/10.1038/s41467-022-32605-5} {\bibfield  {journal} {\bibinfo  {journal} {Nat. Commun.}\ }\textbf {\bibinfo {volume} {13}},\ \bibinfo {pages} {5067} (\bibinfo {year} {2022})}\BibitemShut {NoStop}%
\bibitem [{\citenamefont {Xu}\ \emph {et~al.}(2024{\natexlab{a}})\citenamefont {Xu}, \citenamefont {Liu}, \citenamefont {Wo}, \citenamefont {Liao}, \citenamefont {Yi}, \citenamefont {Li}, \citenamefont {Zhao}, \citenamefont {Qiu}, \citenamefont {Yin},\ and\ \citenamefont {Bernhard}}]{YFe2Ge2}%
  \BibitemOpen
  \bibfield  {author} {\bibinfo {author} {\bibfnamefont {B.}~\bibnamefont {Xu}}, \bibinfo {author} {\bibfnamefont {R.}~\bibnamefont {Liu}}, \bibinfo {author} {\bibfnamefont {H.}~\bibnamefont {Wo}}, \bibinfo {author} {\bibfnamefont {Z.}~\bibnamefont {Liao}}, \bibinfo {author} {\bibfnamefont {S.}~\bibnamefont {Yi}}, \bibinfo {author} {\bibfnamefont {C.}~\bibnamefont {Li}}, \bibinfo {author} {\bibfnamefont {J.}~\bibnamefont {Zhao}}, \bibinfo {author} {\bibfnamefont {X.}~\bibnamefont {Qiu}}, \bibinfo {author} {\bibfnamefont {Z.}~\bibnamefont {Yin}},\ and\ \bibinfo {author} {\bibfnamefont {C.}~\bibnamefont {Bernhard}},\ }\bibfield  {title} {\bibinfo {title} {Unraveling the origin of {Kondo}-like behavior in the 3$d$-electron heavy-fermion compound {YFe$_{2}$Ge$_{2}$}},\ }\href {https://doi.org/10.1073/pnas.2401430121} {\bibfield  {journal} {\bibinfo  {journal} {Proceedings of the National Academy of Sciences}\ }\textbf {\bibinfo {volume} {121}},\ \bibinfo {pages} {e2401430121} (\bibinfo {year}
  {2024}{\natexlab{a}})}\BibitemShut {NoStop}%
\bibitem [{\citenamefont {Wu}\ \emph {et~al.}(2016)\citenamefont {Wu}, \citenamefont {Zhao}, \citenamefont {Wang}, \citenamefont {Wang}, \citenamefont {Xiang}, \citenamefont {Luo}, \citenamefont {Wu},\ and\ \citenamefont {Chen}}]{AFe2}%
  \BibitemOpen
  \bibfield  {author} {\bibinfo {author} {\bibfnamefont {Y.~P.}\ \bibnamefont {Wu}}, \bibinfo {author} {\bibfnamefont {D.}~\bibnamefont {Zhao}}, \bibinfo {author} {\bibfnamefont {A.~F.}\ \bibnamefont {Wang}}, \bibinfo {author} {\bibfnamefont {N.~Z.}\ \bibnamefont {Wang}}, \bibinfo {author} {\bibfnamefont {Z.~J.}\ \bibnamefont {Xiang}}, \bibinfo {author} {\bibfnamefont {X.~G.}\ \bibnamefont {Luo}}, \bibinfo {author} {\bibfnamefont {T.}~\bibnamefont {Wu}},\ and\ \bibinfo {author} {\bibfnamefont {X.~H.}\ \bibnamefont {Chen}},\ }\bibfield  {title} {\bibinfo {title} {Emergent {Kondo} lattice behavior in iron-based superconductors {AFe$_{2}$As$_{2}$} {(A=K, Rb, Cs)}},\ }\href {https://doi.org/10.1103/PhysRevLett.116.147001} {\bibfield  {journal} {\bibinfo  {journal} {Phys. Rev. Lett.}\ }\textbf {\bibinfo {volume} {116}},\ \bibinfo {pages} {147001} (\bibinfo {year} {2016})}\BibitemShut {NoStop}%
\bibitem [{\citenamefont {Sudayama}\ \emph {et~al.}(2009)\citenamefont {Sudayama}, \citenamefont {Wakisaka}, \citenamefont {Takubo}, \citenamefont {Mizokawa}, \citenamefont {Kobayashi}, \citenamefont {Terasaki}, \citenamefont {Tanaka}, \citenamefont {Maeno}, \citenamefont {Arita}, \citenamefont {Namatame},\ and\ \citenamefont {Taniguchi}}]{CaCu3Ru4O12}%
  \BibitemOpen
  \bibfield  {author} {\bibinfo {author} {\bibfnamefont {T.}~\bibnamefont {Sudayama}}, \bibinfo {author} {\bibfnamefont {Y.}~\bibnamefont {Wakisaka}}, \bibinfo {author} {\bibfnamefont {K.}~\bibnamefont {Takubo}}, \bibinfo {author} {\bibfnamefont {T.}~\bibnamefont {Mizokawa}}, \bibinfo {author} {\bibfnamefont {W.}~\bibnamefont {Kobayashi}}, \bibinfo {author} {\bibfnamefont {I.}~\bibnamefont {Terasaki}}, \bibinfo {author} {\bibfnamefont {S.}~\bibnamefont {Tanaka}}, \bibinfo {author} {\bibfnamefont {Y.}~\bibnamefont {Maeno}}, \bibinfo {author} {\bibfnamefont {M.}~\bibnamefont {Arita}}, \bibinfo {author} {\bibfnamefont {H.}~\bibnamefont {Namatame}},\ and\ \bibinfo {author} {\bibfnamefont {M.}~\bibnamefont {Taniguchi}},\ }\bibfield  {title} {\bibinfo {title} {Bulk-sensitive photoemission study of {ACu$_{3}$Ru$_{4}$O$_{12}$} {(A=Ca, Na, and La)} with heavy-fermion behavior},\ }\href {https://doi.org/10.1103/PhysRevB.80.075113} {\bibfield  {journal} {\bibinfo  {journal} {Phys. Rev. B}\ }\textbf {\bibinfo {volume}
  {80}},\ \bibinfo {pages} {075113} (\bibinfo {year} {2009})}\BibitemShut {NoStop}%
\bibitem [{\citenamefont {Kondo}\ \emph {et~al.}(1997)\citenamefont {Kondo}, \citenamefont {Johnston}, \citenamefont {Swenson}, \citenamefont {Borsa}, \citenamefont {Mahajan}, \citenamefont {Miller}, \citenamefont {Gu}, \citenamefont {Goldman}, \citenamefont {Maple}, \citenamefont {Gajewski}, \citenamefont {Freeman}, \citenamefont {Dilley}, \citenamefont {Dickey}, \citenamefont {Merrin}, \citenamefont {Kojima}, \citenamefont {Luke}, \citenamefont {Uemura}, \citenamefont {Chmaissem},\ and\ \citenamefont {Jorgensen}}]{LiV2O4}%
  \BibitemOpen
  \bibfield  {author} {\bibinfo {author} {\bibfnamefont {S.}~\bibnamefont {Kondo}}, \bibinfo {author} {\bibfnamefont {D.~C.}\ \bibnamefont {Johnston}}, \bibinfo {author} {\bibfnamefont {C.~A.}\ \bibnamefont {Swenson}}, \bibinfo {author} {\bibfnamefont {F.}~\bibnamefont {Borsa}}, \bibinfo {author} {\bibfnamefont {A.~V.}\ \bibnamefont {Mahajan}}, \bibinfo {author} {\bibfnamefont {L.~L.}\ \bibnamefont {Miller}}, \bibinfo {author} {\bibfnamefont {T.}~\bibnamefont {Gu}}, \bibinfo {author} {\bibfnamefont {A.~I.}\ \bibnamefont {Goldman}}, \bibinfo {author} {\bibfnamefont {M.~B.}\ \bibnamefont {Maple}}, \bibinfo {author} {\bibfnamefont {D.~A.}\ \bibnamefont {Gajewski}}, \bibinfo {author} {\bibfnamefont {E.~J.}\ \bibnamefont {Freeman}}, \bibinfo {author} {\bibfnamefont {N.~R.}\ \bibnamefont {Dilley}}, \bibinfo {author} {\bibfnamefont {R.~P.}\ \bibnamefont {Dickey}}, \bibinfo {author} {\bibfnamefont {J.}~\bibnamefont {Merrin}}, \bibinfo {author} {\bibfnamefont {K.}~\bibnamefont {Kojima}}, \bibinfo {author}
  {\bibfnamefont {G.~M.}\ \bibnamefont {Luke}}, \bibinfo {author} {\bibfnamefont {Y.~J.}\ \bibnamefont {Uemura}}, \bibinfo {author} {\bibfnamefont {O.}~\bibnamefont {Chmaissem}},\ and\ \bibinfo {author} {\bibfnamefont {J.~D.}\ \bibnamefont {Jorgensen}},\ }\bibfield  {title} {\bibinfo {title} {{LiV$_{2}$O$_{4}$}: A heavy fermion transition metal oxide},\ }\href {https://doi.org/10.1103/PhysRevLett.78.3729} {\bibfield  {journal} {\bibinfo  {journal} {Phys. Rev. Lett.}\ }\textbf {\bibinfo {volume} {78}},\ \bibinfo {pages} {3729} (\bibinfo {year} {1997})}\BibitemShut {NoStop}%
  \bibitem [{\citenamefont {Shimoyamada}\ \emph {et~al.}(2006)\citenamefont {Shimoyamada}, \citenamefont {Tsuda}, \citenamefont {Ishizaka}, \citenamefont {Kiss}, \citenamefont {Shimojima}, \citenamefont {Togashi}, \citenamefont {Watanabe}, \citenamefont {Zhang}, \citenamefont {Chen}, \citenamefont {Matsushita}, \citenamefont {Ueda}, \citenamefont {Ueda},\ and\ \citenamefont {Shin}}]{LiV2O4-2}%
  \BibitemOpen
  \bibfield  {author} {\bibinfo {author} {\bibfnamefont {A.}~\bibnamefont {Shimoyamada}}, 
  \bibinfo {author} {\bibfnamefont {S.}~\bibnamefont {Tsuda}}, 
  \bibinfo {author} {\bibfnamefont {K.}~\bibnamefont {Ishizaka}}, 
  \bibinfo {author} {\bibfnamefont {T.}~\bibnamefont {Kiss}}, 
  \bibinfo {author} {\bibfnamefont {T.}~\bibnamefont {Shimojima}}, 
  \bibinfo {author} {\bibfnamefont {T.}~\bibnamefont {Togashi}}, 
  \bibinfo {author} {\bibfnamefont {S.}~\bibnamefont {Watanabe}}, 
  \bibinfo {author} {\bibfnamefont {C.~Q.}\ \bibnamefont {Zhang}}, 
  \bibinfo {author} {\bibfnamefont {C.~T.}\ \bibnamefont {Chen}}, 
  \bibinfo {author} {\bibfnamefont {Y.}~\bibnamefont {Matsushita}}, 
  \bibinfo {author} {\bibfnamefont {H.}~\bibnamefont {Ueda}}, 
  \bibinfo {author} {\bibfnamefont {Y.}~\bibnamefont {Ueda}},\ and\ 
  \bibinfo {author} {\bibfnamefont {S.}~\bibnamefont {Shin}},\ }
  \bibfield  {title} {\bibinfo {title} {Heavy-Fermion-like State in a Transition Metal Oxide LiV$_{2}$O$_{4}$ Single Crystal: Indication of Kondo Resonance in the Photoemission Spectrum},\ }
  \href {https://doi.org/10.1103/PhysRevLett.96.026403} {\bibfield  {journal} {\bibinfo  {journal} {Phys. Rev. Lett.}\ }\textbf {\bibinfo {volume} {96}},\ \bibinfo {pages} {026403} (\bibinfo {year} {2006})}\BibitemShut {NoStop}%
\bibitem [{\citenamefont {Ding}\ \emph {et~al.}(2021)\citenamefont {Ding}, \citenamefont {Xing}, \citenamefont {Li}, \citenamefont {Balicas}, \citenamefont {Gofryk},\ and\ \citenamefont {Wen}}]{VTe2}%
  \BibitemOpen
  \bibfield  {author} {\bibinfo {author} {\bibfnamefont {X.}~\bibnamefont {Ding}}, \bibinfo {author} {\bibfnamefont {J.}~\bibnamefont {Xing}}, \bibinfo {author} {\bibfnamefont {G.}~\bibnamefont {Li}}, \bibinfo {author} {\bibfnamefont {L.}~\bibnamefont {Balicas}}, \bibinfo {author} {\bibfnamefont {K.}~\bibnamefont {Gofryk}},\ and\ \bibinfo {author} {\bibfnamefont {H.-H.}\ \bibnamefont {Wen}},\ }\bibfield  {title} {\bibinfo {title} {Crossover from {Kondo} to {Fermi}-liquid behavior induced by high magnetic field in {1T}-{VTe$_{2}$} single crystals},\ }\href {https://doi.org/10.1103/PhysRevB.103.125115} {\bibfield  {journal} {\bibinfo  {journal} {Phys. Rev. B}\ }\textbf {\bibinfo {volume} {103}},\ \bibinfo {pages} {125115} (\bibinfo {year} {2021})}\BibitemShut {NoStop}%
\bibitem [{\citenamefont {Kim}\ \emph {et~al.}(2023)\citenamefont {Kim}, \citenamefont {Kim}, \citenamefont {Kim}, \citenamefont {Kim}, \citenamefont {Kim}, \citenamefont {Cheng}, \citenamefont {Choi}, \citenamefont {Jung}, \citenamefont {Lu}, \citenamefont {Kim}, \citenamefont {Cho}, \citenamefont {Song}, \citenamefont {Oh}, \citenamefont {Yu}, \citenamefont {Choi}, \citenamefont {Kim}, \citenamefont {Han}, \citenamefont {Jo}, \citenamefont {Shim}, \citenamefont {Seo}, \citenamefont {Huh},\ and\ \citenamefont {Kim}}]{FeTe}%
  \BibitemOpen
  \bibfield  {author} {\bibinfo {author} {\bibfnamefont {Y.}~\bibnamefont {Kim}}, \bibinfo {author} {\bibfnamefont {M.-S.}\ \bibnamefont {Kim}}, \bibinfo {author} {\bibfnamefont {D.}~\bibnamefont {Kim}}, \bibinfo {author} {\bibfnamefont {M.}~\bibnamefont {Kim}}, \bibinfo {author} {\bibfnamefont {M.}~\bibnamefont {Kim}}, \bibinfo {author} {\bibfnamefont {C.-M.}\ \bibnamefont {Cheng}}, \bibinfo {author} {\bibfnamefont {J.}~\bibnamefont {Choi}}, \bibinfo {author} {\bibfnamefont {S.}~\bibnamefont {Jung}}, \bibinfo {author} {\bibfnamefont {D.}~\bibnamefont {Lu}}, \bibinfo {author} {\bibfnamefont {J.~H.}\ \bibnamefont {Kim}}, \bibinfo {author} {\bibfnamefont {S.}~\bibnamefont {Cho}}, \bibinfo {author} {\bibfnamefont {D.}~\bibnamefont {Song}}, \bibinfo {author} {\bibfnamefont {D.}~\bibnamefont {Oh}}, \bibinfo {author} {\bibfnamefont {L.}~\bibnamefont {Yu}}, \bibinfo {author} {\bibfnamefont {Y.~J.}\ \bibnamefont {Choi}}, \bibinfo {author} {\bibfnamefont {H.-D.}\ \bibnamefont {Kim}}, \bibinfo {author} {\bibfnamefont
  {J.~H.}\ \bibnamefont {Han}}, \bibinfo {author} {\bibfnamefont {Y.}~\bibnamefont {Jo}}, \bibinfo {author} {\bibfnamefont {J.~H.}\ \bibnamefont {Shim}}, \bibinfo {author} {\bibfnamefont {J.}~\bibnamefont {Seo}}, \bibinfo {author} {\bibfnamefont {S.}~\bibnamefont {Huh}},\ and\ \bibinfo {author} {\bibfnamefont {C.}~\bibnamefont {Kim}},\ }\bibfield  {title} {\bibinfo {title} {Kondo interaction in {FeTe} and its potential role in the magnetic order},\ }\href {https://doi.org/10.1038/s41467-023-39827-1} {\bibfield  {journal} {\bibinfo  {journal} {Nat. Commun.}\ }\textbf {\bibinfo {volume} {14}},\ \bibinfo {pages} {4145} (\bibinfo {year} {2023})}\BibitemShut {NoStop}%
\bibitem [{\citenamefont {Sharma}\ \emph {et~al.}(2024)\citenamefont {Sharma}, \citenamefont {Ali}, \citenamefont {Bhatt}, \citenamefont {Chowdhury}, \citenamefont {Patra}, \citenamefont {Singh},\ and\ \citenamefont {Singh}}]{FGT}%
  \BibitemOpen
  \bibfield  {author} {\bibinfo {author} {\bibfnamefont {D.}~\bibnamefont {Sharma}}, \bibinfo {author} {\bibfnamefont {A.}~\bibnamefont {Ali}}, \bibinfo {author} {\bibfnamefont {N.}~\bibnamefont {Bhatt}}, \bibinfo {author} {\bibfnamefont {R.~R.}\ \bibnamefont {Chowdhury}}, \bibinfo {author} {\bibfnamefont {C.}~\bibnamefont {Patra}}, \bibinfo {author} {\bibfnamefont {R.~P.}\ \bibnamefont {Singh}},\ and\ \bibinfo {author} {\bibfnamefont {R.~S.}\ \bibnamefont {Singh}},\ }\bibfield  {title} {\bibinfo {title} {{Manifestation of incoherent-coherent crossover and non-Stoner magnetism in the electronic structure of {Fe$_3$GeTe$_2$}}},\ }\href {https://doi.org/10.1103/PhysRevB.110.125119} {\bibfield  {journal} {\bibinfo  {journal} {Phys. Rev. B}\ }\textbf {\bibinfo {volume} {110}},\ \bibinfo {pages} {125119} (\bibinfo {year} {2024})}\BibitemShut {NoStop}%
\bibitem [{\citenamefont {Rana}\ \emph {et~al.}(2022)\citenamefont {Rana}, \citenamefont {R}, \citenamefont {G}, \citenamefont {Patra}, \citenamefont {Howlader}, \citenamefont {Chowdhury}, \citenamefont {Kabir}, \citenamefont {Singh},\ and\ \citenamefont {Sheet}}]{rana_spin-polarized_2022}%
  \BibitemOpen
  \bibfield  {author} {\bibinfo {author} {\bibfnamefont {D.}~\bibnamefont {Rana}}, \bibinfo {author} {\bibfnamefont {A.}~\bibnamefont {R}}, \bibinfo {author} {\bibfnamefont {B.}~\bibnamefont {G}}, \bibinfo {author} {\bibfnamefont {C.}~\bibnamefont {Patra}}, \bibinfo {author} {\bibfnamefont {S.}~\bibnamefont {Howlader}}, \bibinfo {author} {\bibfnamefont {R.~R.}\ \bibnamefont {Chowdhury}}, \bibinfo {author} {\bibfnamefont {M.}~\bibnamefont {Kabir}}, \bibinfo {author} {\bibfnamefont {R.~P.}\ \bibnamefont {Singh}},\ and\ \bibinfo {author} {\bibfnamefont {G.}~\bibnamefont {Sheet}},\ }\bibfield  {title} {\bibinfo {title} {{Spin-polarized supercurrent through the van der {Waals} {Kondo}-lattice ferromagnet {Fe$_3$GeTe$_2$}}},\ }\href {https://link.aps.org/doi/10.1103/PhysRevB.106.085120} {\bibfield  {journal} {\bibinfo  {journal} {Phys. Rev. B}\ }\textbf {\bibinfo {volume} {106}},\ \bibinfo {pages} {085120} (\bibinfo {year} {2022})}\BibitemShut {NoStop}%
\bibitem [{\citenamefont {Zhu}\ \emph {et~al.}(2016)\citenamefont {Zhu}, \citenamefont {Janoschek}, \citenamefont {Chaves}, \citenamefont {Cezar}, \citenamefont {Durakiewicz}, \citenamefont {Ronning}, \citenamefont {Sassa}, \citenamefont {Mansson}, \citenamefont {Scott}, \citenamefont {Wakeham}, \citenamefont {Bauer},\ and\ \citenamefont {Thompson}}]{zhu_electronic_2016}%
  \BibitemOpen
  \bibfield  {author} {\bibinfo {author} {\bibfnamefont {J.-X.}\ \bibnamefont {Zhu}}, \bibinfo {author} {\bibfnamefont {M.}~\bibnamefont {Janoschek}}, \bibinfo {author} {\bibfnamefont {D.~S.}\ \bibnamefont {Chaves}}, \bibinfo {author} {\bibfnamefont {J.~C.}\ \bibnamefont {Cezar}}, \bibinfo {author} {\bibfnamefont {T.}~\bibnamefont {Durakiewicz}}, \bibinfo {author} {\bibfnamefont {F.}~\bibnamefont {Ronning}}, \bibinfo {author} {\bibfnamefont {Y.}~\bibnamefont {Sassa}}, \bibinfo {author} {\bibfnamefont {M.}~\bibnamefont {Mansson}}, \bibinfo {author} {\bibfnamefont {B.~L.}\ \bibnamefont {Scott}}, \bibinfo {author} {\bibfnamefont {N.}~\bibnamefont {Wakeham}}, \bibinfo {author} {\bibfnamefont {E.~D.}\ \bibnamefont {Bauer}},\ and\ \bibinfo {author} {\bibfnamefont {J.~D.}\ \bibnamefont {Thompson}},\ }\bibfield  {title} {\bibinfo {title} {{Electronic correlation and magnetism in the ferromagnetic metal {Fe$_3$GeTe$_2$}}},\ }\href {https://link.aps.org/doi/10.1103/PhysRevB.93.144404} {\bibfield  {journal} {\bibinfo
  {journal} {Phys. Rev. B}\ }\textbf {\bibinfo {volume} {93}},\ \bibinfo {pages} {144404} (\bibinfo {year} {2016})}\BibitemShut {NoStop}%
\bibitem [{\citenamefont {Xu}\ \emph {et~al.}(2020)\citenamefont {Xu}, \citenamefont {Li}, \citenamefont {Duan}, \citenamefont {Zhang}, \citenamefont {Chen}, \citenamefont {Kang}, \citenamefont {Liang}, \citenamefont {Chen}, \citenamefont {Xia}, \citenamefont {Xu}, \citenamefont {Malinowski}, \citenamefont {Xu}, \citenamefont {Chu}, \citenamefont {Li}, \citenamefont {Guo}, \citenamefont {Liu}, \citenamefont {Yang},\ and\ \citenamefont {Chen}}]{xu_signature_2020}%
  \BibitemOpen
  \bibfield  {author} {\bibinfo {author} {\bibfnamefont {X.}~\bibnamefont {Xu}}, \bibinfo {author} {\bibfnamefont {Y.~W.}\ \bibnamefont {Li}}, \bibinfo {author} {\bibfnamefont {S.~R.}\ \bibnamefont {Duan}}, \bibinfo {author} {\bibfnamefont {S.~L.}\ \bibnamefont {Zhang}}, \bibinfo {author} {\bibfnamefont {Y.~J.}\ \bibnamefont {Chen}}, \bibinfo {author} {\bibfnamefont {L.}~\bibnamefont {Kang}}, \bibinfo {author} {\bibfnamefont {A.~J.}\ \bibnamefont {Liang}}, \bibinfo {author} {\bibfnamefont {C.}~\bibnamefont {Chen}}, \bibinfo {author} {\bibfnamefont {W.}~\bibnamefont {Xia}}, \bibinfo {author} {\bibfnamefont {Y.}~\bibnamefont {Xu}}, \bibinfo {author} {\bibfnamefont {P.}~\bibnamefont {Malinowski}}, \bibinfo {author} {\bibfnamefont {X.~D.}\ \bibnamefont {Xu}}, \bibinfo {author} {\bibfnamefont {J.-H.}\ \bibnamefont {Chu}}, \bibinfo {author} {\bibfnamefont {G.}~\bibnamefont {Li}}, \bibinfo {author} {\bibfnamefont {Y.~F.}\ \bibnamefont {Guo}}, \bibinfo {author} {\bibfnamefont {Z.~K.}\ \bibnamefont {Liu}}, \bibinfo
  {author} {\bibfnamefont {L.~X.}\ \bibnamefont {Yang}},\ and\ \bibinfo {author} {\bibfnamefont {Y.~L.}\ \bibnamefont {Chen}},\ }\bibfield  {title} {\bibinfo {title} {{Signature for non-{Stoner} ferromagnetism in the van der {Waals} ferromagnet {Fe$_3$GeTe$_2$}}},\ }\href {https://link.aps.org/doi/10.1103/PhysRevB.101.201104} {\bibfield  {journal} {\bibinfo  {journal} {Phys. Rev. B}\ }\textbf {\bibinfo {volume} {101}},\ \bibinfo {pages} {201104} (\bibinfo {year} {2020})}\BibitemShut {NoStop}%
\bibitem [{\citenamefont {Zhang}\ \emph {et~al.}(2018)\citenamefont {Zhang}, \citenamefont {Lu}, \citenamefont {Zhu}, \citenamefont {Tan}, \citenamefont {Feng}, \citenamefont {Liu}, \citenamefont {Zhang}, \citenamefont {Chen}, \citenamefont {Liu}, \citenamefont {Luo}, \citenamefont {Xie}, \citenamefont {Luo}, \citenamefont {Zhang},\ and\ \citenamefont {Lai}}]{zhang_emergence_2018}%
  \BibitemOpen
  \bibfield  {author} {\bibinfo {author} {\bibfnamefont {Y.}~\bibnamefont {Zhang}}, \bibinfo {author} {\bibfnamefont {H.}~\bibnamefont {Lu}}, \bibinfo {author} {\bibfnamefont {X.}~\bibnamefont {Zhu}}, \bibinfo {author} {\bibfnamefont {S.}~\bibnamefont {Tan}}, \bibinfo {author} {\bibfnamefont {W.}~\bibnamefont {Feng}}, \bibinfo {author} {\bibfnamefont {Q.}~\bibnamefont {Liu}}, \bibinfo {author} {\bibfnamefont {W.}~\bibnamefont {Zhang}}, \bibinfo {author} {\bibfnamefont {Q.}~\bibnamefont {Chen}}, \bibinfo {author} {\bibfnamefont {Y.}~\bibnamefont {Liu}}, \bibinfo {author} {\bibfnamefont {X.}~\bibnamefont {Luo}}, \bibinfo {author} {\bibfnamefont {D.}~\bibnamefont {Xie}}, \bibinfo {author} {\bibfnamefont {L.}~\bibnamefont {Luo}}, \bibinfo {author} {\bibfnamefont {Z.}~\bibnamefont {Zhang}},\ and\ \bibinfo {author} {\bibfnamefont {X.}~\bibnamefont {Lai}},\ }\bibfield  {title} {\bibinfo {title} {{Emergence of {Kondo} lattice behavior in a van der {Waals} itinerant ferromagnet, {Fe$_3$GeTe$_2$}}},\ }\href
  {https://www.science.org/doi/10.1126/sciadv.aao6791} {\bibfield  {journal} {\bibinfo  {journal} {Sci. Adv.}\ }\textbf {\bibinfo {volume} {4}},\ \bibinfo {pages} {eaao6791} (\bibinfo {year} {2018})}\BibitemShut {NoStop}%
\bibitem [{\citenamefont {Zhao}\ \emph {et~al.}(2021)\citenamefont {Zhao}, \citenamefont {Chen}, \citenamefont {Xi}, \citenamefont {Zhao}, \citenamefont {Xu}, \citenamefont {Zhang}, \citenamefont {Cheng}, \citenamefont {Feng}, \citenamefont {Zhuang}, \citenamefont {Pan}, \citenamefont {Xu}, \citenamefont {Hao}, \citenamefont {Li}, \citenamefont {Zhou}, \citenamefont {Dou},\ and\ \citenamefont {Du}}]{zhao_kondo_2021}%
  \BibitemOpen
  \bibfield  {author} {\bibinfo {author} {\bibfnamefont {M.}~\bibnamefont {Zhao}}, \bibinfo {author} {\bibfnamefont {B.-B.}\ \bibnamefont {Chen}}, \bibinfo {author} {\bibfnamefont {Y.}~\bibnamefont {Xi}}, \bibinfo {author} {\bibfnamefont {Y.}~\bibnamefont {Zhao}}, \bibinfo {author} {\bibfnamefont {H.}~\bibnamefont {Xu}}, \bibinfo {author} {\bibfnamefont {H.}~\bibnamefont {Zhang}}, \bibinfo {author} {\bibfnamefont {N.}~\bibnamefont {Cheng}}, \bibinfo {author} {\bibfnamefont {H.}~\bibnamefont {Feng}}, \bibinfo {author} {\bibfnamefont {J.}~\bibnamefont {Zhuang}}, \bibinfo {author} {\bibfnamefont {F.}~\bibnamefont {Pan}}, \bibinfo {author} {\bibfnamefont {X.}~\bibnamefont {Xu}}, \bibinfo {author} {\bibfnamefont {W.}~\bibnamefont {Hao}}, \bibinfo {author} {\bibfnamefont {W.}~\bibnamefont {Li}}, \bibinfo {author} {\bibfnamefont {S.}~\bibnamefont {Zhou}}, \bibinfo {author} {\bibfnamefont {S.~X.}\ \bibnamefont {Dou}},\ and\ \bibinfo {author} {\bibfnamefont {Y.}~\bibnamefont {Du}},\ }\bibfield  {title} {\bibinfo
  {title} {{Kondo} holes in the two-dimensional itinerant {Ising} ferromagnet {Fe$_3$GeTe$_2$}},\ }\href {https://doi.org/10.1021/acs.nanolett.1c01661} {\bibfield  {journal} {\bibinfo  {journal} {Nano Lett.}\ }\textbf {\bibinfo {volume} {21}},\ \bibinfo {pages} {6117} (\bibinfo {year} {2021})}\BibitemShut {NoStop}%
\bibitem [{\citenamefont {Coleman}\ and\ \citenamefont {Schofield}(2005)}]{Coleman2005}%
  \BibitemOpen
  \bibfield  {author} {\bibinfo {author} {\bibfnamefont {P.}~\bibnamefont {Coleman}}\ and\ \bibinfo {author} {\bibfnamefont {A.~J.}\ \bibnamefont {Schofield}},\ }\bibfield  {title} {\bibinfo {title} {Quantum criticality},\ }\href {https://doi.org/10.1038/nature03279} {\bibfield  {journal} {\bibinfo  {journal} {Nature}\ }\textbf {\bibinfo {volume} {433}},\ \bibinfo {pages} {226} (\bibinfo {year} {2005})}\BibitemShut {NoStop}%
\bibitem [{\citenamefont {Li}\ \emph {et~al.}(2025)\citenamefont {Li}, \citenamefont {Priyadarshi}, \citenamefont {Yang}, \citenamefont {Pohl}, \citenamefont {Stockert}, \citenamefont {von L\"ohneysen}, \citenamefont {Pal}, \citenamefont {Fiebig},\ and\ \citenamefont {Kroha}}]{QCPNT}%
  \BibitemOpen
  \bibfield  {author} {\bibinfo {author} {\bibfnamefont {J.}~\bibnamefont {Li}}, \bibinfo {author} {\bibfnamefont {D.}~\bibnamefont {Priyadarshi}}, \bibinfo {author} {\bibfnamefont {C.-J.}\ \bibnamefont {Yang}}, \bibinfo {author} {\bibfnamefont {U.}~\bibnamefont {Pohl}}, \bibinfo {author} {\bibfnamefont {O.}~\bibnamefont {Stockert}}, \bibinfo {author} {\bibfnamefont {H.}~\bibnamefont {von L\"ohneysen}}, \bibinfo {author} {\bibfnamefont {S.}~\bibnamefont {Pal}}, \bibinfo {author} {\bibfnamefont {M.}~\bibnamefont {Fiebig}},\ and\ \bibinfo {author} {\bibfnamefont {J.}~\bibnamefont {Kroha}},\ }\bibfield  {title} {\bibinfo {title} {Missing spectral weight in a heavy-fermion system far above the {Neel} temperature},\ }\href {https://doi.org/10.1103/PhysRevB.111.035117} {\bibfield  {journal} {\bibinfo  {journal} {Phys. Rev. B}\ }\textbf {\bibinfo {volume} {111}},\ \bibinfo {pages} {035117} (\bibinfo {year} {2025})}\BibitemShut {NoStop}%
\bibitem [{\citenamefont {Gegenwart}\ \emph {et~al.}(2008)\citenamefont {Gegenwart}, \citenamefont {Si},\ and\ \citenamefont {Steglich}}]{QCP1}%
  \BibitemOpen
  \bibfield  {author} {\bibinfo {author} {\bibfnamefont {P.}~\bibnamefont {Gegenwart}}, \bibinfo {author} {\bibfnamefont {Q.}~\bibnamefont {Si}},\ and\ \bibinfo {author} {\bibfnamefont {F.}~\bibnamefont {Steglich}},\ }\bibfield  {title} {\bibinfo {title} {Quantum criticality in heavy-fermion metals},\ }\href {https://doi.org/10.1038/nphys892} {\bibfield  {journal} {\bibinfo  {journal} {Nat. Phys.}\ }\textbf {\bibinfo {volume} {4}},\ \bibinfo {pages} {186} (\bibinfo {year} {2008})}\BibitemShut {NoStop}%
\bibitem [{\citenamefont {Takegami}\ \emph {et~al.}(2022)\citenamefont {Takegami}, \citenamefont {Kuo}, \citenamefont {Kasebayashi}, \citenamefont {Kim}, \citenamefont {Chang}, \citenamefont {Liu}, \citenamefont {Wu}, \citenamefont {Kasinathan}, \citenamefont {Altendorf}, \citenamefont {Hoefer}, \citenamefont {Meneghin}, \citenamefont {Marino}, \citenamefont {Liao}, \citenamefont {Tsuei}, \citenamefont {Chen}, \citenamefont {Ko}, \citenamefont {G\"unther}, \citenamefont {Ebbinghaus}, \citenamefont {Seo}, \citenamefont {Lee}, \citenamefont {Ryu}, \citenamefont {Komarek}, \citenamefont {Sugano}, \citenamefont {Shimakawa}, \citenamefont {Tanaka}, \citenamefont {Mizokawa}, \citenamefont {Kune\ifmmode~\check{s}\else \v{s}\fi{}}, \citenamefont {Tjeng},\ and\ \citenamefont {Hariki}}]{kondoCaCu}%
  \BibitemOpen
  \bibfield  {author} {\bibinfo {author} {\bibfnamefont {D.}~\bibnamefont {Takegami}}, \bibinfo {author} {\bibfnamefont {C.-Y.}\ \bibnamefont {Kuo}}, \bibinfo {author} {\bibfnamefont {K.}~\bibnamefont {Kasebayashi}}, \bibinfo {author} {\bibfnamefont {J.-G.}\ \bibnamefont {Kim}}, \bibinfo {author} {\bibfnamefont {C.~F.}\ \bibnamefont {Chang}}, \bibinfo {author} {\bibfnamefont {C.~E.}\ \bibnamefont {Liu}}, \bibinfo {author} {\bibfnamefont {C.~N.}\ \bibnamefont {Wu}}, \bibinfo {author} {\bibfnamefont {D.}~\bibnamefont {Kasinathan}}, \bibinfo {author} {\bibfnamefont {S.~G.}\ \bibnamefont {Altendorf}}, \bibinfo {author} {\bibfnamefont {K.}~\bibnamefont {Hoefer}}, \bibinfo {author} {\bibfnamefont {F.}~\bibnamefont {Meneghin}}, \bibinfo {author} {\bibfnamefont {A.}~\bibnamefont {Marino}}, \bibinfo {author} {\bibfnamefont {Y.~F.}\ \bibnamefont {Liao}}, \bibinfo {author} {\bibfnamefont {K.~D.}\ \bibnamefont {Tsuei}}, \bibinfo {author} {\bibfnamefont {C.~T.}\ \bibnamefont {Chen}}, \bibinfo {author} {\bibfnamefont
  {K.-T.}\ \bibnamefont {Ko}}, \bibinfo {author} {\bibfnamefont {A.}~\bibnamefont {G\"unther}}, \bibinfo {author} {\bibfnamefont {S.~G.}\ \bibnamefont {Ebbinghaus}}, \bibinfo {author} {\bibfnamefont {J.~W.}\ \bibnamefont {Seo}}, \bibinfo {author} {\bibfnamefont {D.~H.}\ \bibnamefont {Lee}}, \bibinfo {author} {\bibfnamefont {G.}~\bibnamefont {Ryu}}, \bibinfo {author} {\bibfnamefont {A.~C.}\ \bibnamefont {Komarek}}, \bibinfo {author} {\bibfnamefont {S.}~\bibnamefont {Sugano}}, \bibinfo {author} {\bibfnamefont {Y.}~\bibnamefont {Shimakawa}}, \bibinfo {author} {\bibfnamefont {A.}~\bibnamefont {Tanaka}}, \bibinfo {author} {\bibfnamefont {T.}~\bibnamefont {Mizokawa}}, \bibinfo {author} {\bibfnamefont {J.}~\bibnamefont {Kune\ifmmode~\check{s}\else \v{s}\fi{}}}, \bibinfo {author} {\bibfnamefont {L.~H.}\ \bibnamefont {Tjeng}},\ and\ \bibinfo {author} {\bibfnamefont {A.}~\bibnamefont {Hariki}},\ }\bibfield  {title} {\bibinfo {title} {{CaCu$_{3}$Ru$_{4}$O$_{12}$} : A high-{Kondo}-temperature transition-metal oxide},\
  }\href {https://doi.org/10.1103/PhysRevX.12.011017} {\bibfield  {journal} {\bibinfo  {journal} {Phys. Rev. X}\ }\textbf {\bibinfo {volume} {12}},\ \bibinfo {pages} {011017} (\bibinfo {year} {2022})}\BibitemShut {NoStop}%
\bibitem [{\citenamefont {Klein}\ \emph {et~al.}(2008)\citenamefont {Klein}, \citenamefont {Nuber}, \citenamefont {Reinert}, \citenamefont {Kroha}, \citenamefont {Stockert},\ and\ \citenamefont {v.~L\"ohneysen}}]{kondoCeCu}%
  \BibitemOpen
  \bibfield  {author} {\bibinfo {author} {\bibfnamefont {M.}~\bibnamefont {Klein}}, \bibinfo {author} {\bibfnamefont {A.}~\bibnamefont {Nuber}}, \bibinfo {author} {\bibfnamefont {F.}~\bibnamefont {Reinert}}, \bibinfo {author} {\bibfnamefont {J.}~\bibnamefont {Kroha}}, \bibinfo {author} {\bibfnamefont {O.}~\bibnamefont {Stockert}},\ and\ \bibinfo {author} {\bibfnamefont {H.}~\bibnamefont {v.~L\"ohneysen}},\ }\bibfield  {title} {\bibinfo {title} {Signature of quantum criticality in photoemission spectroscopy},\ }\href {https://doi.org/10.1103/PhysRevLett.101.266404} {\bibfield  {journal} {\bibinfo  {journal} {Phys. Rev. Lett.}\ }\textbf {\bibinfo {volume} {101}},\ \bibinfo {pages} {266404} (\bibinfo {year} {2008})}\BibitemShut {NoStop}%
\bibitem [{\citenamefont {Patil}\ \emph {et~al.}(2010{\natexlab{a}})\citenamefont {Patil}, \citenamefont {Medicherla}, \citenamefont {Singh}, \citenamefont {Pandey}, \citenamefont {Sampathkumaran},\ and\ \citenamefont {Maiti}}]{swapnil1}%
  \BibitemOpen
  \bibfield  {author} {\bibinfo {author} {\bibfnamefont {S.}~\bibnamefont {Patil}}, \bibinfo {author} {\bibfnamefont {V.~R.~R.}\ \bibnamefont {Medicherla}}, \bibinfo {author} {\bibfnamefont {R.~S.}\ \bibnamefont {Singh}}, \bibinfo {author} {\bibfnamefont {S.~K.}\ \bibnamefont {Pandey}}, \bibinfo {author} {\bibfnamefont {E.~V.}\ \bibnamefont {Sampathkumaran}},\ and\ \bibinfo {author} {\bibfnamefont {K.}~\bibnamefont {Maiti}},\ }\bibfield  {title} {\bibinfo {title} {Kondo resonance in a magnetically ordered compound {Ce$_{2}$RhSi$_{3}$} : Photoemission spectroscopy and ab initio band structure calculations},\ }\href {https://doi.org/10.1103/PhysRevB.82.104428} {\bibfield  {journal} {\bibinfo  {journal} {Phys. Rev. B}\ }\textbf {\bibinfo {volume} {82}},\ \bibinfo {pages} {104428} (\bibinfo {year} {2010}{\natexlab{a}})}\BibitemShut {NoStop}%
\bibitem [{\citenamefont {Patil}\ \emph {et~al.}(2010{\natexlab{b}})\citenamefont {Patil}, \citenamefont {Pandey}, \citenamefont {Medicherla}, \citenamefont {Singh}, \citenamefont {Bindu}, \citenamefont {Sampathkumaran},\ and\ \citenamefont {Maiti}}]{Patil_2010}%
  \BibitemOpen
  \bibfield  {author} {\bibinfo {author} {\bibfnamefont {S.}~\bibnamefont {Patil}}, \bibinfo {author} {\bibfnamefont {S.~K.}\ \bibnamefont {Pandey}}, \bibinfo {author} {\bibfnamefont {V.~R.~R.}\ \bibnamefont {Medicherla}}, \bibinfo {author} {\bibfnamefont {R.~S.}\ \bibnamefont {Singh}}, \bibinfo {author} {\bibfnamefont {R.}~\bibnamefont {Bindu}}, \bibinfo {author} {\bibfnamefont {E.~V.}\ \bibnamefont {Sampathkumaran}},\ and\ \bibinfo {author} {\bibfnamefont {K.}~\bibnamefont {Maiti}},\ }\bibfield  {title} {\bibinfo {title} {Importance of conduction electron correlation in a {Kondo} lattice, {Ce$_{2}$CoSi$_{3}$}},\ }\href {https://doi.org/10.1088/0953-8984/22/25/255602} {\bibfield  {journal} {\bibinfo  {journal} {J. Phys. Condens. Matter.}\ }\textbf {\bibinfo {volume} {22}},\ \bibinfo {pages} {255602} (\bibinfo {year} {2010}{\natexlab{b}})}\BibitemShut {NoStop}%
\bibitem [{\citenamefont {Patil}\ \emph {et~al.}(2011)\citenamefont {Patil}, \citenamefont {Medicherla}, \citenamefont {Singh}, \citenamefont {Sampathkumaran},\ and\ \citenamefont {Maiti}}]{Patil_2012}%
  \BibitemOpen
  \bibfield  {author} {\bibinfo {author} {\bibfnamefont {S.}~\bibnamefont {Patil}}, \bibinfo {author} {\bibfnamefont {V.~R.~R.}\ \bibnamefont {Medicherla}}, \bibinfo {author} {\bibfnamefont {R.~S.}\ \bibnamefont {Singh}}, \bibinfo {author} {\bibfnamefont {E.~V.}\ \bibnamefont {Sampathkumaran}},\ and\ \bibinfo {author} {\bibfnamefont {K.}~\bibnamefont {Maiti}},\ }\bibfield  {title} {\bibinfo {title} {Evolution of the kondo resonance feature and its relationship to spin-orbit coupling across the quantum critical point in {Ce$_{2}$Rh$_{1-x}$Co$_{x}$Si$_{3}$}},\ }\href {https://doi.org/10.1209/0295-5075/97/17004} {\bibfield  {journal} {\bibinfo  {journal} {EPL}\ }\textbf {\bibinfo {volume} {97}},\ \bibinfo {pages} {17004} (\bibinfo {year} {2011})}\BibitemShut {NoStop}%
\bibitem [{\citenamefont {Chowdhury}\ \emph {et~al.}(2021)\citenamefont {Chowdhury}, \citenamefont {DuttaGupta}, \citenamefont {Patra}, \citenamefont {Tretiakov}, \citenamefont {Sharma}, \citenamefont {Fukami}, \citenamefont {Ohno},\ and\ \citenamefont {Singh}}]{chowdhury_unconventional_2021}%
  \BibitemOpen
  \bibfield  {author} {\bibinfo {author} {\bibfnamefont {R.~R.}\ \bibnamefont {Chowdhury}}, \bibinfo {author} {\bibfnamefont {S.}~\bibnamefont {DuttaGupta}}, \bibinfo {author} {\bibfnamefont {C.}~\bibnamefont {Patra}}, \bibinfo {author} {\bibfnamefont {O.~A.}\ \bibnamefont {Tretiakov}}, \bibinfo {author} {\bibfnamefont {S.}~\bibnamefont {Sharma}}, \bibinfo {author} {\bibfnamefont {S.}~\bibnamefont {Fukami}}, \bibinfo {author} {\bibfnamefont {H.}~\bibnamefont {Ohno}},\ and\ \bibinfo {author} {\bibfnamefont {R.~P.}\ \bibnamefont {Singh}},\ }\bibfield  {title} {\bibinfo {title} {{Unconventional {Hall} effect and its variation with {Co}-doping in van der {Waals} {Fe$_3$GeTe$_2$}}},\ }\href {https://www.nature.com/articles/s41598-021-93402-6} {\bibfield  {journal} {\bibinfo  {journal} {Sci. Rep.}\ }\textbf {\bibinfo {volume} {11}},\ \bibinfo {pages} {14121} (\bibinfo {year} {2021})}\BibitemShut {NoStop}%
\bibitem [{\citenamefont {Chowdhury}\ \emph {et~al.}(2025)\citenamefont {Chowdhury}, \citenamefont {Kurebayashi}, \citenamefont {Lustikova}, \citenamefont {Tretiakov}, \citenamefont {Fukami}, \citenamefont {Singh},\ and\ \citenamefont {DuttaGupta}}]{FGTarxiv}%
  \BibitemOpen
  \bibfield  {author} {\bibinfo {author} {\bibfnamefont {R.~R.}\ \bibnamefont {Chowdhury}}, \bibinfo {author} {\bibfnamefont {D.}~\bibnamefont {Kurebayashi}}, \bibinfo {author} {\bibfnamefont {J.}~\bibnamefont {Lustikova}}, \bibinfo {author} {\bibfnamefont {O.~A.}\ \bibnamefont {Tretiakov}}, \bibinfo {author} {\bibfnamefont {S.}~\bibnamefont {Fukami}}, \bibinfo {author} {\bibfnamefont {R.~P.}\ \bibnamefont {Singh}},\ and\ \bibinfo {author} {\bibfnamefont {S.}~\bibnamefont {DuttaGupta}},\ }\bibfield  {title} {\bibinfo {title} {Unraveling effects of competing interactions and frustration in vdw ferromagnetic {Fe$_{3}$GeTe$_2$} nanoflake devices},\ }\href {https://arxiv.org/abs/2502.05018} {\  (\bibinfo {year} {2025})},\ \Eprint {https://arxiv.org/abs/2502.05018} {arXiv:2502.05018} \BibitemShut {NoStop}%
\bibitem [{\citenamefont {Wu}\ \emph {et~al.}(2023)\citenamefont {Wu}, \citenamefont {Niu}, \citenamefont {Li}, \citenamefont {Yang}, \citenamefont {Gu}, \citenamefont {Liu}, \citenamefont {Wang}, \citenamefont {Chang}, \citenamefont {Wei}, \citenamefont {Li}, \citenamefont {Liu}, \citenamefont {Zhang}, \citenamefont {Ma}, \citenamefont {He}, \citenamefont {Xu},\ and\ \citenamefont {Pu}}]{Cobalt_doping}%
  \BibitemOpen
  \bibfield  {author} {\bibinfo {author} {\bibfnamefont {Z.}~\bibnamefont {Wu}}, \bibinfo {author} {\bibfnamefont {W.}~\bibnamefont {Niu}}, \bibinfo {author} {\bibfnamefont {W.}~\bibnamefont {Li}}, \bibinfo {author} {\bibfnamefont {J.}~\bibnamefont {Yang}}, \bibinfo {author} {\bibfnamefont {K.}~\bibnamefont {Gu}}, \bibinfo {author} {\bibfnamefont {X.}~\bibnamefont {Liu}}, \bibinfo {author} {\bibfnamefont {X.}~\bibnamefont {Wang}}, \bibinfo {author} {\bibfnamefont {S.}~\bibnamefont {Chang}}, \bibinfo {author} {\bibfnamefont {L.}~\bibnamefont {Wei}}, \bibinfo {author} {\bibfnamefont {F.}~\bibnamefont {Li}}, \bibinfo {author} {\bibfnamefont {P.}~\bibnamefont {Liu}}, \bibinfo {author} {\bibfnamefont {X.}~\bibnamefont {Zhang}}, \bibinfo {author} {\bibfnamefont {J.}~\bibnamefont {Ma}}, \bibinfo {author} {\bibfnamefont {L.}~\bibnamefont {He}}, \bibinfo {author} {\bibfnamefont {Y.}~\bibnamefont {Xu}},\ and\ \bibinfo {author} {\bibfnamefont {Y.}~\bibnamefont {Pu}},\ }\bibfield  {title} {\bibinfo {title} {Cobalt doping
  induced emergent humps of hall resistance in van der waals ferromagnetic nanodevices of {(Fe$_{0.74}$Co$_{0.26}$)$_3$GeTe$_2$}},\ }\href {https://doi.org/10.1063/5.0173456} {\bibfield  {journal} {\bibinfo  {journal} {Appl. Phys. Lett.}\ }\textbf {\bibinfo {volume} {123}},\ \bibinfo {pages} {192403} (\bibinfo {year} {2023})}\BibitemShut {NoStop}%
\bibitem [{\citenamefont {Hwang}\ \emph {et~al.}(2019)\citenamefont {Hwang}, \citenamefont {Coak}, \citenamefont {Lee}, \citenamefont {Ko}, \citenamefont {Oh}, \citenamefont {Jeon}, \citenamefont {Son}, \citenamefont {Zhang}, \citenamefont {Kim},\ and\ \citenamefont {Park}}]{FeCo3GeTe2}%
  \BibitemOpen
  \bibfield  {author} {\bibinfo {author} {\bibfnamefont {I.}~\bibnamefont {Hwang}}, \bibinfo {author} {\bibfnamefont {M.~J.}\ \bibnamefont {Coak}}, \bibinfo {author} {\bibfnamefont {N.}~\bibnamefont {Lee}}, \bibinfo {author} {\bibfnamefont {D.-S.}\ \bibnamefont {Ko}}, \bibinfo {author} {\bibfnamefont {Y.}~\bibnamefont {Oh}}, \bibinfo {author} {\bibfnamefont {I.}~\bibnamefont {Jeon}}, \bibinfo {author} {\bibfnamefont {S.}~\bibnamefont {Son}}, \bibinfo {author} {\bibfnamefont {K.}~\bibnamefont {Zhang}}, \bibinfo {author} {\bibfnamefont {J.}~\bibnamefont {Kim}},\ and\ \bibinfo {author} {\bibfnamefont {J.-G.}\ \bibnamefont {Park}},\ }\bibfield  {title} {\bibinfo {title} {Hard ferromagnetic van-der-waals metal {(Fe,Co)$_3$GeTe$_2$}: a new platform for the study of low-dimensional magnetic quantum criticality},\ }\href {https://doi.org/10.1088/1361-648X/ab3135} {\bibfield  {journal} {\bibinfo  {journal} {J. Phys. Condens. Matter.}\ }\textbf {\bibinfo {volume} {31}},\ \bibinfo {pages} {50LT01} (\bibinfo {year}
  {2019})}\BibitemShut {NoStop}%
\bibitem [{\citenamefont {Blaha}\ \emph {et~al.}(2020)\citenamefont {Blaha}, \citenamefont {Schwarz}, \citenamefont {Tran}, \citenamefont {Laskowski}, \citenamefont {Madsen},\ and\ \citenamefont {Marks}}]{Wien2k2020}%
  \BibitemOpen
  \bibfield  {author} {\bibinfo {author} {\bibfnamefont {P.}~\bibnamefont {Blaha}}, \bibinfo {author} {\bibfnamefont {K.}~\bibnamefont {Schwarz}}, \bibinfo {author} {\bibfnamefont {F.}~\bibnamefont {Tran}}, \bibinfo {author} {\bibfnamefont {R.}~\bibnamefont {Laskowski}}, \bibinfo {author} {\bibfnamefont {G.~K.~H.}\ \bibnamefont {Madsen}},\ and\ \bibinfo {author} {\bibfnamefont {L.~D.}\ \bibnamefont {Marks}},\ }\bibfield  {title} {\bibinfo {title} {{WIEN2k: An {APW}+lo program for calculating the properties of solids}},\ }\href {https://doi.org/10.1063/1.5143061} {\bibfield  {journal} {\bibinfo  {journal} {J. Chem. Phys.}\ }\textbf {\bibinfo {volume} {152}},\ \bibinfo {pages} {074101} (\bibinfo {year} {2020})}\BibitemShut {NoStop}%
\bibitem [{\citenamefont {Perdew}\ \emph {et~al.}(1996)\citenamefont {Perdew}, \citenamefont {Burke},\ and\ \citenamefont {Ernzerhof}}]{perdew_generalized_1996}%
  \BibitemOpen
  \bibfield  {author} {\bibinfo {author} {\bibfnamefont {J.~P.}\ \bibnamefont {Perdew}}, \bibinfo {author} {\bibfnamefont {K.}~\bibnamefont {Burke}},\ and\ \bibinfo {author} {\bibfnamefont {M.}~\bibnamefont {Ernzerhof}},\ }\bibfield  {title} {\bibinfo {title} {{Generalized {Gradient} {Approximation} {Made} {Simple}}},\ }\href {https://link.aps.org/doi/10.1103/PhysRevLett.77.3865} {\bibfield  {journal} {\bibinfo  {journal} {Phys. Rev. Lett.}\ }\textbf {\bibinfo {volume} {77}},\ \bibinfo {pages} {3865} (\bibinfo {year} {1996})}\BibitemShut {NoStop}%
\bibitem [{\citenamefont {Xu}\ \emph {et~al.}(2024{\natexlab{b}})\citenamefont {Xu}, \citenamefont {Wang}, \citenamefont {Jin}, \citenamefont {Liu}, \citenamefont {Liu}, \citenamefont {Song},\ and\ \citenamefont {Tian}}]{Liu}%
  \BibitemOpen
  \bibfield  {author} {\bibinfo {author} {\bibfnamefont {Y.}~\bibnamefont {Xu}}, \bibinfo {author} {\bibfnamefont {Y.-C.}\ \bibnamefont {Wang}}, \bibinfo {author} {\bibfnamefont {X.}~\bibnamefont {Jin}}, \bibinfo {author} {\bibfnamefont {H.}~\bibnamefont {Liu}}, \bibinfo {author} {\bibfnamefont {Y.}~\bibnamefont {Liu}}, \bibinfo {author} {\bibfnamefont {H.}~\bibnamefont {Song}},\ and\ \bibinfo {author} {\bibfnamefont {F.}~\bibnamefont {Tian}},\ }\bibfield  {title} {\bibinfo {title} {{Mechanism of magnetic phase transition in correlated magnetic metal: insight into itinerant ferromagnet {Fe$_{3-\delta}$GeTe$_2$}}},\ }\href {https://doi.org/10.1038/s42005-024-01874-5} {\bibfield  {journal} {\bibinfo  {journal} {Comm. Phys}\ }\textbf {\bibinfo {volume} {7}},\ \bibinfo {pages} {381} (\bibinfo {year} {2024}{\natexlab{b}})}\BibitemShut {NoStop}%
\bibitem [{\citenamefont {Lee}\ \emph {et~al.}(2023)\citenamefont {Lee}, \citenamefont {Yan}, \citenamefont {Oh}, \citenamefont {Hwang}, \citenamefont {Denlinger}, \citenamefont {Hwang}, \citenamefont {Lei}, \citenamefont {Mo}, \citenamefont {Park},\ and\ \citenamefont {Ryu}}]{Lee}%
  \BibitemOpen
  \bibfield  {author} {\bibinfo {author} {\bibfnamefont {J.-E.}\ \bibnamefont {Lee}}, \bibinfo {author} {\bibfnamefont {S.}~\bibnamefont {Yan}}, \bibinfo {author} {\bibfnamefont {S.}~\bibnamefont {Oh}}, \bibinfo {author} {\bibfnamefont {J.}~\bibnamefont {Hwang}}, \bibinfo {author} {\bibfnamefont {J.~D.}\ \bibnamefont {Denlinger}}, \bibinfo {author} {\bibfnamefont {C.}~\bibnamefont {Hwang}}, \bibinfo {author} {\bibfnamefont {H.}~\bibnamefont {Lei}}, \bibinfo {author} {\bibfnamefont {S.-K.}\ \bibnamefont {Mo}}, \bibinfo {author} {\bibfnamefont {S.~Y.}\ \bibnamefont {Park}},\ and\ \bibinfo {author} {\bibfnamefont {H.}~\bibnamefont {Ryu}},\ }\bibfield  {title} {\bibinfo {title} {Electronic structure of above-room-temperature van der waals ferromagnet {Fe$_3$GaTe$_2$}},\ }\href {https://doi.org/10.1021/acs.nanolett.3c03203} {\bibfield  {journal} {\bibinfo  {journal} {Nano Lett.}\ }\textbf {\bibinfo {volume} {23}},\ \bibinfo {pages} {11526} (\bibinfo {year} {2023})}\BibitemShut {NoStop}%
\bibitem [{\citenamefont {Jang}\ \emph {et~al.}(2020)\citenamefont {Jang}, \citenamefont {Yoon}, \citenamefont {Jeong}, \citenamefont {Ryee}, \citenamefont {Kim},\ and\ \citenamefont {Han}}]{C9NR10171C}%
  \BibitemOpen
  \bibfield  {author} {\bibinfo {author} {\bibfnamefont {S.~W.}\ \bibnamefont {Jang}}, \bibinfo {author} {\bibfnamefont {H.}~\bibnamefont {Yoon}}, \bibinfo {author} {\bibfnamefont {M.~Y.}\ \bibnamefont {Jeong}}, \bibinfo {author} {\bibfnamefont {S.}~\bibnamefont {Ryee}}, \bibinfo {author} {\bibfnamefont {H.-S.}\ \bibnamefont {Kim}},\ and\ \bibinfo {author} {\bibfnamefont {M.~J.}\ \bibnamefont {Han}},\ }\bibfield  {title} {\bibinfo {title} {{Origin of ferromagnetism and the effect of doping on {Fe$_3$GeTe$_2$}}},\ }\href {https://doi.org/10.1039/C9NR10171C} {\bibfield  {journal} {\bibinfo  {journal} {Nanoscale}\ }\textbf {\bibinfo {volume} {12}},\ \bibinfo {pages} {13501} (\bibinfo {year} {2020})}\BibitemShut {NoStop}%
\bibitem [{\citenamefont {Haule}\ \emph {et~al.}(2010)\citenamefont {Haule}, \citenamefont {Yee},\ and\ \citenamefont {Kim}}]{haule_dynamical_2010}%
  \BibitemOpen
  \bibfield  {author} {\bibinfo {author} {\bibfnamefont {K.}~\bibnamefont {Haule}}, \bibinfo {author} {\bibfnamefont {C.-H.}\ \bibnamefont {Yee}},\ and\ \bibinfo {author} {\bibfnamefont {K.}~\bibnamefont {Kim}},\ }\bibfield  {title} {\bibinfo {title} {{Dynamical mean-field theory within the full-potential methods: {Electronic} structure of {CeIrIn}$_5$ , {CeCoIn}$_5$ , and {CeRhIn}$_5$}},\ }\href {https://link.aps.org/doi/10.1103/PhysRevB.81.195107} {\bibfield  {journal} {\bibinfo  {journal} {Phys. Rev. B}\ }\textbf {\bibinfo {volume} {81}},\ \bibinfo {pages} {195107} (\bibinfo {year} {2010})}\BibitemShut {NoStop}%
\bibitem [{\citenamefont {Haule}(2007)}]{haule_quantum_2007}%
  \BibitemOpen
  \bibfield  {author} {\bibinfo {author} {\bibfnamefont {K.}~\bibnamefont {Haule}},\ }\bibfield  {title} {\bibinfo {title} {{Quantum {Monte} {Carlo} impurity solver for cluster dynamical mean-field theory and electronic structure calculations with adjustable cluster base}},\ }\href {https://link.aps.org/doi/10.1103/PhysRevB.75.155113} {\bibfield  {journal} {\bibinfo  {journal} {Phys. Rev. B}\ }\textbf {\bibinfo {volume} {75}},\ \bibinfo {pages} {155113} (\bibinfo {year} {2007})}\BibitemShut {NoStop}%
\bibitem [{\citenamefont {Haule}(2015)}]{haule_exact_2015}%
  \BibitemOpen
  \bibfield  {author} {\bibinfo {author} {\bibfnamefont {K.}~\bibnamefont {Haule}},\ }\bibfield  {title} {\bibinfo {title} {Exact double counting in combining the dynamical mean field theory and the density functional theory},\ }\href {https://link.aps.org/doi/10.1103/PhysRevLett.115.196403} {\bibfield  {journal} {\bibinfo  {journal} {Phys. Rev. Lett.}\ }\textbf {\bibinfo {volume} {115}},\ \bibinfo {pages} {196403} (\bibinfo {year} {2015})}\BibitemShut {NoStop}%
\bibitem [{\citenamefont {Kim}\ \emph {et~al.}(2018)\citenamefont {Kim}, \citenamefont {Seo}, \citenamefont {Lee}, \citenamefont {Ko}, \citenamefont {Kim}, \citenamefont {Jang}, \citenamefont {Ok}, \citenamefont {Lee}, \citenamefont {Jo}, \citenamefont {Kang}, \citenamefont {Shim}, \citenamefont {Kim}, \citenamefont {Yeom}, \citenamefont {I~Min}, \citenamefont {Yang},\ and\ \citenamefont {Kim}}]{kim_large_2018}%
  \BibitemOpen
  \bibfield  {author} {\bibinfo {author} {\bibfnamefont {K.}~\bibnamefont {Kim}}, \bibinfo {author} {\bibfnamefont {J.}~\bibnamefont {Seo}}, \bibinfo {author} {\bibfnamefont {E.}~\bibnamefont {Lee}}, \bibinfo {author} {\bibfnamefont {K.-T.}\ \bibnamefont {Ko}}, \bibinfo {author} {\bibfnamefont {B.~S.}\ \bibnamefont {Kim}}, \bibinfo {author} {\bibfnamefont {B.~G.}\ \bibnamefont {Jang}}, \bibinfo {author} {\bibfnamefont {J.~M.}\ \bibnamefont {Ok}}, \bibinfo {author} {\bibfnamefont {J.}~\bibnamefont {Lee}}, \bibinfo {author} {\bibfnamefont {Y.~J.}\ \bibnamefont {Jo}}, \bibinfo {author} {\bibfnamefont {W.}~\bibnamefont {Kang}}, \bibinfo {author} {\bibfnamefont {J.~H.}\ \bibnamefont {Shim}}, \bibinfo {author} {\bibfnamefont {C.}~\bibnamefont {Kim}}, \bibinfo {author} {\bibfnamefont {H.~W.}\ \bibnamefont {Yeom}}, \bibinfo {author} {\bibfnamefont {B.}~\bibnamefont {I~Min}}, \bibinfo {author} {\bibfnamefont {B.-J.}\ \bibnamefont {Yang}},\ and\ \bibinfo {author} {\bibfnamefont {J.~S.}\ \bibnamefont {Kim}},\ }\bibfield
   {title} {\bibinfo {title} {{Large anomalous {Hall} current induced by topological nodal lines in a ferromagnetic van der {Waals} semimetal}},\ }\href {https://www.nature.com/articles/s41563-018-0132-3} {\bibfield  {journal} {\bibinfo  {journal} {Nat. Mater.}\ }\textbf {\bibinfo {volume} {17}},\ \bibinfo {pages} {794} (\bibinfo {year} {2018})}\BibitemShut {NoStop}%
\bibitem [{\citenamefont {Ghosh}\ \emph {et~al.}(2023)\citenamefont {Ghosh}, \citenamefont {Ershadrad}, \citenamefont {Borisov},\ and\ \citenamefont {Sanyal}}]{ghosh_unraveling_2023}%
  \BibitemOpen
  \bibfield  {author} {\bibinfo {author} {\bibfnamefont {S.}~\bibnamefont {Ghosh}}, \bibinfo {author} {\bibfnamefont {S.}~\bibnamefont {Ershadrad}}, \bibinfo {author} {\bibfnamefont {V.}~\bibnamefont {Borisov}},\ and\ \bibinfo {author} {\bibfnamefont {B.}~\bibnamefont {Sanyal}},\ }\bibfield  {title} {\bibinfo {title} {{Unraveling effects of electron correlation in two-dimensional {Fe$_n$GeTe$_2$} (n = 3, 4, 5) by dynamical mean field theory}},\ }\href {https://www.nature.com/articles/s41524-023-01024-5} {\bibfield  {journal} {\bibinfo  {journal} {NPJ Comput. Mater.}\ }\textbf {\bibinfo {volume} {9}},\ \bibinfo {pages} {1} (\bibinfo {year} {2023})}\BibitemShut {NoStop}%
\bibitem [{\citenamefont {Zhang}\ \emph {et~al.}(2021)\citenamefont {Zhang}, \citenamefont {Liang}, \citenamefont {Zhao}, \citenamefont {Chen}, \citenamefont {He}, \citenamefont {Liu}, \citenamefont {Lv}, \citenamefont {Yang}, \citenamefont {Wu},\ and\ \citenamefont {Chen}}]{ZHANG2021127219}%
  \BibitemOpen
\bibfield  {journal} {  }\bibfield  {author} {\bibinfo {author} {\bibfnamefont {S.}~\bibnamefont {Zhang}}, \bibinfo {author} {\bibfnamefont {X.}~\bibnamefont {Liang}}, \bibinfo {author} {\bibfnamefont {H.}~\bibnamefont {Zhao}}, \bibinfo {author} {\bibfnamefont {Y.}~\bibnamefont {Chen}}, \bibinfo {author} {\bibfnamefont {Q.}~\bibnamefont {He}}, \bibinfo {author} {\bibfnamefont {J.}~\bibnamefont {Liu}}, \bibinfo {author} {\bibfnamefont {L.}~\bibnamefont {Lv}}, \bibinfo {author} {\bibfnamefont {J.}~\bibnamefont {Yang}}, \bibinfo {author} {\bibfnamefont {H.}~\bibnamefont {Wu}},\ and\ \bibinfo {author} {\bibfnamefont {L.}~\bibnamefont {Chen}},\ }\bibfield  {title} {\bibinfo {title} {Tuning the magnetic properties of {Fe$_3$GeTe$_2$} by doping with 3$d$ transition-metals},\ }\href {https://doi.org/https://doi.org/10.1016/j.physleta.2021.127219} {\bibfield  {journal} {\bibinfo  {journal} {Phys. Lett. A.}\ }\textbf {\bibinfo {volume} {396}},\ \bibinfo {pages} {127219} (\bibinfo {year} {2021})}\BibitemShut {NoStop}%
\bibitem [{\citenamefont {Roy~Chowdhury}\ \emph {et~al.}(2022)\citenamefont {Roy~Chowdhury}, \citenamefont {Patra}, \citenamefont {DuttaGupta}, \citenamefont {Satheesh}, \citenamefont {Dan}, \citenamefont {Fukami},\ and\ \citenamefont {Singh}}]{roy_chowdhury_modification_2022}%
  \BibitemOpen
  \bibfield  {author} {\bibinfo {author} {\bibfnamefont {R.}~\bibnamefont {Roy~Chowdhury}}, \bibinfo {author} {\bibfnamefont {C.}~\bibnamefont {Patra}}, \bibinfo {author} {\bibfnamefont {S.}~\bibnamefont {DuttaGupta}}, \bibinfo {author} {\bibfnamefont {S.}~\bibnamefont {Satheesh}}, \bibinfo {author} {\bibfnamefont {S.}~\bibnamefont {Dan}}, \bibinfo {author} {\bibfnamefont {S.}~\bibnamefont {Fukami}},\ and\ \bibinfo {author} {\bibfnamefont {R.~P.}\ \bibnamefont {Singh}},\ }\bibfield  {title} {\bibinfo {title} {{Modification of unconventional {Hall} effect with doping at the nonmagnetic site in a two-dimensional van der {Waals} ferromagnet}},\ }\href {https://link.aps.org/doi/10.1103/PhysRevMaterials.6.014002} {\bibfield  {journal} {\bibinfo  {journal} {Phys. Rev. Mater.}\ }\textbf {\bibinfo {volume} {6}},\ \bibinfo {pages} {014002} (\bibinfo {year} {2022})}\BibitemShut {NoStop}%
\bibitem [{\citenamefont {Chang}\ \emph {et~al.}(2025)\citenamefont {Chang}, \citenamefont {Wang}, \citenamefont {Wu}, \citenamefont {Yuan}, \citenamefont {Gao}, \citenamefont {Wang}, \citenamefont {Wang}, \citenamefont {Liu}, \citenamefont {Gu}, \citenamefont {Liu}, \citenamefont {Zhang},\ and\ \citenamefont {Niu}}]{CHANG2025177837}%
  \BibitemOpen
  \bibfield  {author} {\bibinfo {author} {\bibfnamefont {S.}~\bibnamefont {Chang}}, \bibinfo {author} {\bibfnamefont {M.}~\bibnamefont {Wang}}, \bibinfo {author} {\bibfnamefont {Z.}~\bibnamefont {Wu}}, \bibinfo {author} {\bibfnamefont {K.}~\bibnamefont {Yuan}}, \bibinfo {author} {\bibfnamefont {J.}~\bibnamefont {Gao}}, \bibinfo {author} {\bibfnamefont {S.}~\bibnamefont {Wang}}, \bibinfo {author} {\bibfnamefont {Z.}~\bibnamefont {Wang}}, \bibinfo {author} {\bibfnamefont {K.}~\bibnamefont {Liu}}, \bibinfo {author} {\bibfnamefont {K.}~\bibnamefont {Gu}}, \bibinfo {author} {\bibfnamefont {P.}~\bibnamefont {Liu}}, \bibinfo {author} {\bibfnamefont {X.}~\bibnamefont {Zhang}},\ and\ \bibinfo {author} {\bibfnamefont {W.}~\bibnamefont {Niu}},\ }\bibfield  {title} {\bibinfo {title} {{Critical behavior of cobalt-doped van der Waals ferromagnet (Fe$_{0.74}$Co$_{0.26}$)$_3$GeTe$_2$}},\ }\href {https://doi.org/https://doi.org/10.1016/j.jallcom.2024.177837} {\bibfield  {journal} {\bibinfo  {journal} {J. Alloys Compd.}\
  }\textbf {\bibinfo {volume} {1010}},\ \bibinfo {pages} {177837} (\bibinfo {year} {2025})}\BibitemShut {NoStop}%
\bibitem [{\citenamefont {Doniach}\ and\ \citenamefont {Sunjic}(1970)}]{SDoniach_1970}%
  \BibitemOpen
  \bibfield  {author} {\bibinfo {author} {\bibfnamefont {S.}~\bibnamefont {Doniach}}\ and\ \bibinfo {author} {\bibfnamefont {M.}~\bibnamefont {Sunjic}},\ }\bibfield  {title} {\bibinfo {title} {{Many-electron singularity in X-ray photoemission and X-ray line spectra from metals}},\ }\href {https://doi.org/10.1088/0022-3719/3/2/010} {\bibfield  {journal} {\bibinfo  {journal} {J. Phys. C: Solid State Phys}\ }\textbf {\bibinfo {volume} {3}},\ \bibinfo {pages} {285} (\bibinfo {year} {1970})}\BibitemShut {NoStop}%
\bibitem [{SM()}]{SM}%
  \BibitemOpen
  \href@noop {} {\bibinfo  {journal} {See Supplemental Material at [URL will be inserted by publisher] for further experimental, DFT and DFT+DMFT results and analysis.}\ }\BibitemShut {NoStop}%
\bibitem [{\citenamefont {Moulder}\ and\ \citenamefont {Chastain}(1992)}]{Handbook}%
  \BibitemOpen
  \bibfield  {author} {\bibinfo {author} {\bibfnamefont {J.}~\bibnamefont {Moulder}}\ and\ \bibinfo {author} {\bibfnamefont {J.}~\bibnamefont {Chastain}},\ }\href {https://books.google.co.in/books?id=A_XGQgAACAAJ} {\emph {\bibinfo {title} {{Handbook of X-ray Photoelectron Spectroscopy: A Reference Book of Standard Spectra for Identification and Interpretation of XPS Data}}}}\ (\bibinfo  {publisher} {Physical Electronics Division, Perkin-Elmer Corporation},\ \bibinfo {year} {1992})\BibitemShut {NoStop}%
\bibitem [{\citenamefont {Klebanoff}\ \emph {et~al.}(1994)\citenamefont {Klebanoff}, \citenamefont {Van~Campen},\ and\ \citenamefont {Pouliot}}]{Coexchangesplitt}%
  \BibitemOpen
  \bibfield  {author} {\bibinfo {author} {\bibfnamefont {L.~E.}\ \bibnamefont {Klebanoff}}, \bibinfo {author} {\bibfnamefont {D.~G.}\ \bibnamefont {Van~Campen}},\ and\ \bibinfo {author} {\bibfnamefont {R.~J.}\ \bibnamefont {Pouliot}},\ }\bibfield  {title} {\bibinfo {title} {Spin-resolved and high-energy-resolution xps studies of cobalt metal and a cobalt magnetic glass},\ }\href {https://doi.org/10.1103/PhysRevB.49.2047} {\bibfield  {journal} {\bibinfo  {journal} {Phys. Rev. B}\ }\textbf {\bibinfo {volume} {49}},\ \bibinfo {pages} {2047} (\bibinfo {year} {1994})}\BibitemShut {NoStop}%
\bibitem [{\citenamefont {Rossi}\ \emph {et~al.}(1997)\citenamefont {Rossi}, \citenamefont {Panaccione}, \citenamefont {Sirotti}, \citenamefont {Lizzit}, \citenamefont {Baraldi},\ and\ \citenamefont {Paolucci}}]{PhysRevB.55.11488}%
  \BibitemOpen
  \bibfield  {author} {\bibinfo {author} {\bibfnamefont {G.}~\bibnamefont {Rossi}}, \bibinfo {author} {\bibfnamefont {G.}~\bibnamefont {Panaccione}}, \bibinfo {author} {\bibfnamefont {F.}~\bibnamefont {Sirotti}}, \bibinfo {author} {\bibfnamefont {S.}~\bibnamefont {Lizzit}}, \bibinfo {author} {\bibfnamefont {A.}~\bibnamefont {Baraldi}},\ and\ \bibinfo {author} {\bibfnamefont {G.}~\bibnamefont {Paolucci}},\ }\bibfield  {title} {\bibinfo {title} {{Magnetic dichroism in the angular distribution of {F}e 2$p$ and 3$p$ photoelectrons: Empirical support to Zeeman-like analysis}},\ }\href {https://doi.org/10.1103/PhysRevB.55.11488} {\bibfield  {journal} {\bibinfo  {journal} {Phys. Rev. B}\ }\textbf {\bibinfo {volume} {55}},\ \bibinfo {pages} {11488} (\bibinfo {year} {1997})}\BibitemShut {NoStop}%
\bibitem [{\citenamefont {Sarkar}\ \emph {et~al.}(2021)\citenamefont {Sarkar}, \citenamefont {Sadhukhan}, \citenamefont {Singh}, \citenamefont {Gloskovskii}, \citenamefont {Deguchi}, \citenamefont {Fujita},\ and\ \citenamefont {Barman}}]{PhysRevResearch.3.013151}%
  \BibitemOpen
  \bibfield  {author} {\bibinfo {author} {\bibfnamefont {S.}~\bibnamefont {Sarkar}}, \bibinfo {author} {\bibfnamefont {P.}~\bibnamefont {Sadhukhan}}, \bibinfo {author} {\bibfnamefont {V.~K.}\ \bibnamefont {Singh}}, \bibinfo {author} {\bibfnamefont {A.}~\bibnamefont {Gloskovskii}}, \bibinfo {author} {\bibfnamefont {K.}~\bibnamefont {Deguchi}}, \bibinfo {author} {\bibfnamefont {N.}~\bibnamefont {Fujita}},\ and\ \bibinfo {author} {\bibfnamefont {S.~R.}\ \bibnamefont {Barman}},\ }\bibfield  {title} {\bibinfo {title} {{Bulk electronic structure of high-order quaternary approximants}},\ }\href {https://doi.org/10.1103/PhysRevResearch.3.013151} {\bibfield  {journal} {\bibinfo  {journal} {Phys. Rev. Res.}\ }\textbf {\bibinfo {volume} {3}},\ \bibinfo {pages} {013151} (\bibinfo {year} {2021})}\BibitemShut {NoStop}%
\bibitem [{\citenamefont {Panda}\ \emph {et~al.}(2024)\citenamefont {Panda}, \citenamefont {Behera}, \citenamefont {Madhur}, \citenamefont {Rana}, \citenamefont {Gloskovskii}, \citenamefont {Otani}, \citenamefont {Barman},\ and\ \citenamefont {Sarkar}}]{IndranilSarkar}%
  \BibitemOpen
  \bibfield  {author} {\bibinfo {author} {\bibfnamefont {D.}~\bibnamefont {Panda}}, \bibinfo {author} {\bibfnamefont {K.~K.}\ \bibnamefont {Behera}}, \bibinfo {author} {\bibfnamefont {S.}~\bibnamefont {Madhur}}, \bibinfo {author} {\bibfnamefont {B.}~\bibnamefont {Rana}}, \bibinfo {author} {\bibfnamefont {A.}~\bibnamefont {Gloskovskii}}, \bibinfo {author} {\bibfnamefont {Y.}~\bibnamefont {Otani}}, \bibinfo {author} {\bibfnamefont {A.}~\bibnamefont {Barman}},\ and\ \bibinfo {author} {\bibfnamefont {I.}~\bibnamefont {Sarkar}},\ }\bibfield  {title} {\bibinfo {title} {{Role of the nonmagnetic underlayer in controlling the electronic structure of ferromagnet/nonmagnetic-metal heterostructures}},\ }\href {https://doi.org/10.1103/PhysRevB.110.094424} {\bibfield  {journal} {\bibinfo  {journal} {Phys. Rev. B}\ }\textbf {\bibinfo {volume} {110}},\ \bibinfo {pages} {094424} (\bibinfo {year} {2024})}\BibitemShut {NoStop}%
\bibitem [{\citenamefont {Yeh}\ and\ \citenamefont {Lindau}(1985)}]{yeh_atomic_1985}%
  \BibitemOpen
  \bibfield  {author} {\bibinfo {author} {\bibfnamefont {J.~J.}\ \bibnamefont {Yeh}}\ and\ \bibinfo {author} {\bibfnamefont {I.}~\bibnamefont {Lindau}},\ }\bibfield  {title} {\bibinfo {title} {{Atomic subshell photoionization cross sections and asymmetry parameters: 1 $\leq$ {Z} $\leq$ 103}},\ }\href {https://www.sciencedirect.com/science/article/pii/0092640X85900166} {\bibfield  {journal} {\bibinfo  {journal} {Atomic Data and Nuclear Data Tables}\ }\textbf {\bibinfo {volume} {32}},\ \bibinfo {pages} {1} (\bibinfo {year} {1985})}\BibitemShut {NoStop}%
\bibitem [{\citenamefont {Maiti}\ \emph {et~al.}(2002)\citenamefont {Maiti}, \citenamefont {Malagoli}, \citenamefont {Dallmeyer},\ and\ \citenamefont {Carbone}}]{maiti_finite_2002}%
  \BibitemOpen
  \bibfield  {author} {\bibinfo {author} {\bibfnamefont {K.}~\bibnamefont {Maiti}}, \bibinfo {author} {\bibfnamefont {M.~C.}\ \bibnamefont {Malagoli}}, \bibinfo {author} {\bibfnamefont {A.}~\bibnamefont {Dallmeyer}},\ and\ \bibinfo {author} {\bibfnamefont {C.}~\bibnamefont {Carbone}},\ }\bibfield  {title} {\bibinfo {title} {{Finite {Temperature} {Magnetism} in {Gd}: {Evidence} against a {Stoner} {Behavior}}},\ }\href {https://link.aps.org/doi/10.1103/PhysRevLett.88.167205} {\bibfield  {journal} {\bibinfo  {journal} {Phys. Rev. Lett.}\ }\textbf {\bibinfo {volume} {88}},\ \bibinfo {pages} {167205} (\bibinfo {year} {2002})}\BibitemShut {NoStop}%
\bibitem [{\citenamefont {Singh}\ \emph {et~al.}(2007)\citenamefont {Singh}, \citenamefont {Medicherla},\ and\ \citenamefont {Maiti}}]{APL}%
  \BibitemOpen
  \bibfield  {author} {\bibinfo {author} {\bibfnamefont {R.~S.}\ \bibnamefont {Singh}}, \bibinfo {author} {\bibfnamefont {V.~R.~R.}\ \bibnamefont {Medicherla}},\ and\ \bibinfo {author} {\bibfnamefont {K.}~\bibnamefont {Maiti}},\ }\bibfield  {title} {\bibinfo {title} {{Role of long range ferromagnetic order in the electronic structure of {Sr$_{1 - x}$Ca$_x$RuO$_3$}}},\ }\href {https://doi.org/10.1063/1.2789731} {\bibfield  {journal} {\bibinfo  {journal} {Appl. Phys. Lett.}\ }\textbf {\bibinfo {volume} {91}},\ \bibinfo {pages} {132503} (\bibinfo {year} {2007})}\BibitemShut {NoStop}%
  \bibitem [{\citenamefont {Krishna-murthy}\ \emph {et~al.}(1980)\citenamefont
  {Krishna-murthy}, \citenamefont {Wilkins},\ and\ \citenamefont
  {Wilson}}]{PhysRevB.21.1003}%
  \BibitemOpen
  \bibfield  {author} {\bibinfo {author} {\bibfnamefont {H.~R.}\ \bibnamefont
  {Krishna-murthy}}, \bibinfo {author} {\bibfnamefont {J.~W.}\ \bibnamefont
  {Wilkins}},\ and\ \bibinfo {author} {\bibfnamefont {K.~G.}\ \bibnamefont
  {Wilson}},\ }\bibfield  {title} {\bibinfo {title} {Renormalization-group
  approach to the {Anderson} model of dilute magnetic alloys. {I. Static}
  properties for the symmetric case},\ }\href
  {https://doi.org/10.1103/PhysRevB.21.1003} {\bibfield  {journal} {\bibinfo
  {journal} {Phys. Rev. B}\ }\textbf {\bibinfo {volume} {21}},\ \bibinfo
  {pages} {1003} (\bibinfo {year} {1980})}\BibitemShut {NoStop}%
\bibitem [{\citenamefont {Bansal}\ \emph {et~al.}(2023)\citenamefont {Bansal}, \citenamefont {Maurya}, \citenamefont {Ali}, \citenamefont {Reddy},\ and\ \citenamefont {Singh}}]{LSNO}%
  \BibitemOpen
  \bibfield  {author} {\bibinfo {author} {\bibfnamefont {S.}~\bibnamefont {Bansal}}, \bibinfo {author} {\bibfnamefont {R.~K.}\ \bibnamefont {Maurya}}, \bibinfo {author} {\bibfnamefont {A.}~\bibnamefont {Ali}}, \bibinfo {author} {\bibfnamefont {B.~H.}\ \bibnamefont {Reddy}},\ and\ \bibinfo {author} {\bibfnamefont {R.~S.}\ \bibnamefont {Singh}},\ }\bibfield  {title} {\bibinfo {title} {{Role of electron correlation and disorder on the electronic structure of layered nickelate ${({\mathrm{La}}_{0.5}{\mathrm{Sr}}_{0.5})}_{2} ${N}i{O}$_4$}},\ }\href {https://link.aps.org/doi/10.1103/PhysRevMaterials.7.064007} {\bibfield  {journal} {\bibinfo  {journal} {Phys. Rev. Mater.}\ }\textbf {\bibinfo {volume} {7}},\ \bibinfo {pages} {064007} (\bibinfo {year} {2023})}\BibitemShut {NoStop}%
\bibitem [{\citenamefont {Reddy}\ \emph {et~al.}(2019)\citenamefont {Reddy}, \citenamefont {Ali},\ and\ \citenamefont {Singh}}]{Reddy_2019}%
  \BibitemOpen
  \bibfield  {author} {\bibinfo {author} {\bibfnamefont {B.~H.}\ \bibnamefont {Reddy}}, \bibinfo {author} {\bibfnamefont {A.}~\bibnamefont {Ali}},\ and\ \bibinfo {author} {\bibfnamefont {R.~S.}\ \bibnamefont {Singh}},\ }\bibfield  {title} {\bibinfo {title} {Role of disorder and strong 5{$d$} electron correlation in the electronic structure of {Sr$_2$TiIrO$_6$}},\ }\href {https://doi.org/10.1209/0295-5075/127/47003} {\bibfield  {journal} {\bibinfo  {journal} {EPL}\ }\textbf {\bibinfo {volume} {127}},\ \bibinfo {pages} {47003} (\bibinfo {year} {2019})}\BibitemShut {NoStop}%
  \bibitem [{\citenamefont {Hüfner}(2007)}]{Hufner2007}%
  \BibitemOpen
  \bibfield  {author} {\bibinfo {author} {\bibfnamefont {S.}~\bibnamefont {Hüfner}},\ }
  \bibfield  {title} {\bibinfo {title} {Very High Resolution Photoelectron Spectroscopy},\ } \href {https://doi.org/10.1007/3-540-68133-7} {\bibfield  {journal} {\bibinfo  {journal} {Springer Berlin}\ } \bibinfo {pages} {XIV, 397} (\bibinfo {year} {2007})}\BibitemShut {NoStop}%
\bibitem [{\citenamefont {Lee}\ and\ \citenamefont {Ramakrishnan}(1985)}]{Disordered}%
  \BibitemOpen
  \bibfield  {author} {\bibinfo {author} {\bibfnamefont {P.~A.}\ \bibnamefont {Lee}}\ and\ \bibinfo {author} {\bibfnamefont {T.~V.}\ \bibnamefont {Ramakrishnan}},\ }\bibfield  {title} {\bibinfo {title} {Disordered electronic systems},\ }\href {https://doi.org/10.1103/RevModPhys.57.287} {\bibfield  {journal} {\bibinfo  {journal} {Rev. Mod. Phys.}\ }\textbf {\bibinfo {volume} {57}},\ \bibinfo {pages} {287} (\bibinfo {year} {1985})}\BibitemShut {NoStop}%
\bibitem [{\citenamefont {Jiang}\ \emph {et~al.}(2021)\citenamefont {Jiang}, \citenamefont {Lee}, \citenamefont {Lee}, \citenamefont {Lau}, \citenamefont {Li}, \citenamefont {Pedersen}, \citenamefont {Liu}, \citenamefont {Gorovikov}, \citenamefont {Zhdanovich}, \citenamefont {Damascelli}, \citenamefont {Zou}, \citenamefont {Walker}, \citenamefont {Ismail-Beigi},\ and\ \citenamefont {Ahn}}]{Electronic}%
  \BibitemOpen
  \bibfield  {author} {\bibinfo {author} {\bibfnamefont {J.}~\bibnamefont {Jiang}}, \bibinfo {author} {\bibfnamefont {A.~T.}\ \bibnamefont {Lee}}, \bibinfo {author} {\bibfnamefont {S.}~\bibnamefont {Lee}}, \bibinfo {author} {\bibfnamefont {C.}~\bibnamefont {Lau}}, \bibinfo {author} {\bibfnamefont {M.}~\bibnamefont {Li}}, \bibinfo {author} {\bibfnamefont {T.~M.}\ \bibnamefont {Pedersen}}, \bibinfo {author} {\bibfnamefont {C.}~\bibnamefont {Liu}}, \bibinfo {author} {\bibfnamefont {S.}~\bibnamefont {Gorovikov}}, \bibinfo {author} {\bibfnamefont {S.}~\bibnamefont {Zhdanovich}}, \bibinfo {author} {\bibfnamefont {A.}~\bibnamefont {Damascelli}}, \bibinfo {author} {\bibfnamefont {K.}~\bibnamefont {Zou}}, \bibinfo {author} {\bibfnamefont {F.~J.}\ \bibnamefont {Walker}}, \bibinfo {author} {\bibfnamefont {S.}~\bibnamefont {Ismail-Beigi}},\ and\ \bibinfo {author} {\bibfnamefont {C.~H.}\ \bibnamefont {Ahn}},\ }\bibfield  {title} {\bibinfo {title} {Electronic properties of epitaxial {La$_{1-x}$Sr$_x$RhO$_3$} thin films},\
  }\href {https://doi.org/10.1103/PhysRevB.103.195153} {\bibfield  {journal} {\bibinfo  {journal} {Phys. Rev. B}\ }\textbf {\bibinfo {volume} {103}},\ \bibinfo {pages} {195153} (\bibinfo {year} {2021})}\BibitemShut {NoStop}%
\bibitem [{\citenamefont {Kobayashi}\ \emph {et~al.}(2007)\citenamefont {Kobayashi}, \citenamefont {Tanaka}, \citenamefont {Fujimori}, \citenamefont {Ray},\ and\ \citenamefont {Sarma}}]{Altshuler}%
  \BibitemOpen
  \bibfield  {author} {\bibinfo {author} {\bibfnamefont {M.}~\bibnamefont {Kobayashi}}, \bibinfo {author} {\bibfnamefont {K.}~\bibnamefont {Tanaka}}, \bibinfo {author} {\bibfnamefont {A.}~\bibnamefont {Fujimori}}, \bibinfo {author} {\bibfnamefont {S.}~\bibnamefont {Ray}},\ and\ \bibinfo {author} {\bibfnamefont {D.~D.}\ \bibnamefont {Sarma}},\ }\bibfield  {title} {\bibinfo {title} {Critical test for altshuler-aronov theory: Evolution of the density of states singularity in double perovskite {Sr$_2$FeMoO$_6$} with controlled disorder},\ }\href {https://doi.org/10.1103/PhysRevLett.98.246401} {\bibfield  {journal} {\bibinfo  {journal} {Phys. Rev. Lett.}\ }\textbf {\bibinfo {volume} {98}},\ \bibinfo {pages} {246401} (\bibinfo {year} {2007})}\BibitemShut {NoStop}%
\bibitem [{\citenamefont {Altshuler}\ and\ \citenamefont {Aronov}(1979)}]{ALTSHULER1979115}%
  \BibitemOpen
  \bibfield  {author} {\bibinfo {author} {\bibfnamefont {B.}~\bibnamefont {Altshuler}}\ and\ \bibinfo {author} {\bibfnamefont {A.}~\bibnamefont {Aronov}},\ }\bibfield  {title} {\bibinfo {title} {Zero bias anomaly in tunnel resistance and electron-electron interaction},\ }\href {https://doi.org/https://doi.org/10.1016/0038-1098(79)90967-0} {\bibfield  {journal} {\bibinfo  {journal} {Solid State Commun.}\ }\textbf {\bibinfo {volume} {30}},\ \bibinfo {pages} {115} (\bibinfo {year} {1979})}\BibitemShut {NoStop}%
\bibitem [{\citenamefont {Stewart}(1984)}]{Heavy-fermion}%
  \BibitemOpen
  \bibfield  {author} {\bibinfo {author} {\bibfnamefont {G.~R.}\ \bibnamefont {Stewart}},\ }\bibfield  {title} {\bibinfo {title} {Heavy-fermion systems},\ }\href {https://doi.org/10.1103/RevModPhys.56.755} {\bibfield  {journal} {\bibinfo  {journal} {Rev. Mod. Phys.}\ }\textbf {\bibinfo {volume} {56}},\ \bibinfo {pages} {755} (\bibinfo {year} {1984})}\BibitemShut {NoStop}%
\bibitem [{\citenamefont {Pfleiderer}(2009)}]{f-electron}%
  \BibitemOpen
  \bibfield  {author} {\bibinfo {author} {\bibfnamefont {C.}~\bibnamefont {Pfleiderer}},\ }\bibfield  {title} {\bibinfo {title} {Superconducting phases of $f$-electron compounds},\ }\href {https://doi.org/10.1103/RevModPhys.81.1551} {\bibfield  {journal} {\bibinfo  {journal} {Rev. Mod. Phys.}\ }\textbf {\bibinfo {volume} {81}},\ \bibinfo {pages} {1551} (\bibinfo {year} {2009})}\BibitemShut {NoStop}%
\bibitem [{\citenamefont {Wirth}\ and\ \citenamefont {Steglich}(2016)}]{heavyfermionsreview}%
  \BibitemOpen
  \bibfield  {author} {\bibinfo {author} {\bibfnamefont {S.}~\bibnamefont {Wirth}}\ and\ \bibinfo {author} {\bibfnamefont {F.}~\bibnamefont {Steglich}},\ }\bibfield  {title} {\bibinfo {title} {Exploring heavy fermions from macroscopic to microscopic length scales},\ }\href {https://doi.org/10.1038/natrevmats.2016.51} {\bibfield  {journal} {\bibinfo  {journal} {Nat. Rev. Mater.}\ }\textbf {\bibinfo {volume} {1}},\ \bibinfo {pages} {16051} (\bibinfo {year} {2016})}\BibitemShut {NoStop}%
\bibitem [{\citenamefont {Calvo}\ \emph {et~al.}(2009)\citenamefont {Calvo}, \citenamefont {Fern{\'a}ndez-Rossier}, \citenamefont {Palacios}, \citenamefont {Jacob}, \citenamefont {Natelson},\ and\ \citenamefont {Untiedt}}]{kondoFCN}%
  \BibitemOpen
  \bibfield  {author} {\bibinfo {author} {\bibfnamefont {M.~R.}\ \bibnamefont {Calvo}}, \bibinfo {author} {\bibfnamefont {J.}~\bibnamefont {Fern{\'a}ndez-Rossier}}, \bibinfo {author} {\bibfnamefont {J.~J.}\ \bibnamefont {Palacios}}, \bibinfo {author} {\bibfnamefont {D.}~\bibnamefont {Jacob}}, \bibinfo {author} {\bibfnamefont {D.}~\bibnamefont {Natelson}},\ and\ \bibinfo {author} {\bibfnamefont {C.}~\bibnamefont {Untiedt}},\ }\bibfield  {title} {\bibinfo {title} {The kondo effect in ferromagnetic atomic contacts},\ }\href {https://doi.org/10.1038/nature07878} {\bibfield  {journal} {\bibinfo  {journal} {Nature}\ }\textbf {\bibinfo {volume} {458}},\ \bibinfo {pages} {1150} (\bibinfo {year} {2009})}\BibitemShut {NoStop}%
\bibitem [{\citenamefont {Tian}\ \emph {et~al.}(2019)\citenamefont {Tian}, \citenamefont {Wang}, \citenamefont {Ji}, \citenamefont {Wang}, \citenamefont {Xia}, \citenamefont {Wang}, \citenamefont {Liu}, \citenamefont {Zhang},\ and\ \citenamefont {Cheng}}]{tian_domain_2019}%
  \BibitemOpen
  \bibfield  {author} {\bibinfo {author} {\bibfnamefont {C.-K.}\ \bibnamefont {Tian}}, \bibinfo {author} {\bibfnamefont {C.}~\bibnamefont {Wang}}, \bibinfo {author} {\bibfnamefont {W.}~\bibnamefont {Ji}}, \bibinfo {author} {\bibfnamefont {J.-C.}\ \bibnamefont {Wang}}, \bibinfo {author} {\bibfnamefont {T.-L.}\ \bibnamefont {Xia}}, \bibinfo {author} {\bibfnamefont {L.}~\bibnamefont {Wang}}, \bibinfo {author} {\bibfnamefont {J.-J.}\ \bibnamefont {Liu}}, \bibinfo {author} {\bibfnamefont {H.-X.}\ \bibnamefont {Zhang}},\ and\ \bibinfo {author} {\bibfnamefont {P.}~\bibnamefont {Cheng}},\ }\bibfield  {title} {\bibinfo {title} {{Domain wall pinning and hard magnetic phase in {Co}-doped bulk single crystalline {Fe$_3$GeTe$_2$}}},\ }\href {https://link.aps.org/doi/10.1103/PhysRevB.99.184428} {\bibfield  {journal} {\bibinfo  {journal} {Phys. Rev. B}\ }\textbf {\bibinfo {volume} {99}},\ \bibinfo {pages} {184428} (\bibinfo {year} {2019})}\BibitemShut {NoStop}%
\end{thebibliography}
%

\end{document}


\title{Supplemental Material for "Spectroscopic evidence of Kondo resonance in 3$d$ van der Waals ferromagnets"}
	
\author{Deepali Sharma\orcidlink{0009-0000-2309-4386}}
\affiliation{Department of Physics, Indian Institute of Science Education and Research Bhopal, Bhopal Bypass Road, Bhauri, Bhopal 462066, India}%

\author{Neeraj Bhatt\orcidlink{0009-0002-8012-139X}}
\affiliation{Department of Physics, Indian Institute of Science Education and Research Bhopal, Bhopal Bypass Road, Bhauri, Bhopal 462066, India}%

\author{Asif Ali\orcidlink{0000-0001-7471-8654}}
\affiliation{Department of Physics, Indian Institute of Science Education and Research Bhopal, Bhopal Bypass Road, Bhauri, Bhopal 462066, India}%

\author{Rajeswari Roy Chowdhury\orcidlink{0000-0001-9693-9255}}
\affiliation{Department of Physics, Indian Institute of Science Education and Research Bhopal, Bhopal Bypass Road, Bhauri, Bhopal 462066, India}%

\author{Chandan Patra\orcidlink{0009-0001-1338-4977}}
\affiliation{Department of Physics, Indian Institute of Science Education and Research Bhopal, Bhopal Bypass Road, Bhauri, Bhopal 462066, India}%

\author{Ravi Prakash Singh\orcidlink{0000-0003-2548-231X}}
\affiliation{Department of Physics, Indian Institute of Science Education and Research Bhopal, Bhopal Bypass Road, Bhauri, Bhopal 462066, India}%

\author{Ravi Shankar Singh\orcidlink{0000-0003-3654-2023}}
\email{rssingh@iiserb.ac.in}
\affiliation{Department of Physics, Indian Institute of Science Education and Research Bhopal, Bhopal Bypass Road, Bhauri, Bhopal 462066, India}%
		

\maketitle


\begin{figure}
	\centering
	\includegraphics[width=.78\textwidth]{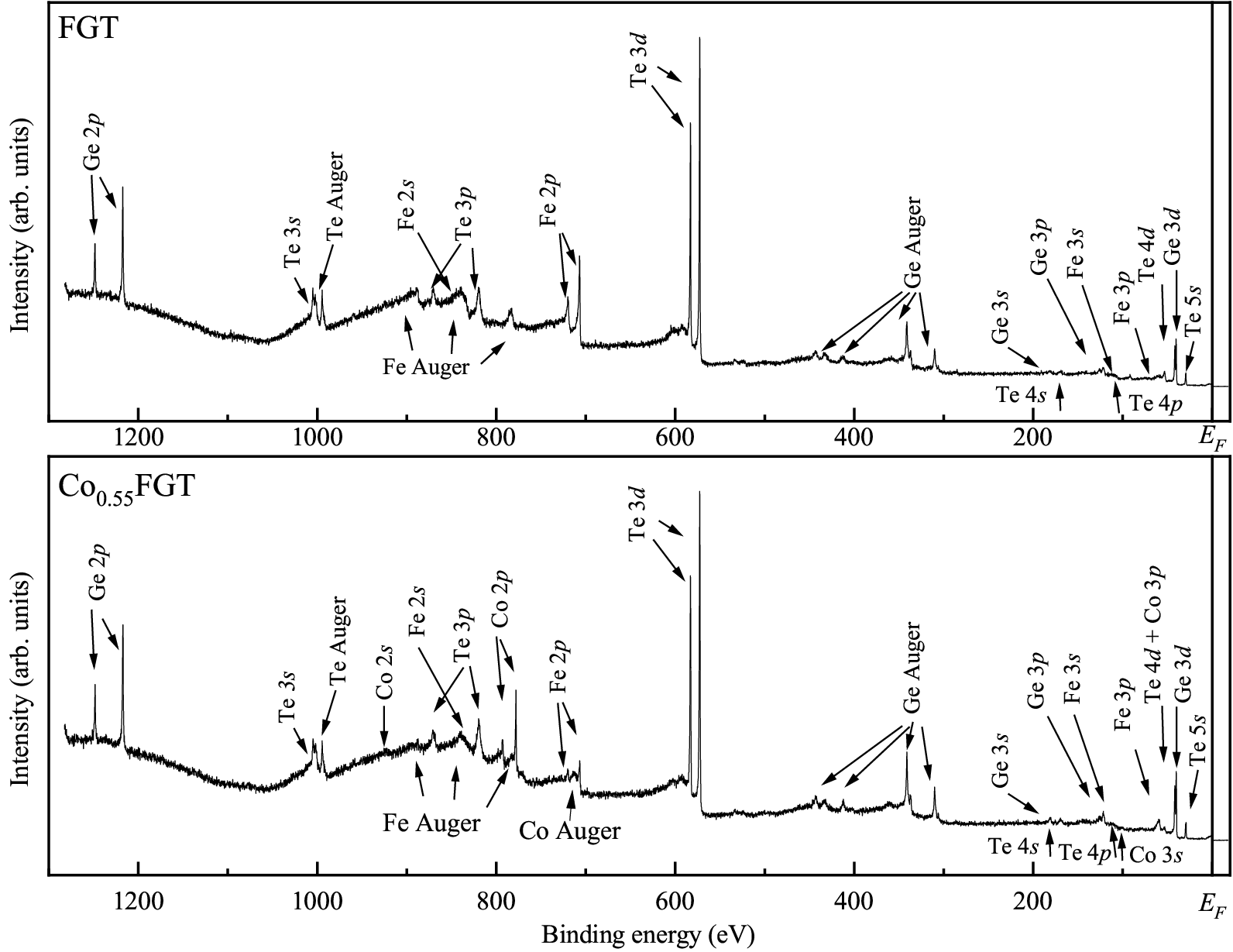}
	\caption{Survey scan of FGT and Co$_{0.55}$FGT collected using Al $K_{\alpha}$ radiation at 30 K.} \label{fig:SFig1}
\end{figure}

\textbf{X-ray photoemission survey scan} : \\

The survey scan of FGT and Co$_{0.55}$FGT collected using Al $K_{\alpha}$ radiation at 30 K is shown in Fig. \ref{fig:SFig1}. All the core levels and Auger features are marked with arrows. The high quality of the samples and clean surface obtained by \textit{in situ} cleaving are ascertained by the sharpness of peaks and absence of any oxide-related features in the core level spectra. \\

\textbf{X-ray photoemission core levels} : \\

\begin{figure}
	\centering
	\includegraphics[width=.55\textwidth]{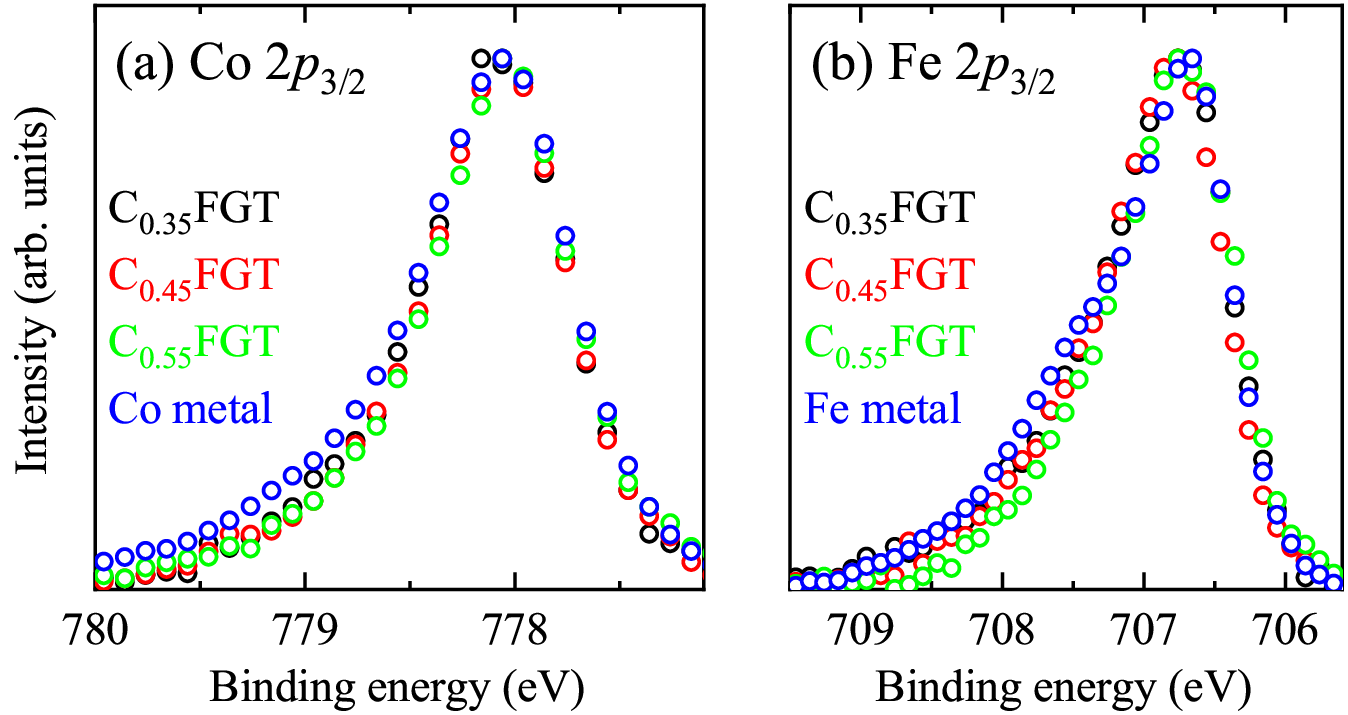}
    	\caption{Shirley-type integral background subtracted core level photoemission spectra of (a) Co 2\textit{p}$_{3/2}$, and (b) Fe 2\textit{p}$_{3/2}$, collected at 300 K using Al $K_{\alpha}$ for Co$_{x}$FGT, for \textit{x} = 0.35, 0.45 and 0.55. For comparison spectra corresponding to Co metal and Fe metal has also been shown.}\label{fig:SFig2}
\end{figure}

A Shirley-type integral background subtracted core level spectral region corresponding to Co 2\textit{p}$_{3/2}$ and Fe 2\textit{p}$_{3/2}$ for Co$_{x}$FGT collected using Al $K_{\alpha}$ radiation at 300 K are shown in Fig. \ref{fig:SFig2}. All the spectra were normalized by the peak intensity. The width and peak position of Co 2\textit{p}$_{3/2}$ core level remains very similar with increasing Co content in Co$_{x}$FGT. Notably, the Co 2\textit{p}$_{3/2}$ core level has slightly larger width of ferromagnetic cobalt metal as compared to Co$_{x}$FGT, this is attributed to the difference in exchange splitting (hence, magnetic moment) resulting in observed dichroism in the core level spectra, specially in the case of magnetic systems \cite{FGT,PhysRevB.55.11488,PhysRevResearch.3.013151,IndranilSarkar,Coexchangesplitt}. Similar to Co 2\textit{p}$_{3/2}$ core level, the Fe 2\textit{p}$_{3/2}$ core level spectra also remains very similar with increasing Co content in Co$_{x}$FGT. The width of Fe 2\textit{p}$_{3/2}$ core level spectra of ferromagnet iron shows slightly larger width as compare to Co$_{x}$FGT.\\

\textbf{Non-magnetic and ferromagnetic DFT results} :\\

 \begin{figure}[b]
	\centering
	\includegraphics[width=.55\textwidth]{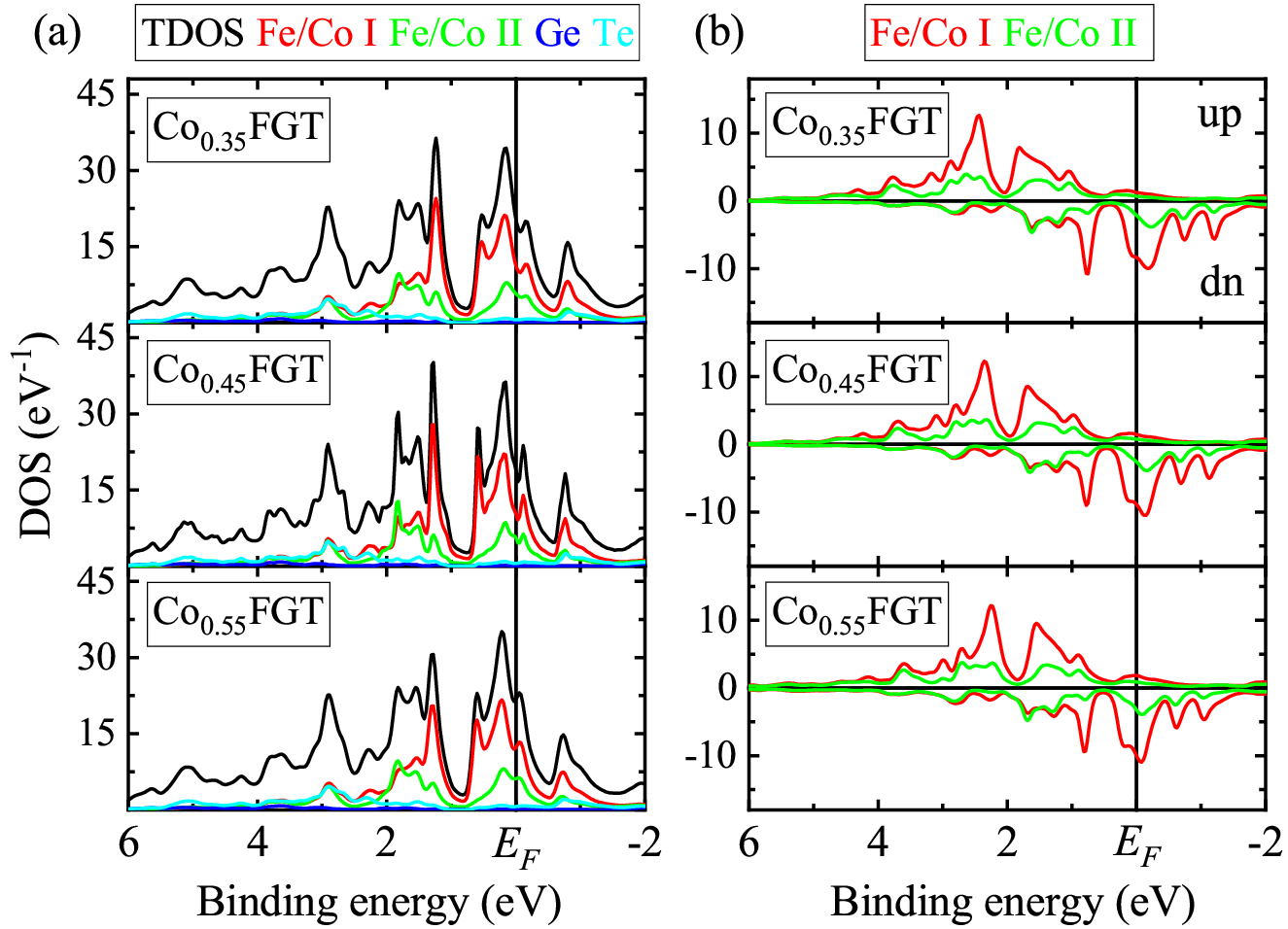}
	\caption{(a) Total DOS and partial DOSs of Fe/Co I 3$d$ and Fe/Co II 3$d$, Ge 4$p$ and Te 5$p$ obtained from non-magnetic DFT calculations, and (b) spin-polarized partial DOSs for Fe/Co I 3\textit{d} and Fe/Co II 3\textit{d} using ferromagnetic DFT calculations, for Co$_{x}$FGT.} \label{fig:SFig3}
\end{figure}

Figure \ref{fig:SFig3} (a) shows the total density of states (DOS) and partial DOSs of Fe/Co I 3$d$, Fe/Co II 3$d$, Ge 4$p$ and Te 5$p$ obtained from the non-magnetic (NM) DFT calculations for Co$_{x}$FGT. The Co substituted Co$_{x}$FGT reveals finite intensity at $E_{F}$, indicating metallic character. Predominant contribution of Fe/Co 3\textit{d} states appear within $\pm$ 2 eV BE, and Te 5\textit{p} states  contribution appearing at higher BE with negligibly small contribution of Ge 4\textit{p} states. The overall TDOS remain very similar with increasing \textit{x} in Co$_{x}$FGT, however, a closer look around $E_{F}$ (within 2 eV BE) reveals gradual shift in TDOS towards higher BE with increase in \textit{x}, suggesting rigid band shift with electron doping, in contrast to experimental observations, as shown in main text Fig. 1 (c). As expected, the total energy per unit cell reduces for the ferromagnetic (FM) phase for all the compounds, and the resulting spin-polarized 3\textit{d} states corresponding to Fe/Co I and Fe/Co II are shown in the Fig. \ref{fig:SFig3} (b) for Co$_{x}$FGT obtained from FM DFT calculations. Large exchange splitting of spin-polarized DOS is clearly evident for the inequivalent Fe/Co sites along with dominant contributions of 3\textit{d} states appearing beyond 2 eV BE, in contrast to the experimental results. Further, the obtained magnetic moment per f.u. of about 5.55 $\mu_B$, 5.03 $\mu_B$, and 4.44 $\mu_B$, for Co$_{0.35}$FGT, Co$_{0.45}$FGT, and Co$_{0.55}$FGT is an overestimation from the value obtained from magnetic measurements \cite{chowdhury_unconventional_2021}. \\ 

\begin{figure}
	\centering
	\includegraphics[width=.92\textwidth]{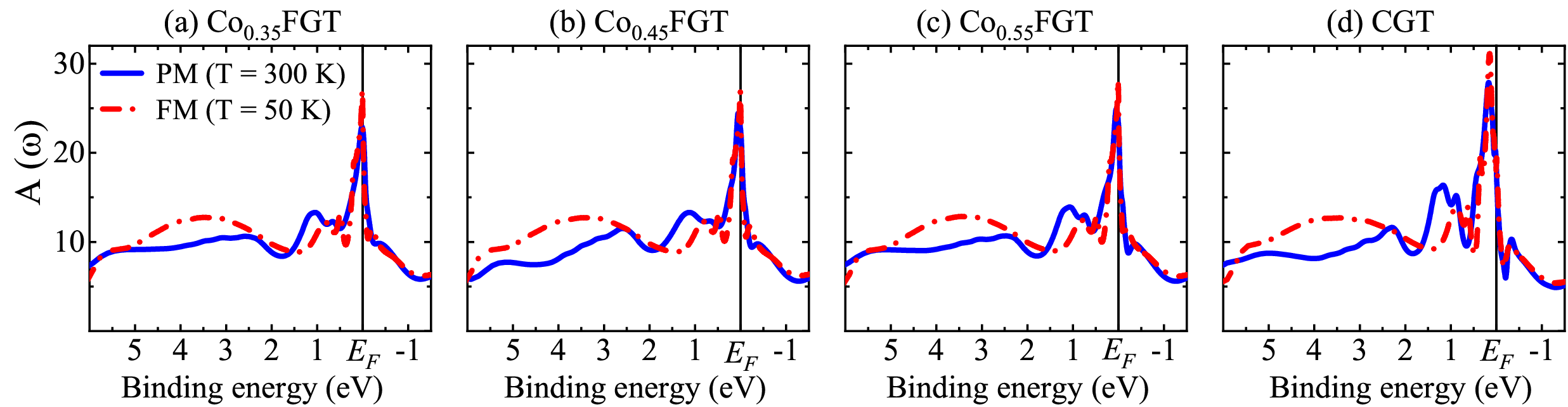}
	\caption{(a) Total spectral function from PM DFT+DMFT (\textit{T} = 300 K) and FM DFT+DMFT (\textit{T} = 50 K) for (a) Co$_{0.35}$FGT, (b) Co$_{0.45}$FGT, (c) Co$_{0.55}$FGT, and (d) CGT.} \label{fig:SFig4}
\end{figure}

\textbf{Spectral functions from paramagnetic and ferromagnetic DFT+DMFT} :\\

 \begin{figure}[b]
	\centering
	\includegraphics[width=.92\textwidth]{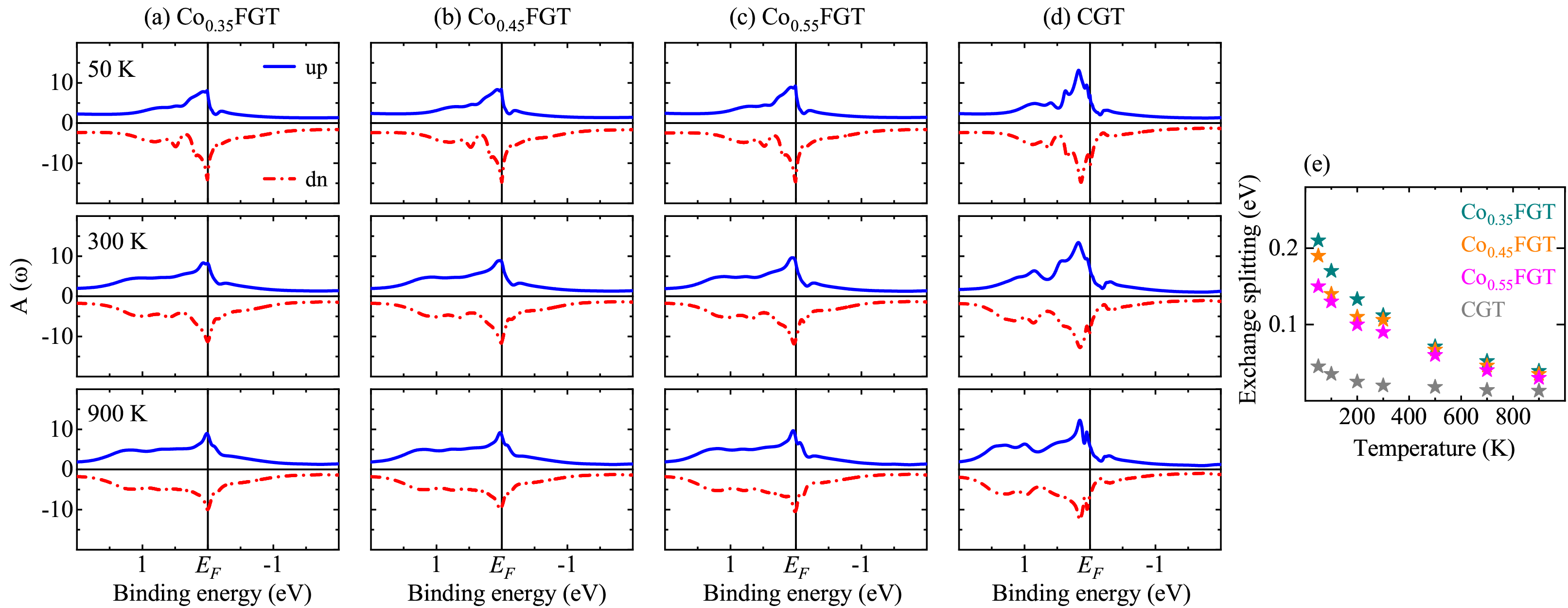}
	\caption{ Spin-polarized 3\textit{d} spectral functions for (a) Co$_{0.35}$FGT, (b) Co$_{0.45}$FGT, (c) Co$_{0.55}$FGT, and (d) CGT, at various temperatures using FM DFT+DMFT. (e) Exchange splitting at various temperatures for Co$_{x}$FGT.} \label{fig:SFig5}
\end{figure}

\begin{figure*}
\centerline{\includegraphics[width=0.7\textwidth]{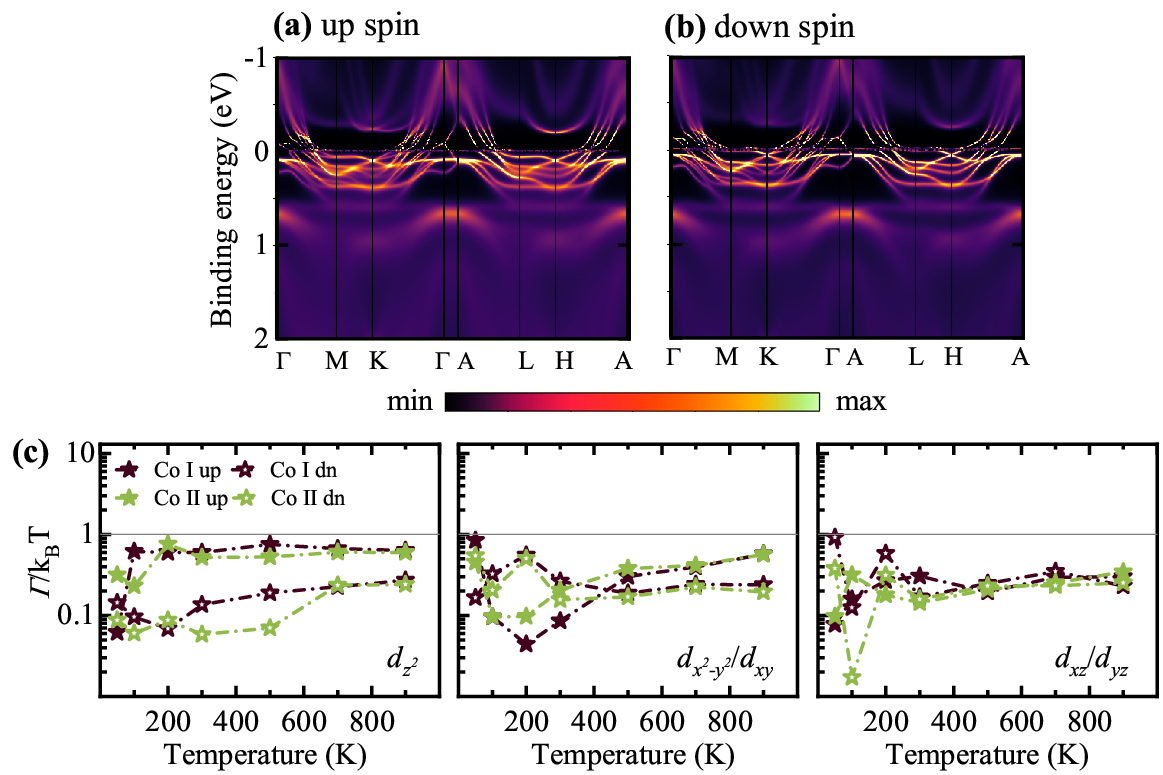}}
\caption{ $k$-resolved spectral functions for (a) spin-up and (b) spin-down, along high symmetry directions at \textit{T} = 50 K obtained using FM DFT+DMFT calculations for CGT. (c) $\varGamma$/\textit{k}$_{B}T$ for \textit{d}$_{z^2}$, $d_{{{x}^{2}}-{{y}^{2}}}$/$d_{xy}$ and $d_{xz}$/$d_{yz}$ orbital for both the spin channels and for both the non-degenerate Co sites at various temperatures. Spin-up and spin-down are shown in closed markers and open markers, respectively for Co I (brown) and Co II (green).}\label{fig:SFig6}
\end{figure*}

Total spectral function, A(\textit{w}) obtained from PM DFT+DMFT (\textit{T} = 300 K) and FM DFT+DMFT (\textit{T} = 50 K) for \textit{x} = 0.35, 0.45, and 0.55 in Co$_{x}$FGT has been shown in Fig. \ref{fig:SFig4} (a), (b), and (c), respectively. The overall spectral lineshape remains very similar for Co$_{x}$FGT and aligns well with no significant change observed in the valence band spectra across the magnetic transition (Fig. 1(c) of the main text). For comprehensive understanding, Co$_{3}$GeTe$_{2}$ (CGT) was also explored within DFT+DMFT framework (as shown in Fig. \ref{fig:SFig4} (d)) where, the total spectral function in PM phase revealed finite intensity at $E_{F}$ indicative of metallic character with predominant Co 3\textit{d} states appearing within $\pm$ 2 eV BE, similar to Co$_{x}$FGT. The overall spectral function in PM (as well as FM) phase remains very similar while going from Co$_{0.35}$FGT to CGT. However, the feature at $E_{F}$ shows a small shift towards higher binding energy accompanied by an increase in its intensity for the case of CGT. The local spin moment was found to be 3.92 $\mu_{B}$/f.u. for CGT in PM DFT+DMFT. The overall spectral lineshape remains very similar in PM and FM phases for CGT as well. 
Temperature dependent spin-polarized spectral functions from FM DFT+DMFT for Co$_{x}$FGT (\textit{x} = 0.35, 0.45, and 0.55) are shown in Fig. \ref{fig:SFig5} (a)-(c). The exchange splitting between the spin polarized bands decreases with increase in temperature where both the spin-polarized spectral functions tries to acquire a similar structure at high temperature. However, the exchange splitting remains finite even upto 900 K for all Co$_{x}$FGT. We estimated the exchange splitting in the Fe/Co bands from the energy difference of the center of the weight of spin-up and spin-down from \textit{k}-integrated spectral function and is shown in Fig. \ref{fig:SFig5} (e) revealing finite splitting even upto 900 K, followed by similar trend for magnetic moment. It is to be noted that for CGT, the PM and FM phases reveal similar ground state energy for all the temperatures. In FM phase at low temperatures both spin-up and spin-down channels show similar structure with very small exchange splitting and magnetic moment compared to Co$_{x}$FGT, which further reduces at high temperatures suggesting possible paramagnetic ground state for CGT. \\

\begin{figure*}[b]
\centerline{\includegraphics[width=0.52\textwidth]{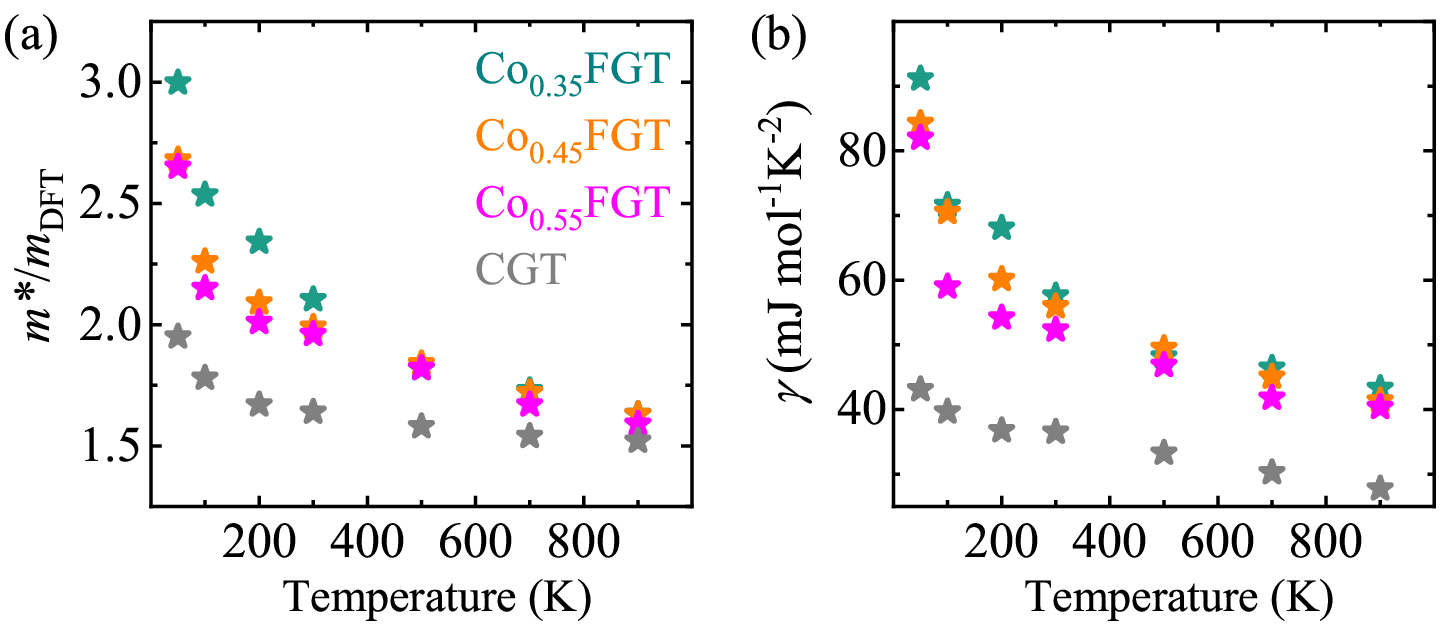}}
\caption{Calculated (a) mass enhancement factor (${m^*}$/${m_{\text{DFT}}}$) and (b) Sommerfeld coefficient ($\gamma$), at various temperatures for Co$_{x}$FGT using FM DFT+DMFT.}\label{fig:SFig7}
\end{figure*}

\textbf{k-resolved spectral function and quasiparticle scattering rate for CGT} :\\

$k$-resolved spin-polarized spectral function for CGT of both the spin channels along high symmetry directions obtained from FM DFT+DMFT calculations at \textit{T} = 50 K, is shown in Fig. \ref{fig:SFig6} (a) and (b). Both the spin bands overall look very similar and are  sharper/coherent in the vicinity of $E_F$ in CGT. Quasiparticle scattering rate, $\varGamma$ (inverse of the lifetime) was calculated, which revealed $\varGamma$ remains below $k_BT$ throughout, as shown in Fig. \ref{fig:SFig6} (c), suggesting CGT always remains in coherent regime for both the spins and all the orbitals.

\begin{figure*}
\centerline{\includegraphics[width=0.67\textwidth]{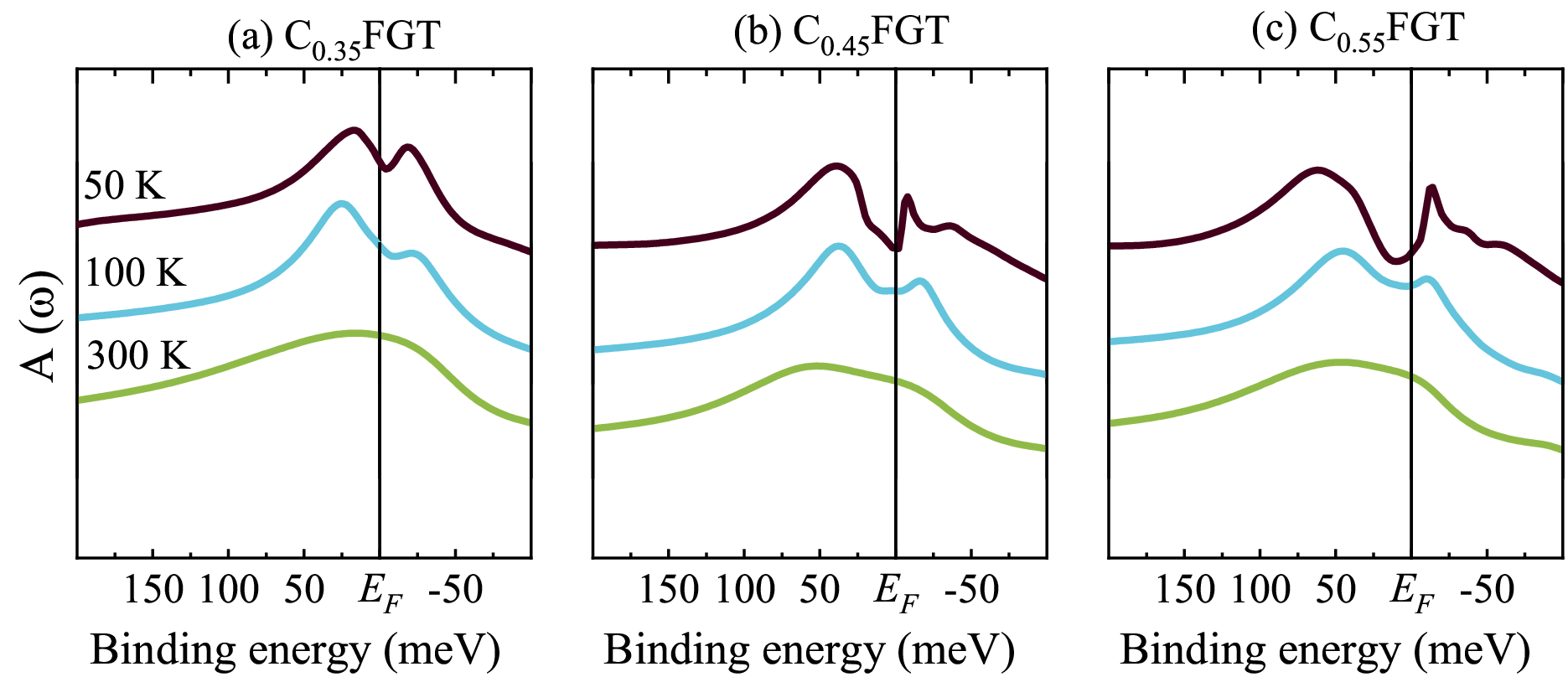}}
\caption{Temperature dependent spectral function for (a) Co$_{0.35}$FGT, (b) Co$_{0.45}$FGT, and (c) Co$_{0.55}$FGT, using PM DFT+DMFT.}\label{fig:SFig8}
\end{figure*}

Within DFT+DMFT framework, $\frac{m^*}{m_{\text{DFT}}}$ is weighted sum of contributions arising from all the orbital ($l$) and spin ($s$), weighted by their local Green’s function ($\propto$ partial DOS) at $E_{F}$. The $\frac{m^*}{m_{\text{DFT}}}$ obtained for all \textit{x} in Co$_{x}$FGT from FM DFT+DMFT calculations have been shown in Fig. \ref{fig:SFig7} (a), unveiling increasing mass enhancement with lowering temperature. The Sommerfeld coefficient$, \gamma_{_{\text{DMFT}}}$ is obtained by linear sum of each contribution, $\gamma_{_{l,s}}$, ($\gamma_{_{\text{DMFT}}}$ = $\sum_{l,s} \gamma_{_{l,s}}$ = $\sum_{l,s}$ [($\pi^2 k_B^2$/3) $\frac{m^*}{m_b}$$_{_{l,s}}$ PDOS$_{l,s}$ ] $\approx$  [($\pi^2 k_B^2$/3) $\frac{m^*}{m_\text{DFT}}$ A$(\omega)$] and show an increasing trend with lowering temperature, as shown in Fig. \ref{fig:SFig7} (b). Extrapolated values at 0 K \cite{FGT} for $\frac{m^*}{m_{\text{DFT}}}$ are 3.65, 3.27, 3.23 and 2.22; $\gamma_{_{\text{DMFT}}}$ are 111 mJ mol$^{-1}$ K$^{-2}$, 102.83 mJ mol$^{-1}$ K$^{-2}$, 99.9 mJ mol$^{-1}$ K$^{-2}$ and 49.1 mJ mol$^{-1}$ K$^{-2}$, for \textit{x} = 0.35, 0.45, 0.55 and 1.0, suggesting Co$_{x}$FGT to be heavy fermionic system. \\

\textbf{Evolution of DFT+DMFT spectral function} :\\

Fig. \ref{fig:SFig8} shows temperature dependent total spectral function obtained from PM DFT+DMFT for Co$_{x}$FGT. At 300 K, the spectral function for Co$_{0.35}$FGT remains relatively flat around $E_F$, as shown in Fig. \ref{fig:SFig8} (a). Upon lowering the temperature, the spectral function develops a dip like structure accompanied by a hump just above $E_F$, which becomes more pronounced at low temperature. Similar trend is also observed for Co$_{0.45}$FGT and Co$_{0.55}$FGT, as shown in Fig. \ref{fig:SFig8} (b) and (c), respectively. Interestingly, the feature above $E_F$ becomes more pronounced with increasing Co concentration. 
Similar behavior is well reflected in the experimental results where increasing \textit{x} in Co$_{x}$FGT, leading to higher Kondo temperature, and is manifested by pronounced Kondo resonance peak at low temperature in case of Co$_{0.55}$FGT due to enhanced screening and reduced magnetic moment. Our DFT+DMFT calculations quantitatively reproduces the experimental Kondo resonance in Co$_{x}$FGT, with the characteristic low-temperature peak emerging clearly in the spectral functions.


%